\documentclass[format=acmsmall, review=false, screen=true]{acmart}

\usepackage{booktabs} 
\usepackage{lmodern}
\usepackage{multirow}
\usepackage{subcaption}
\usepackage{xspace}
\usepackage{listings}
\usepackage{newfloat,caption}
\usepackage{color}
\usepackage{xcolor}

\usepackage[ruled]{algorithm2e} 

\SetAlFnt{\small}
\SetAlCapFnt{\small}
\SetAlCapNameFnt{\small}
\SetAlCapHSkip{0pt}
\IncMargin{-\parindent}

\newcommand{{\rnn}}{{\sc rnn}}
\newcommand{{\cnn}}{{\sc cnn-1d}}
\newcommand{{\cnntwo}}{{\sc cnn-2d}}

\newcommand{\cm}{\textit{complex method}}
\newcommand{\ecb}{\textit{empty catch block}}
\newcommand{\mn}{\textit{magic number}}
\newcommand{\ma}{\textit{multifaceted abstraction}}

\newcommand{\MA}{\textit{Multifaceted abstraction}}
\newcommand{\etal}{\textit{et al.}}
\usepackage{xcolor}

\lstdefinestyle{sharpc}{language=[Sharp]C, frame=lr, rulecolor=\color{blue!80!black}}
\DeclareFloatingEnvironment[fileext=frm,placement={!ht},name=Listing,within=section]{listing}

\acmJournal{TOSEM}
\acmVolume{1}
\acmNumber{1}
\acmYear{2019}
\acmMonth{5}

\received{May 2019}

\begin{document}
\title[On the Feasibility of Transfer-learning Code Smells using Deep Learning] {On the Feasibility of Transfer-learning Code Smells using Deep Learning}

\author{Tushar Sharma}
\orcid{1234-5678-9012-3456}
\affiliation{%
  \institution{Athens University of Economics and Business}
  \streetaddress{76, Patission street}
  \city{Athens}
  \country{Greece}}
\email{tushar@aueb.gr}
\author{Vasiliki Efstathiou}
\orcid{1234-5678-9012-3456}
\affiliation{%
	\institution{Athens University of Economics and Business}
	\streetaddress{76, Patission street}
	\city{Athens}
	\country{Greece}}
\email{vefstathiou@aueb.gr}
\author{Panos Louridas}
\orcid{1234-5678-9012-3456}
\affiliation{%
	\institution{Athens University of Economics and Business}
	\streetaddress{76, Patission street}
	\city{Athens}
	\country{Greece}}
\email{louridas@aueb.gr}
\author{Diomidis Spinellis}
\orcid{1234-5678-9012-3456}
\affiliation{%
	\institution{Athens University of Economics and Business}
	\streetaddress{76, Patission street}
	\city{Athens}
	\country{Greece}}
\email{dds@aueb.gr}

\begin{abstract}
\textbf{Context:}
An excessive number of code smells make a software system hard to evolve and maintain.
Machine learning methods, in addition to metric-based and heuristic-based methods,
have been recently applied to detect code smells;
however, current practices are considered far from mature.  
\\
\textbf{Objective:}
First, explore the feasibility of applying deep learning models to detect smells
without extensive feature engineering.
Second, investigate the possibility of applying transfer-learning in the context of detecting code smells.
\\
\textbf{Method:}
We train smell detection models based on Convolution Neural Networks
and Recurrent Neural Networks as their principal hidden layers.
For the first objective, we perform training and evaluation on C\# samples,
whereas for the second objective, we train the models from C\# code and
evaluate the models over Java code samples and vice-versa.
\\
\textbf{Results:}
We find it feasible to detect smells using deep learning methods though
performance of the models is smell-specific. 
Our experiments show that transfer-learning is definitely feasible for implementation
smells with performance comparable to that of direct-learning. 
This work opens up a new paradigm to detect code smells by transfer-learning
especially for the programming languages where the comprehensive code smell
detection tools are not available.

\end{abstract}

%
%


%
%


\maketitle

\renewcommand{\shortauthors}{Sharma \etal{}}

\section{Introduction} 

The metaphor of code smells is used to indicate the presence of quality issues in
source code \cite{Fowler1999, Sharma2018b}.
A large number of smells in a software system is associated with a high level of
technical debt \cite{Philippe2012}
hampering the system's evolution.
Given the practical significance of code smells, software engineering researchers have studied
the concept in detail and explored various aspects associated with it including
causes, impacts, and detection methods \cite{Sharma2018b}.

A large body of work has been carried out to detect smells in source code.
Traditionally, metric-based \cite{Radu2005, Mazeiar2006} and
rule/heuristic-based \cite{Naouel2010, Tushar2016} smell detection techniques
are commonly used  \cite{Sharma2018b}.
In recent years, machine-learning based smell detection 
techniques \cite{Abdou2012, Gabriela2015} have emerged
as a potent alternative
as they not only have the potential to bring human judgment
in the smell detection but also
provide the grounds for transferring results from one problem to another.
Researchers have used Bayesian belief networks \cite{Foutse2009b, Foutse2011},
support vector machines \cite{Abdou2012b}, and binary logistic regression \cite{Sergio2010}
to identify smells. 

In particular, transfer-learning refers to the technique where
a learning algorithm exploits the commonalities between different learning tasks 
to enable knowledge transfer across the tasks \cite{Bengio2013}.
In this context, it would be interesting to explore the possibility of leveraging the availability of tools and data related to code smell detection in a programming 
language in order to train machine learning models that address the same problem on another language. 
The cross-application of a machine learning model could provide opportunities for 
detecting smells without actually developing a language-specific smell detection tool from scratch.

Despite the potential prospects, existing approaches for applying machine learning
techniques for smell detection are considered far from mature. 
In a recent study, Di Nucci \etal{} \cite{Nucci2018} note that the problem of
detecting smells still requires extensive research to attain a maturity that 
would produce results of practical use. 
In addition, machine learning techniques 
(such as Bayesian networks, support vector machines, and logistic regression)
that have been applied so far require considerable
pre-processing to generate features for the source code, 
a substantial effort that hinders their adoption in practice. 
Traditionally, researchers use machine-learning methods that require
extracting feature-sets from source code.
Typically, code metrics are used as the feature set for smell detection purposes.
We perceive two shortcomings in such usage of machine-learning methods
for detecting smells.
First, we need an external tool to compute metrics for the target programming
language on which we would like to apply the machine learning model.
Those that have a metrics computation tool may deduce many smells directly by
combining these metrics \cite{Radu2004, Sharma2018b}
and thus applying a machine-learning method is redundant.
Second and more importantly, we are limiting the machine learning algorithm to
use only the metrics that we are computing and feeding as feature-set.
Therefore, the machine learning algorithm cannot observe any pattern that is
not captured by the provided set of metrics.


In this context, deep learning models, specifically neural networks, offer a compelling alternative.
The Convolution Neural Network ({\sc cnn}) and the Recurrent Neural Network (\rnn{}) are 
state-of-the-art supervised learning methods currently employed in many practical applications,
including image recognition \cite{Krizhevsky2012, Szegedy2015},
voice recognition and processing \cite{Sainath2015},
and natural language processing \cite{Johnson2015}.
These advanced models are capable of inferring features during training and can learn to
classify samples based
on these inferred features.

In this paper, we present experiments with deep learning models
with two specific goals:
\begin{itemize}
	\item To investigate whether deep learning methods\textemdash particularly  architectures
	that include layers of {\sc cnn}s and {\rnn{}}s\textemdash can effectively detect code smells.
	In addition, how different models perform on detecting diverse code smells and how model
	performance is affected by tweaking the learning hyper-parameters.
	\item To investigate whether results on smell detection through deep learning are transferable; specifically, to explore whether models trained for detecting  smells on a programming language can be re-used to detect smells on another language. 

\end{itemize}

Keeping these goals in mind, we define research questions and 
prepare an experimental setup to detect four smells
\textit{viz.} \cm{}, \ecb{}, \mn{}, and \ma{}
using deep learning models in different configurations.
We develop a set of tools and scripts to automate the experiment and collate the results.
Based on the results, we derive conclusions to our addressed research questions.

The contributions of this paper are summarized below.
\begin{itemize}
	\item An extensive study that applies deep learning models in detecting code smells and
	compares the performance of different methods; to the best of our knowledge this is the
	first study of this kind and scale.
	\item An exploration regarding the feasibility of employing deep learning models in transfer-learning.
	This exploration potentially will open a new paradigm to detect smells
	specifically for programming languages for which comprehensive code smells
	detection tools are not available.
	\item Openly available tools, scripts, and data used in this
	experiment\footnote{\url{https://github.com/tushartushar/DeepLearningSmells}}
	to promote replication as well as extension studies.
\end{itemize}

The rest of the paper is organized as follows.
Section \ref{sec:background} sets up the stage by presenting background and related work.
We define our research objective in Section \ref{sec:objectives} and research method in
Section \ref{sec:method}.
Section \ref{sec:results} presents our findings, discussion, and further research opportunities.
We present threats to validity of this work in Section \ref{sec:threats} and conclude in Section \ref{sec:conclusions}.

\section{Background and Related Work}
\label{sec:background}
In this section, we present the background about the topic of code smells as well as machine
learning and elaborate on the related literature.

\subsection{Code Smells}
\label{sec:smells}
Kent Beck coined the term ``code smell'' \cite{Fowler1999}
and defined it as \textit{``certain structures
	in the code that suggest (or sometimes scream) for refactoring.''}
Code smells indicate the presence of quality problems impacting
many facets of quality \cite{Sharma2018b}
of a software system \cite{Fowler1999, Girish2014}.
The presence of an excessive number of smells in a software system
makes it hard to maintain and evolve.

Smells are categorized as implementation \cite{Fowler1999},
design \cite{Girish2014}, and architecture smells \cite{Joshua2009b}
based on their scope, granularity, and impact.
Implementation smells are typically confined to a limited scope and impact (\textit{e.g.,} a method).
Examples of implementation smells are \textit{long method}, 
\cm{}, \textit{long parameter list},
and \textit{complex conditional }\cite{Fowler1999}.
Design smells occur at higher granularity, \textit{i.e.,} abstractions, and hence
are confined to a class or a set of classes.
\textit{God class}, \ma{}, \textit{cyclic-dependency modularization}, 
and \textit{rebellious
hierarchy} are examples of design smells \cite{Girish2014}.
Along the similar lines, architecture smells span across multiple components and
have a system-wide impact.
Some examples of architecture smells are \textit{god component} \cite{Lippert2006}, 
\textit{feature concentration} \cite{Hugo2014},
and \textit{scattered functionality} \cite{Joshua2009}.

A plethora of work related to code smell detection exists in the software engineering literature. 
Researchers have proposed methods for detecting smells that can be largely
divided into five categories \cite{Sharma2018b}.
\textit{Metric-based smell detection methods} \cite{Radu2005, Santiago2014, Mazeiar2006}
take source code as input, prepare a source code model,
such as an Abstract Syntax Tree ({\sc ast}),
compute a set of source code metrics, and detect smells by applying 
appropriate thresholds \cite{Radu2005}.
\textit{Rule/Heuristic-based smell detection} methods  \cite{Naouel2010, Tushar2016, Venera2013, Nikolaos2011}
typically take source code models  
and sometimes additional software metrics
as inputs.
They detect a set of smells when the defined rules/heuristics get satisfied.
\textit{History-based smell detection} methods use 
source code evolution information \cite{Fabio2015c, Shizhe2015}.
Such methods extract structural information of the code and how it has changed over a period of time.
This information is used by a detection model to infer smells in the code.
\textit{Optimization-based smell detection}
approaches \cite{Dilan2014, Ali2015, Wael2014}
apply an optimization algorithm on 
computed software metrics
and, in some cases, existing examples of smells
to detect new smells in the source code.

In recent times, \textit{machine learning-based smell detection} methods have
attracted software engineering researchers.
Machine learning is a subfield of artificial intelligence
that \textit{trains}
solutions to problems rather than modeling them through hard-coded rules. 
In this approach, the rules that solve a problem are not set a-priori; 
rather, they are inferred in a data-driven manner. 
In supervised learning,
a model is trained by being exposed to examples 
of instances of the problem along with their expected answers and 
statistical regularities are drawn. 
The representations that are learned from 
the data can in turn be applied and generalized to new, unseen data in a similar context.

A typical machine learning smell detection method starts with a mathematical model representing the
smell detection problem.
Existing examples and source code models
are used to train the model.
The trained model is used to classify or predict the code fragments into smelly or non-smelly
instances.
Foutse \etal{} \cite{Foutse2009b, Foutse2011} use a Bayesian approach for the detection
of smells.
Their study forms a Bayesian graph using a set of metrics
and determines the probability whether a class
belongs to a smell or not.
Similarly, Abdou \etal{} \cite{Abdou2012, Abdou2012b} employ support vector machine-based
classifiers trained using a set of $60$
object-oriented metrics for each class
to detect design smells (\textit{blob}, \textit{feature concentration},
\textit{spaghetti code}, and \textit{swiss army knife}).
Furthermore, S{\'e}rgio \etal{} \cite{Sergio2010} detect \textit{long method} smell instances
by employing binary logistic
regression.
They use commonly used method metrics, such as  Method Lines of Code ({\sc mloc}) and cyclomatic
complexity as regressors.
Bardez \etal{} \cite{Barbez2019} presents an ensemble method that combine outcome
of multiple tools to detect \textit{god class} and \textit{feature envy} smells.
They identify a set of key metrics for each smell and feed them to a {\sc cnn}-based architecture.
Fontana \etal{} \cite{Fontana2016} compare performance of 
various machine learning algorithms
in detecting \textit{data class, god class, feature envy}, and \textit{long method}.

Performance of a machine learning task heavily depends on the formation of the evaluation
samples.
As the ratio of positive and negative samples becomes more balanced, the classification task
of the models
becomes easier.
Hence, a network would perform significantly better when classifying data from balanced datasets.
Most of the above-mentioned approaches do not explicitly mention the ratio of positive and negative
samples used for the evaluation.
Fontana \etal{} \cite{Fontana2016} carry out the evaluation using $140$ positive and $280$ negative
samples for each smell which is considerably balanced compared to a realistic case.
The realistic samples that we used for evaluation are highly imbalanced (refer Section \ref{sec:results});
for example, the ratio of negative over positive samples for {\sc 1d} evaluation 
is $182$ to $1$ for \ma{} smell.

\subsection {Deep Learning} 
\label{sec:ml}

Deep learning is a subfield of machine learning that allows computational models 
composed of multiple processing layers to learn representations of 
data with multiple levels of abstraction \cite{Lecun2015, Goodfellow2016}. 
Even though the idea of layered neural networks with internal ``hidden'' units 
was already introduced in the 80s \cite{Rumelhart1986}, a breakthrough in 
the field came in 2006 by Hinton \etal{} \cite{Hinton2006} who introduced 
the idea of learning a hierarchy of features one level at a time. 
Ever since, and particularly during the course of the last decade, 
the field has taken off due to the advances in hardware, the release of benchmark 
datasets \cite{imagenet_cvpr09, cifar2009, mnist2010},  
and a growing research focus on optimization methods \cite{Martens2010,Kingma2014}. 
Although deep learning architectures often consist of tens or hundreds 
of successive layers, 
much shallower architectures may also fall under the category of deep learning, 
as long as at least one hidden layer exists between the input and the output layer. 

Deep learning architectures are being used extensively for addressing a multitude 
of detection, classification, and prediction problems. 
Architectures involving layers of
{\sc cnn}s are inspired by the hierarchical organization of the visual  
cortex in animals, which consists of alternating layers of simple and complex
cells  \cite{Felleman1991, Hubel1962}. 
{\sc cnn}s have been proven particularly effective for problems of optical 
recognition and are widely used for image classification and detection \cite{Krizhevsky2012, Szegedy2015, Lecun1998},  
segmentation of regions of interest in biological images \cite{Kraus2016},
and face recognition \cite{Lawrence1997, Parkhi2015}.  
Besides recognition of directly interpretable visual features 
of an image, {\sc cnn}s have also been used for pattern 
recognition in signal spectograms, with applications in speech recognition \cite{Sainath2015}. 
In these applications the input data are given in the form of matrices ({\sc 2d} arrays) 
for representing the {\sc 2d} grid layout of pixels in an image. 
{\sc 1d} representations of data have been used for applying {\sc 1d} convolutions in sequential 
data such as textual patterns \cite{Johnson2015} or temporal event patterns \cite{Lee2017, Abdeljaber2017}. 
However, when it comes to sequential data, Recurrent Neural Networks ({\rnn{}}s)
\cite{Rumelhart1986} have been proven superior due to their capability 
to dynamically  ``memorize'' information provided in previous states and 
incorporate it to a current state. 
Long Short Term Memory ({\sc lstm}) networks are a special kind of
 \rnn{}
that can connect information spanning long-term intervals, thus capturing long-term 
dependencies. 
{\sc lstm}s have been found to perform reasonably well on various data sets  
within the context of representative applications that exhibit sequential patterns, 
such as  speech recognition and music modeling \cite{Greff2017,Graves2013}. 
In addition, they have been established as state-of-the-art networks for 
a variety of natural language processing tasks; indicative applications include 
natural language generation \cite{Wen2015}, sentiment classification \cite{Wang2016,Baziotis2017} and neural machine translation \cite{Cho2014}, 
among others. 




\subsection{Machine Learning Techniques on Source Code}
\label{sec:mlcode}
The emergence of online open-source repository hosting platforms such as GitHub in recent years  
has led to an explosion on the volumes of openly available source code along with metadata 
related to software development activities; 
this bulk of data is often referred to as ``Big Code'' \cite{Allamanis2018}. 
As an effect, software maintenance activities have
started leveraging the
wealth of openly available data, the availability of computational 
resources, and the recent advances in machine learning research.  
In this context, statistical regularities observed in source code 
have revealed the repetitive and predictable nature of programming languages, 
which has been compared to that of natural languages \cite{Hindle2012, Ernst2017}. 
To this end, problems of automation in natural language processing,  
such as identification of semantic similarity between texts, translation, 
text summarisation, word prediction and language generation have been 
examined in parallel with the automation of software development tasks. 
Relevant problems in software development include 
clone detection \cite{White2016, Wei2017}, 
de-obfuscation \cite{Vasilescu2017},  
language migration \cite{Nguyen2013}, 
source code summarisation \cite{Iyer2016}, 
auto-correction \cite{Pu2016, Gupta2017},  
auto-completion \cite{Foster2012}, 
generation \cite{Oda2015, Ling2016, Yin2017}, 
and comprehension \cite{alexandru2017}. 

On a par with equivalent problems in natural language processing, 
the methods employed to address these software engineering problems have switched from 
traditional rule-based and probabilistic n-gram models to deep learning 
methods. 
The majority of the proposed deep learning solutions rely on the use of {\rnn{}}s 
which provide sophisticated mechanisms for capturing long term dependencies 
in sequential data, and specifically
{\sc lstm}s \cite{Hochreiter1997} 
that have demonstrated particularly effective performance on 
natural language processing problems. 

Alternative approaches to mining source code have employed {\sc cnn}s in 
order to learn features from source code. 
Li \etal{} \cite{Li2017} have  used a single-dimension {\sc cnn}s to learn 
semantic and structural features of programs by working at 
the {\sc ast} level of granularity and combining the learned features with 
traditional hand-crafted features to predict software defects. 
This method however incorporates hand-crafted features in the learning 
process and is not proven to yield transferable results.
Similarly, a one-dimensional  {\sc cnn}-based architecture
has been used by Allamanis \etal{} \cite{Allamanis2016} 
in order to detect patterns in source code and identify ``interesting'' locations 
where attention should be focused.
The objective of the study is to predict 
short and descriptive names of source code snippets (\textit{e.g.,} a method body) 
given solely its tokens. 
{\sc cnn}s have also been used by Huo \etal{} \cite{Huo2016} in order to address 
the problem of bug localization. This approach leverages both the 
lexical information expressed in the natural language of a bug report and 
the structural information of source code in order to learn unified features. 
A more coarse-grain approach that also employs {\sc cnn}s has been proposed in the 
context of program comprehension \cite{Ott2018} where the authors use imagery 
rather than text in order to discriminate between scripts written in two 
programming languages, namely Java and Python.

\subsection{Challenges in Applying Deep Learning on Source Code}
Applying deep learning techniques on source code is non-trivial.
In this section, we present challenges that we face in the process of applying deep
learning techniques on source code.

\subsubsection{Analogies with other problems}
Deep learning is advancing rapidly in domains that address problems of 
image, video, audio, text, and speech processing \cite{Lecun2015}. 
Consequently, these advances drive current trends in deep learning and 
inspire applications across disciplines. 
As such, studies that apply deep learning on source code rely heavily on results 
from these domains, and particularly that of text mining. 

Based on prior observations that demonstrate similarity between 
source code and natural language \cite{Hindle2012}, 
the research community has largely addressed relevant problems on
mining source code
by adopting latest state-of-the-art natural language processing 
methods \cite{Allamanis2016,Palomba2016,Iyer2016,Vasilescu2017,Yin2017}. 
However, besides similarities, there also exist major differences 
that need to be taken into consideration when designing such studies. 
First of all, source code, unlike natural language, is semantically 
fragile; minor syntactic changes can drastically change the meaning of code 
\cite{Allamanis2018}.
As an effect, treating code as text by ignoring the underlying formal 
semantics carries the risk of not preserving the appropriate meaning. 
Besides formal semantics, the syntax of source code obviously presents 
substantial differences compared to the syntax found in text. 
As a result, methods that perform well on text are likely to under-perform 
on source code. 
Architectures involving
\cnn{} layers, for instance, have been proven effective for matching 
subsequences of short lengths \cite{Chollet2017}, which are often found in natural 
language where the length of sentences is limited. This however 
does not necessarily apply on self-contained fragments of source code, such as method 
definitions, which tend to be longer. 
Finally, even though good practices dictate naming conventions in coding, 
unlike natural language, there is no universal vocabulary of source code. 
This results to a diversity in the artificial vocabulary found in source code 
that may affect the quality of the models learned. 

Approaches that treat code as text mainly focus on the mining of 
sequential patterns of source code tokens. 
Other emerging approaches look into structural characteristics of the code with 
the objective of extracting visual patterns delineated on code \cite{Ott2018}. 
Even though there are features in source code, such as nesting, which demonstrate 
distinctive visual patterns, treating source code in terms of such patterns and 
ignoring the rich intertwined semantics carries the risk of oversimplifying the 
problem.


\subsubsection {Lack of Resources}
Research employing deep learning techniques on software engineering data, including 
source code as well as other relevant artifacts, is still young. 
Consequently, results against traditional baseline techniques are very 
limited \cite{Fu2017, Hellendoorn2017}.
Especially when it comes to processing solely source code artifacts, 
relevant studies are scarce and mostly address the problem of 
drawing out semantics related to the functionality of a piece of code 
\cite{Allamanis2016,White2015,White2016,Mou2016,Piech2015}. 
To the best of our knowledge, our study is the first to thoroughly investigate 
the application of deep learning techniques with the objective of 
examining characteristics of source code quality. 
Therefore, a major challenge in studies of this kind is that there is 
no prior knowledge that would guide this investigation, 
a challenge reflected on all stages of the inquiry. 
At the level of designing an experiment, there exist no rules of thumb 
indicating a set up for a deep learning architecture that adequately 
models the fine-grained features required for the problem in hand. 
Furthermore, at the level of training a model, there is no prior baseline  
for hyper-parameters that would lead to an optimal solution. 
Finally, at the level of evaluating a trained model, there exist no benchmarks 
to compare against; 
there is no prior concrete indication on the expected outcomes in terms of 
reported metrics. 
Hence, a result that would appear sub-optimal in another domain such as 
natural language processing, may actually account for a significant advance 
in software quality assessment. 

Besides challenges that relate to the know-how of applying deep learning 
techniques on source code, there are technical difficulties that arise due  
to the paucity of curated data in the field. 
The need for openly available data that can serve for replicating data-driven 
studies in software engineering has been long stressed \cite{Robles2010}. 
The release of curated data in the field is encouraged through 
badging artifact-evaluated papers as well as dedicated data showcase venues 
for publication. 
However, the software engineering domain is still far from providing 
benchmark datasets, whereas the available datasets are limited to curated 
collections of repositories with associated metadata that lack 
ground truth annotation that is essential for a multitude of supervised 
machine learning tasks. Therefore, unlike domains such as image processing  
and natural language processing where an abundance of annotated data exist 
\cite{cifar2009,imagenet_cvpr09,mnist2010,imdb2011}, in the field of 
software engineering the lack of gold standards induces the inherent difficulty 
of collecting and curating data from scratch.

\section{Research Objectives}
\label{sec:objectives}
The goal of this research is to explore the possibility of applying
state-of-the-art deep learning methods to detect smells.
Further, within the same context, this work inquires into the feasibility of applying transfer-learning.
Based on the stated goals, we define the following research questions that this 
work aims to explore. 
\begin{description}
	\item[RQ1] Is it possible to use deep learning methods to detect code smells? 
	If yes, which deep learning method performs superior?
\end{description}
We use {\sc cnn} and \rnn{} models
in this exploration.
For the {\sc cnn}-based architecture,
we provide input samples in {\sc 1d} and {\sc 2d} format to observe the difference
in their capabilities due to the added dimension;
we refer to them as \cnn{} and \cnntwo{} respectively.
In the context of this research question, we define the following hypotheses.

\begin{description}
	\item[RQ1.H1]\textit{ It is feasible to detect smells using deep learning methods.}\\
	The considered deep learning models are powerful mechanisms that have the ability to detect
	complex patterns with sufficient training.
	These models have demonstrated high performance in the domain of 
	image processing \cite{Krizhevsky2012, Szegedy2015}
	and natural language processing \cite{Luong2015}.
	We believe we can leverage these models in the presented context.
	
	\item [RQ1.H2] \textit{\cnntwo{} performs better than \cnn{} in
		the context of detecting smells.}\\
	The rationale behind this hypothesis is the added dimensionality in \cnntwo{}.
	The {\sc 2d} model might observe inherent patterns when input data is presented in 
	two dimensions that may possibly be hidden in one dimensional format.
	For instance, a {\sc 2-d} variant could possibly identify the nesting depth of a method
	easier than its {\sc 1-d} counterpart when detecting \cm{} smell.
	
	\item [RQ1.H3] \textit{An \rnn{} model performs better than {\sc cnn} models in the smell detection context.}\\
	{\rnn{}} are considered better for capturing sequential patterns
	and have the reputation to work well with text. 
	Thus, taking into account the similarities that source code and natural language share, \rnn{} could prove superior than {\sc cnn} models.
	
\end{description}

\begin{description}
	\item[RQ2] Is transfer-learning feasible in the context of detecting smells?
	If yes, which deep learning model exhibits superior performance in
	detecting smells when applied in transfer-learning setting?
\end{description}

Transfer-learning is the capability of an algorithm to exploit the similarities between different
learning tasks and offering a solution of a task by transferring knowledge acquired while
solving another task. 
We would like to explore whether it is feasible to train a deep learning model 
from samples of C\# and predict the smells using this trained model in samples of Java
programming language and vice-versa.

We derive the following hypotheses.
\begin{description}
	\item[RQ2.H1] \textit{It is feasible to apply transfer-learning in the context of code smell detection.}\\
	We train the deep learning models using C\# code fragments
	and evaluate the trained model using Java fragments.
	Given the high similarity in the syntax between the two programming languages, 
	we believe that we may train the model
	from training samples and use the trained model to classify smelly and non-smelly
	fragments from our evaluation samples. 
	\item[RQ2.H2] \textit{Transfer-learning performs inferior compared to direct-learning.}\\
	Direct-learning in the context of our study refers to the case where training and evaluation samples belong to
	the same programming language.
	We expect that the performance of the models in the transfer-learning could be
	inferior to that compared to direct-learning given both the problems are equally hard
	\textit{i.e.,} negative and positive sample showing similar distribution.
%
\end{description}

\section{Research Method}
\label{sec:method}
This section describes the employed research method by
first providing an overview and
then elaborating on the data curation process.
We also discuss the selection protocol of smells and 
architecture of the deep
learning models.

\subsection{Overview of the Method}

\begin{figure}[h]
	\centering
	\includegraphics[width=\linewidth]{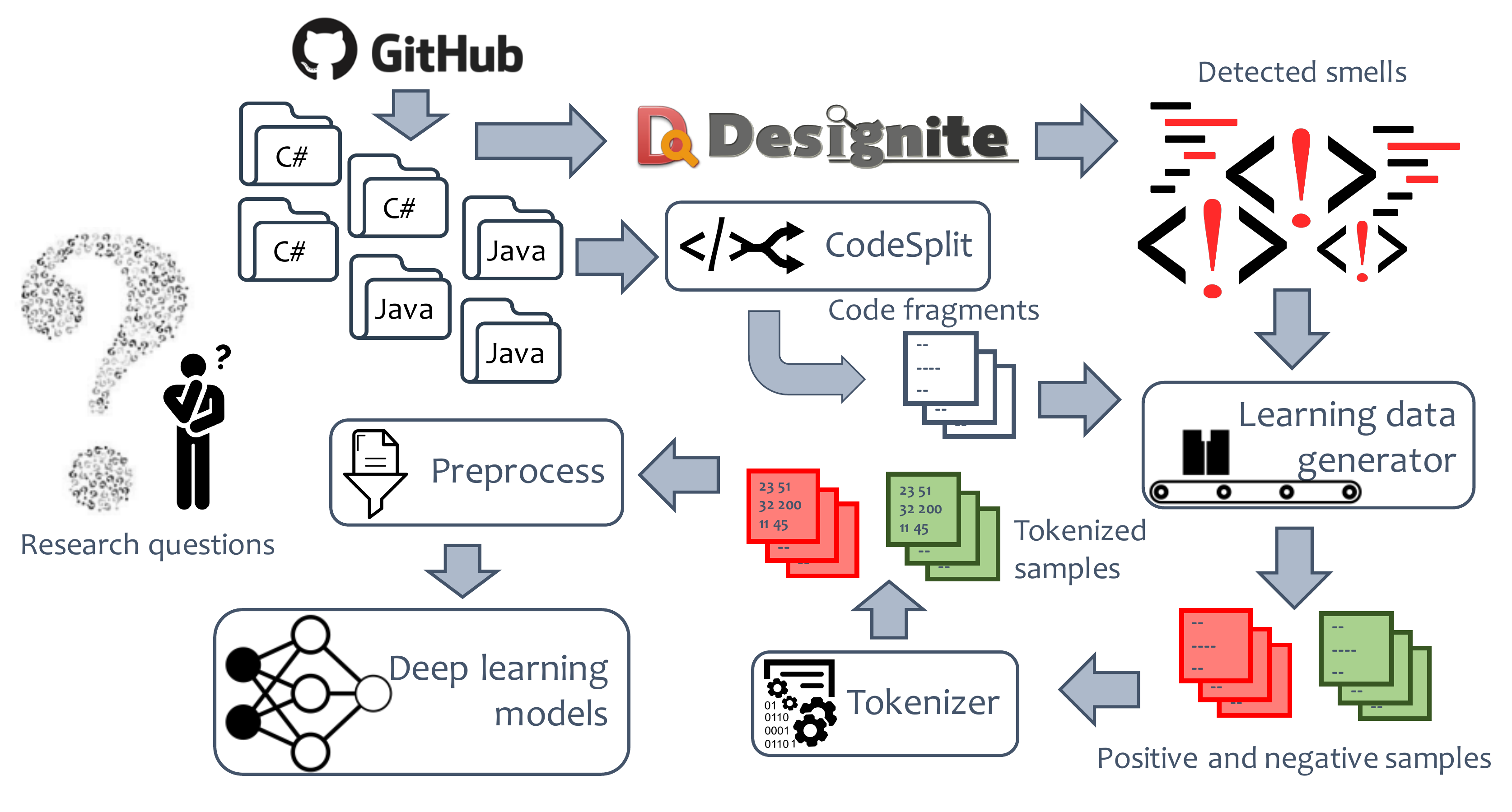}
	\caption{Overview of the Proposed Method}
	\label{fig:overview}
\end{figure}

Figure \ref{fig:overview} provides an overview of the experiment.
We download $1,072$ C\# and $100$ Java repositories
from GitHub.
We use Designite and DesigniteJava 
to analyze C\# and Java code respectively.
We use CodeSplit to extract each method and class definition into separate files
from C\# and Java programs.
Then the learning data generator uses the detected smells to bifurcate code
fragments into positive or negative samples for a smell\textemdash positive samples contain the smell
while the negative samples are free from that smell.
Tokenizer takes a method or class definition and generates integer tokens for each
token in the source code.
We apply preprocessing operation, specifically duplicates removal, on the output of Tokenizer.
The processed output of Tokenizer is ready to feed to the neural networks.

\subsection{Data Curation} \label{data_curation}
In this section, we elaborate on the process of generating training and evaluation
samples along with the tools used in the process.

\subsubsection{Downloading Repositories}
We use the following protocol to identify and download our subject systems.
\begin{itemize}
	\item We download repositories containing C\# and Java code from GitHub.
	We use RepoReapers \cite{Munaiah2017} to filter out low-quality repositories. 
	RepoReapers analyzes GitHub repositories and provides scores for eight dimensions
	of their quality.
	These dimensions are architecture, community, continuous integration,
	documentation, history, license, issues, and unit tests.
	
	\item We select all the repositories where at least six out of eight and seven out of eight
	RepoReapers' dimensions have suitable scores for C\# and Java repositories
	respectively.	
	We consider a score suitable if it has a value greater than zero.
	\item We ensure that RepoReapers results do not include forked repositories.
	\item We discard repositories with fewer than five stars and less than $1,000$ {\sc loc}.
	
	\item Following these criteria, we get a filtered list of $1,072$ C\# and $2,528$ Java repositories.
	We select $100$ repositories randomly from the filtered list of Java repositories.
	Finally, we download and analyze the $1,072$ C\# and $100$ Java repositories. 
\end{itemize}

\subsubsection{Smell Detection}
We use Designite to detect smells in C\# code.
Designite \cite{Tushar2016, Sharma2016} is a software design quality assessment tool for code written in C\#.
It supports detection of eleven implementation, $19$ design, and seven architecture smells.
It also provides commonly used code metrics and other features such as trend analysis,
code clone detection,
and dependency structure matrix to help developers assess the software quality.
A free academic license of Designite can be requested.

Similar to the C\# version, we have developed 
DesigniteJava \cite{Sharma2018c},
which is an open-source tool for analyzing and detecting smells in a Java codebase.
The tool supports detection of $17$ design and ten implementation smells.

We use the console version of
Designite (version $2.5.10$) and DesigniteJava (version $1.1.0$) to
analyze C\# and Java code respectively
and detect the specified design and implementation smells in each of the downloaded repositories.

\subsubsection{Splitting Code Fragments}
CodeSplit is a set of two utility programs, one for each programming language,
that split methods or classes written in C\# and Java source code into individual files. 
Hence, given a C\# or Java project, the utilities can parse the code correctly
(using Roslyn for C\# and Eclipse {\sc jdt} for Java), and emit the individual method or class
fragments into separate files following hierarchical structure (\textit{i.e.,} namespaces/packages becomes folders).
%
CodeSplit for Java is an open-source project that can be found on
GitHub \cite{Sharma2019}.
CodeSplit for C\# can be downloaded freely 
online \cite{Sharma2019b}.

\subsubsection{Generating Training and Evaluation Data}
The learning data generator requires information from two sources\textemdash
a list of detected smells for each analyzed repository and a path to the folder
where the code fragments corresponding to the repository are stored. 
The program takes a method (or class in case of design smells) at a time and
checks whether the given smell has been detected in the method (or class) by Designite.
If the method (or class) suffers from the smell, the program puts the code fragment into a
``positive'' folder corresponding to the smell
otherwise into a ``negative'' folder.
	
\subsubsection{Tokenizing Learning Data}
\begin{figure}[h!]
	\centering
	\includegraphics[width=0.6\linewidth]{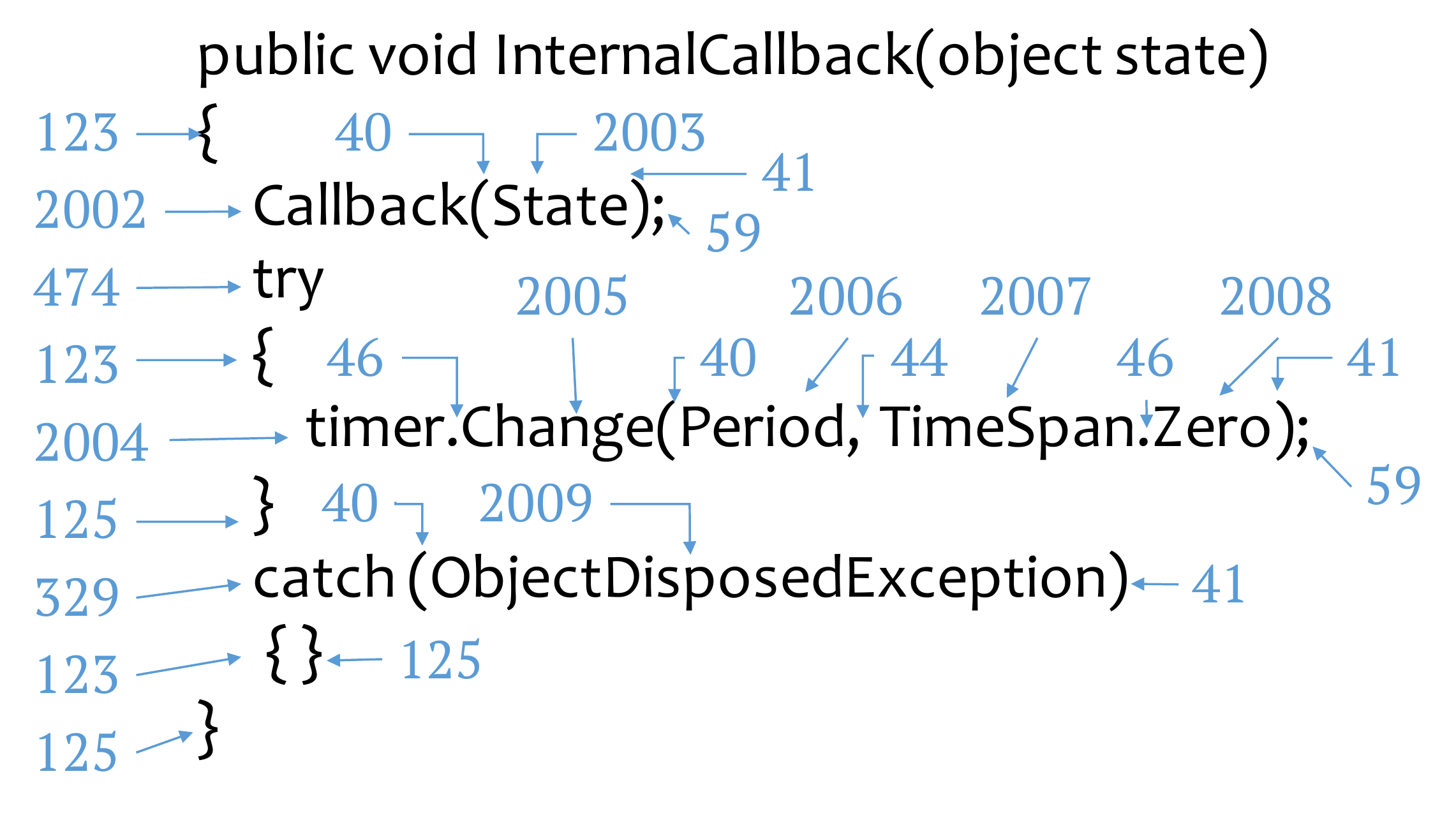}
	\caption{Tokens generated by Tokenizer for an example}
	\label{fig:tokenizer}
\end{figure}

Machine learning algorithms require the inputs to be given in a representation appropriate for
extracting the features of interest, given the problem in hand. 
For a multitude of machine learning tasks it is a common practice to convert data into 
numerical representations before feeding them to a machine learning algorithm. 
In the context of this study, we need to convert source code into vectors of numbers
honoring the language keywords and other semantics. 
Tokenizer \cite{Spinellis2019}
is an open-source tool that provides, among others, functionality for tokenizing 
source code elements into integers where different ranges of integers map to 
different types of elements in source code. 
Figure \ref{fig:tokenizer} shows a small C\# method and corresponding
tokens generated by Tokenizer.
Currently, it supports six programming languages, including C\# and Java.

\subsubsection{Data Preparation}
The stored samples are read into \textit{numpy} arrays, preprocessed, and filtered.
We first perform bare minimum preprocessing to clean the data\textemdash
for both {\sc 1d} and {\sc 2d}
samples, we scan all the samples for each smell and remove duplicates if any exist.
	
We split the samples in the ratio of 70-30 for training; \textit{i.e.,} 70\% of the samples are
used for training a model while 30\% samples are used for evaluation.
We limit the maximum number of positive/negative training samples to $5,000$.
Therefore, for instance, if negative samples are more than $5,000$, we drop the rest of the samples.
We perform model training using balanced samples, \textit{i.e.,}
we balance the number of samples for training by choosing the smaller number 
from the positive and negative sample count;
we discard the remaining training samples from the larger side. 
Table \ref{table:samples} presents an example of data preparation.

\begin{table}[h!]
	\caption{Number of samples in each step of preparing input data}
	\begin{tabular}{ll|rrrr}
	&&Initial samples&70-30 split&Applying max limit&Balancing\\\hline
	\multirow{2}{*}{Positive}&Training&\multirow{2}{*}{4,961}&3,472&3,472&3,472\\
	&Evaluation&&1,489&1,489&1,489\\
	\multirow{2}{*}{Negative}&Training&\multirow{2}{*}{178,048}&121,161&5,000&3,472\\
	&Evaluation&&51,926&51,926&51,926\\\hline
	\end{tabular}
	\label{table:samples}
\end{table}

Each individual input instance, either a method in the case of implementation smells,
or a class in the case of design smells, 
is stored in the appropriate data structure depending upon the model that will use it. 
In {\sc 1d} representation, each individual input instance 
 is represented by a flat {\sc 1d} array of sequences of tokens, compatible for use with 
 the \rnn{} and the \cnn{} models. 
In the  {\sc 2d} representation, each input instance is represented by a {\sc 2d} array of tokens,  
preserving the original statement-by-statement delineation of source code thus providing the grid-like
input format that is required by \cnntwo{} models. 
All the individual samples are stored in a few files (where each file size is approximately $50$ {\sc mb})
to optimize the I/O operations due to a large number of files.
We read all the samples into a \textit{numpy} array and we filter out the outliers. 
In particular, we compute the mean input size and discard all the samples with length
over one standard deviation away from the mean. 
This filtering helps us keep the training set in reasonable bounds and
	avoids waste of memory and processing resources. 
	We pad the input array with zeros to the extent of the longest remaining input in order to 
	create vectors of uniform length and bring the data in the appropriate format for 
	using with the deep learning models. 
Finally, we shuffle the array of input samples along with its corresponding labels array.


\subsection{Selection of Smells}
Over the last two decades, the software engineering community has documented many smells
associated with different granularities, scope, and domains \cite{Sharma2018b}.
A comprehensive taxonomy of the software smells can be found 
online.\footnote{\url{http://www.tusharma.in/smells}}
For this study, selection of smells is a crucial decision.
The scope of the higher granularity smells, such as design and architecture smells,
is large, often spanning to multiple classes and components.
It is essential to provide all the intertwined source code fragments to the deep learning
model to make sure that the model captures the key deciding elements from the provided input source code.
Hence, it is naturally difficult to detect them using deep learning approaches, 
unless extensive feature engineering is performed beforehand in order to attain an
appropriate representation of the data. 
We started with implementation smells because they can be detected typically just by
looking at a method.
However, we would like to avoid very simple smells (such as \textit{long method}) 
which can be easily detected by less sophisticated techniques.

We chose \cm{} ({\sc cm}\textemdash\textit{i.e.,}
the method has high cyclomatic complexity),
\mn{} ({\sc mn}\textemdash \textit{i.e.,} an unexplained numeric literal is used in an expression),
and\textit{ empty catch block }({\sc ecb}\textemdash \textit{i.e.,} a catch block of an exception is empty).
These three smells represent three different kinds of smells where neural networks have to spot
specific features.
For instance, to detect  \mn{}, the neural networks must spot a specific range of tokens
representing magic numbers.
On the other hand, detection of \cm{} requires looking at the entire method and the
structural property within it (\textit{i.e.,} nesting depth of the method). 
For the detection of \ecb{} the neural network has to recognize
a sequence of a try block followed by an empty catch block. 

To expand the horizon of the experiment, we also select \ma{}
({\sc ma}\textemdash \textit{i.e.,} a class has more than one responsibility assigned to it) design smell.
The scope of this smell is larger (\textit{i.e.,} the whole class) and detection is not trivial
since the neural network has to capture cohesion aspect (typically captured by the
Lack of Cohesion of Methods
({\sc lcom}) metric
in deterministic tools) among the methods to detect it accurately.
This smell not only allows us to compare the capabilities of neural networks
in detecting implementation smells
with  design smells but also sets the stage for the future work to build on.

\subsection{Architecture of Deep Learning Models}\label{sec:arch-models}
In this section, we present the architecture of the neural network models that we use in this study.
The Python implementation of the experiments using the Keras library can be found
online.\footnote{\url{https://github.com/tushartushar/DeepLearningSmells}}


\subsubsection{{\sc cnn} Model}
\begin{figure}[h!]
	\centering
	\includegraphics[width=0.5\linewidth]{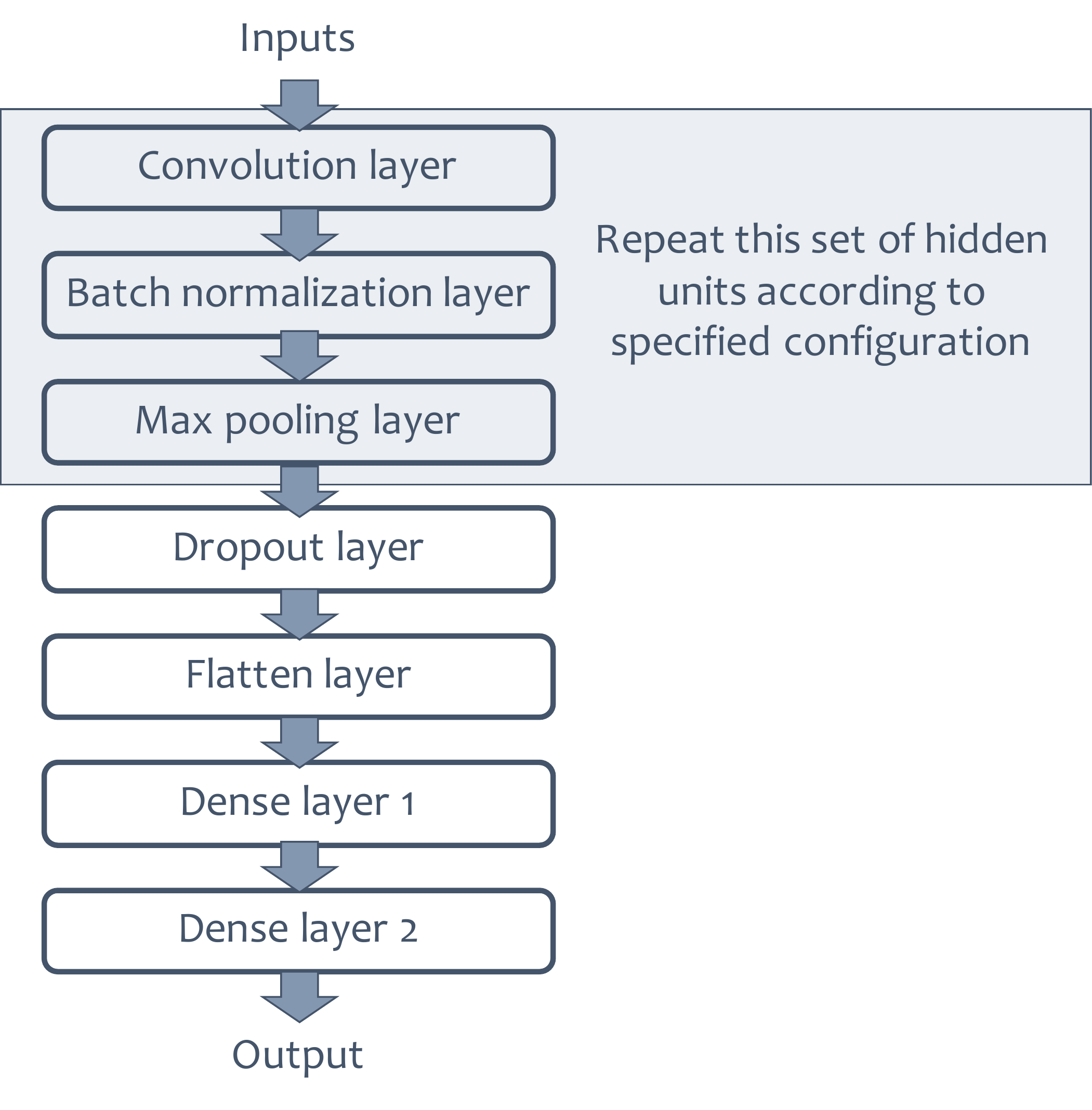}
	\caption{Architecture of employed CNN}
	\label{fig:cnn-arch}
\end{figure}

Figure \ref{fig:cnn-arch} presents the architecture of {\sc cnn} model used to detect smells.
This architecture is inspired by typical {\sc cnn} architectures used in image classification 
\cite{Krizhevsky2012} and consists of a feature extraction part followed by a classification part. 
The feature extraction part is composed of an ensemble of layers, specifically,
convolution, batch normalization, and max pooling layers.
This set of layers form the hidden layers of the architecture.
The convolution layer performs convolution operations based on the specified filter and 
kernel parameters and computes accordingly the network weights to the next layer, 
whereas the max pooling layer effectuates a reduction on the dimensionality of the feature space. 
Batch normalization \cite{Ioffe2015}  mitigates the effects of varied input distributions for 
each training mini-batch, thus optimizing training. 
In order to experiment with different configurations, we use one, two, and three hidden layers.

The output of the last max pooling layer is connected to a dropout layer.
Dropout performs another type of regularization by ignoring some randomly selected nodes during training 
in order to prevent over-fitting \cite{Srivastava2014}. 
In our experiments we set the dropout rate for the layer to be equal to $0.1$ which means that the 
nodes to be ignored are randomly selected with probability $0.1$. 

The output of the last dropout layer is fed into a densely connected classifier network that 
consists of a stack of two dense layers. 
These classifiers process {\sc 1d} vectors, whereas the incoming output from the last hidden 
layer is a 3D tensor (that corresponds to height and width of an input sample,
and channel; in this case, the number of channels is one). 
For this reason, a flatten layer is used first, to transform the data 
in the appropriate format before feeding them to the first dense layer 
with $32$ units and \textit{relu} activation.
This is followed by 
the second dense layer with one unit and \textit{sigmoid} activation. 
This last dense layer comprises the output layer and contains a single neuron in order to make predictions  
on whether a given instance belongs to the positive or negative class in terms of the smell under investigation.
The layer 
uses the sigmoid activation function in order to produce a probability within the 
range of $0$ to $1.$ 


We use dynamic batch size depending upon the size of samples to train.
We divide the training sample size by $512$ and use the result as the index to choose
one of the items in the possible batch size array ($32,$ $64,$ $128,$ $256$).
For instance, we use $32$ as batch size when the training sample size is $500$ and
$256$ when the training sample
size is $2000.$

The hyper-parameters are set to different values in order to experiment with different configurations of the model. 
Table \ref{table:hyper-params-cnn} lists all the different values chosen for the hyper-parameters.
We execute {\sc cnn} models for $144$ configurations that result from generating combinations
of different values of hyper-parameters and number of repetitions of the set of hidden units.
We label each configuration between $1$ and $144$ where configuration $1$ refers to
number of repetitions of the set of hidden units = $1,$
number of filters = $8,$
kernel size = $5,$
and pooling window size = $2.$
Similarly, configuration $144$ refers to 
number of repetitions of the set of hidden units = $3,$
number of filters = $64,$
kernel size = $11,$
and pooling window size = $5.$
Both the {\sc 1d} and {\sc 2d} variants use the same architecture replacing the {\sc 2d} version of
Keras layers for their {\sc 1d} counterparts.
\begin{table}[h]
	
	\caption{Chosen values of hyper-parameters for the {\sc cnn} model}
	\begin{tabular}{lr}
		\hline
		\textbf{Hyper-parameter}                          & \textbf{Values}            \\\hline
		Filters in convolution layer           & \{8, 16, 32, 64\} \\
		Kernel size in convolution layer         & \{5, 7, 11\}      \\
		Pooling window size in max pooling layer & \{2, 3, 4, 5\}   \\
		Maximum epochs                                 & 50             \\   
		\hline
	\end{tabular}
\label{table:hyper-params-cnn}
\end{table}

We ensure the best attainable performance and avoid over-fitting by using \textit{early stopping\footnote{\label{keras_callback}\url{https://keras.io/callbacks/}}}
as a regularization method.
It implies that the model may reach a maximum of $50$ epochs during training.
However, if there is no improvement in the validation loss of the trained model
for five consecutive epochs (since patience, a parameter to early stopping mechanism,
is set to five),
the training is interrupted.
Along with it, we also use 
\textit{model check point}
to restore the best weights of the trained model.

For each experiment, we compute the following performance metrics ---
accuracy, {\sc roc-auc} (Receiver Operating Curve-Area Under Curve),
precision, recall, F1, and average precision score.
We also record the actual epoch count where the models stopped training (due to early stopping).
After we complete all the experiments with all the chosen hyper-parameters,
we choose the best performing configuration and the corresponding number of epochs used
by the experiment and retrain the model and record the final and best performance of the model.

\subsubsection{\rnn{} Model}
\begin{figure}[h!]
	\centering
	\includegraphics[width=0.5\linewidth]{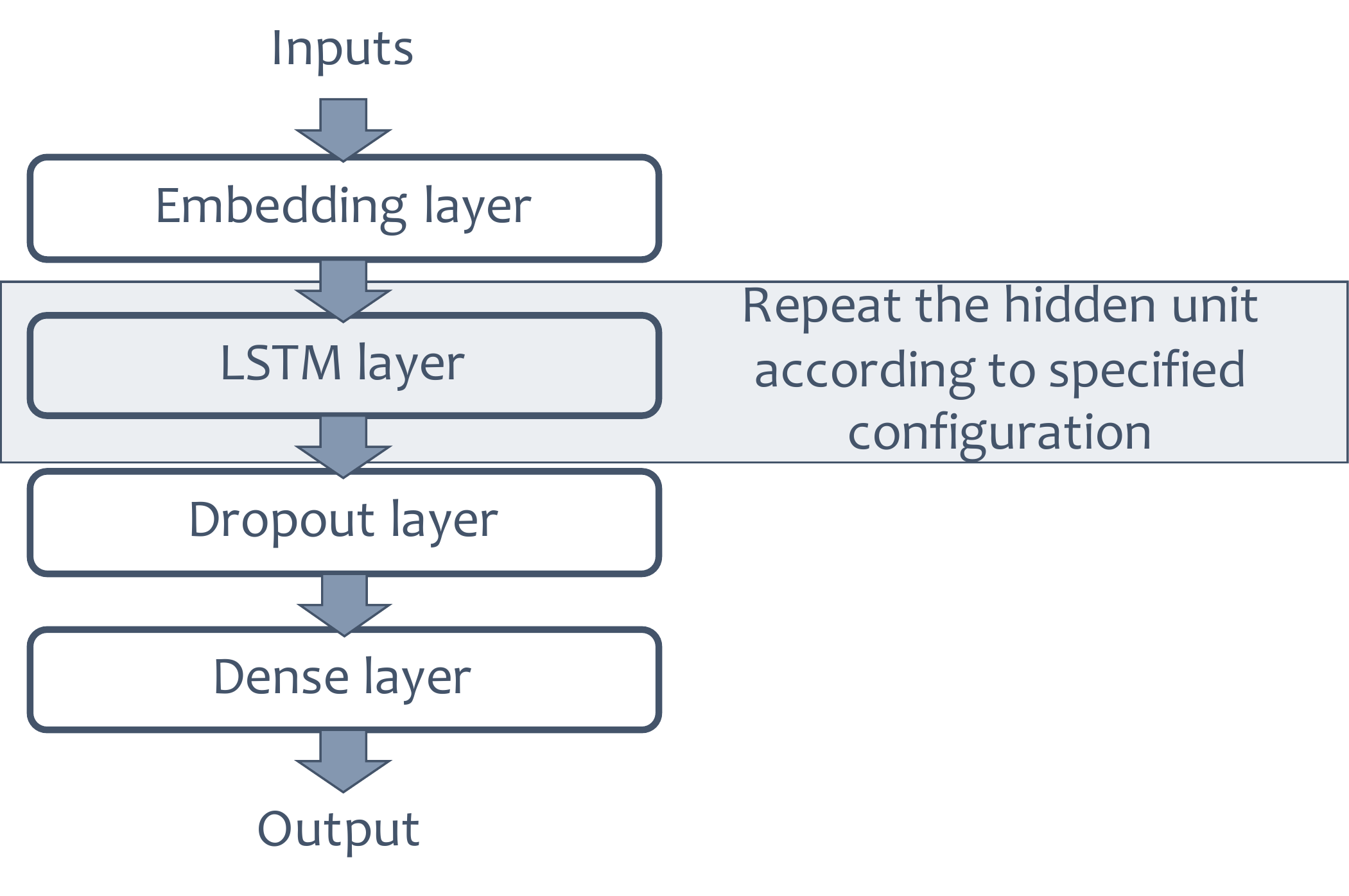}
	\caption{Architecture of employed RNN}
	\label{fig:rnn-arch}
\end{figure}

Figure \ref{fig:rnn-arch} presents the
architecture of the employed \rnn{} model which
 is inspired by state-of-the-art models in natural language modeling that employ an
 {\sc lstm} network as a recurrent layer \cite{Sundermeyer2012}. 
The model consists of an embedding layer followed by the feature learning part --- a
hidden {\sc lstm} layer.
It is succeeded by the regularization (realized by a dropout layer) and
classification (consisting of a dense layer) part. 

The embedding layer maps discrete tokens into compact dense vector representations.
One of the advantages of the {\sc lstm} networks is that they can effectively
handle sequences of varying lengths.
To this end, in order to avoid the noise produced by the 
padded zeros in the input array,
we set the \textit{mask\_zero} parameter to \textit{True} provided by the Keras
embedding layer implementation.
Thus the padding is ignored and only the meaningful part of the input data is taken into account.
We set \textit{dropout} and \textit{recurrent\_dropout} parameters of {\sc lstm} layer to $0.1$.
The regular dropouts mask (or drop) network units at inputs and/or outputs
whereas recurrent dropouts drop the connections between the recurrent units along with
dropping units at inputs and/or outputs \cite{Yarin2015}.
The output from the embedding layer in fed into the
{\sc lstm} layer, which in turn outputs to the dropout layer. 
As in the case of the {\sc cnn} model, we experiment for different depths of the \rnn{}
model by repeating multiple instances of the hidden layer.

The dropout layer uses a dropout rate equal to $0.2$, which we empirically found 
effective for preventing over-training, yet conservative enough for avoiding under-training. 
The dense layer, which comprises the classification output layer, is configured with
one unit and \textit{sigmoid} activation as in the case of the {\sc cnn} model. 
Similarly to the {\sc cnn} model, we use \textit{early stopping}
(with maximum epochs = $50$ and patience = $2$) and
\textit{model check point} callbacks.
Also, we use the dynamic batch size selection as explained in the previous subsection.

We try different values for the model hyper-parameters.
Table \ref{table:hyper-params-rnn} presents different values selected for each hyper-parameter.
We measure the performance of the \rnn{} model in $18$ configurations by forming
the  combinations produced by the different chosen values of hyper-parameters
and the number of repetitions of the set of hidden units.

\begin{table}[h!]
	
	\caption{Chosen values of hyper-parameters for the \rnn{} model}
	\begin{tabular}{lr}\hline
		\textbf{Hyper-parameter }         & \textbf{Values}          \\\hline
		Dimensionality of embedding layer & \{16, 32\}  \\
		{\sc lstm} units             & \{32, 64, 128\} \\
		Maximum epochs                 & 50         \\\hline    
	\end{tabular}
\label{table:hyper-params-rnn}
\end{table}

As described earlier, we pick the best performing hyper-parameters and
number of epochs and retrain the model to obtain the final and best performance of the model.

\subsection{Hardware Specification}
We perform all the experiments on the super-computing facility offered by {\sc grnet}
(Greek Research and Technology Network).
The experiments were run on {\sc gpu} nodes (8x NVidia V100). 
Each {\sc gpu} incorporates 5120 {\sc cuda} cores.
We request 1 {\sc gpu} node with $64$ {\sc gb} of memory for most of the experiments
while submitting the job to the super computing facility.
Some \rnn{} experiments require more memory
to perform the training; we request $128$ {\sc gb} of memory for them.

\section{Results and Discussion}
\label{sec:results}
As elaborated in this section,
we found that it is feasible to detect smells using deep learning models
without extensive feature engineering.
Our results also indicate that performance of deep learning models is highly
smell-specific.
Furthermore, we found that it is feasible to apply transfer-learning in the context
of code smells detection.
In the rest of the section, we discuss the results in detail.
 
\subsection{Results of RQ1}
\begin{description}
	\item[RQ1] Is it possible to use deep learning methods to detect code smells? 
	If yes, which deep learning method performs superior?
\end{description}
\subsubsection{Approach}
We prepare the input samples as described in Section \ref{data_curation}.
Table \ref{table:rq1.samples} presents the number of positive and negative samples used for each smell
for training and evaluation;
\cnn{} and \rnn{} use {\sc 1d} samples and \cnntwo{} uses {\sc 2d} samples.
As mentioned earlier, we train our models with the same number of positive and negative samples.
Sample size for \ma{} ({\sc ma}) is considerably low compared to other smells
because each sample in this smell is a class (other smells use method fragments).
The one-dimensional sample counts are different from their two-dimensional counterparts because
we apply additional constraint for outlier exclusion, on permissible height, in addition to the width. 

\begin{table}[h!]
	\caption{Number of positive (P) and negative (N) samples used for training and evaluation for RQ1}
	\begin{tabular}{l|rrr|rrr}
		&\multicolumn{3}{c}{\cnn{} and \rnn{}}&\multicolumn{3}{c}{\cnntwo{}}\\
		& Training & \multicolumn{2}{c}{Evaluation}  
		& Training & \multicolumn{2}{c}{Evaluation}                              \\
		& {\sc p} and {\sc n}     & {\sc p} & {\sc n} & {\sc p} and {\sc n}     & {\sc p} & {\sc n}\\\hline
		{\sc cm}           & 3,472    & 1,489    & 51,926 & 2,641 & 1,132 & 45,204  \\
		{\sc ecb}        & 1,200      & 515     & 52,900   & 982 & 422 & 45,915       \\
		{\sc mn}             & 5,000     & 5,901& 47,514 & 5,000 & 5,002 & 41,334         \\
		{\sc ma}  & 290      & 125    & 22,727 & 284 & 122 & 17,362 \\\hline         
	\end{tabular}
	\label{table:rq1.samples}
\end{table}

\subsubsection{Results}
\begin{figure}[h!]
	\centering
	\begin{subfigure}[b]{0.5\textwidth}
		\includegraphics[width=\textwidth]{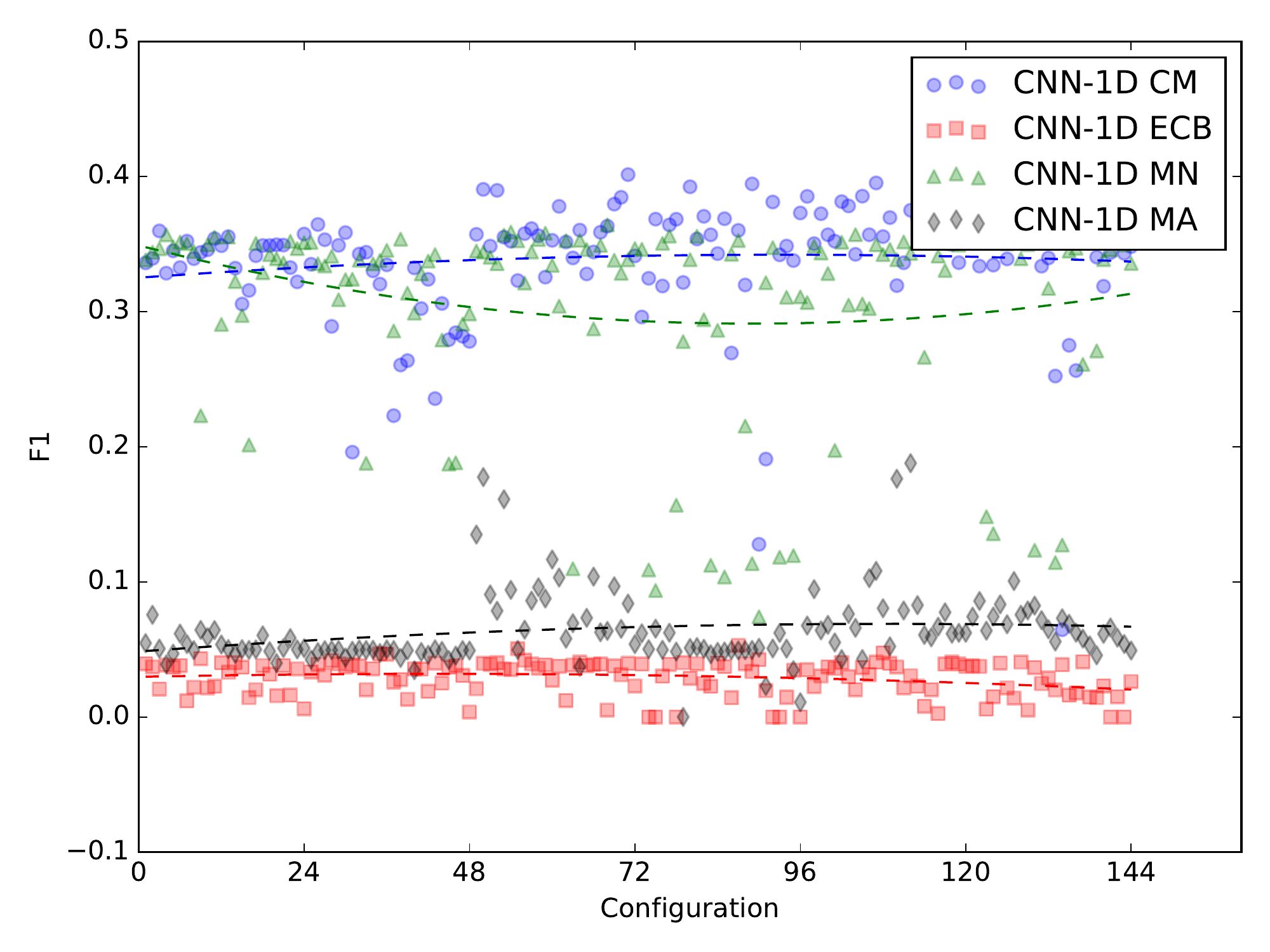}
		\caption{Performance of CNN-1D }
	\end{subfigure}
	~ 
	\begin{subfigure}[b]{0.5\textwidth}
		\includegraphics[width=\textwidth]{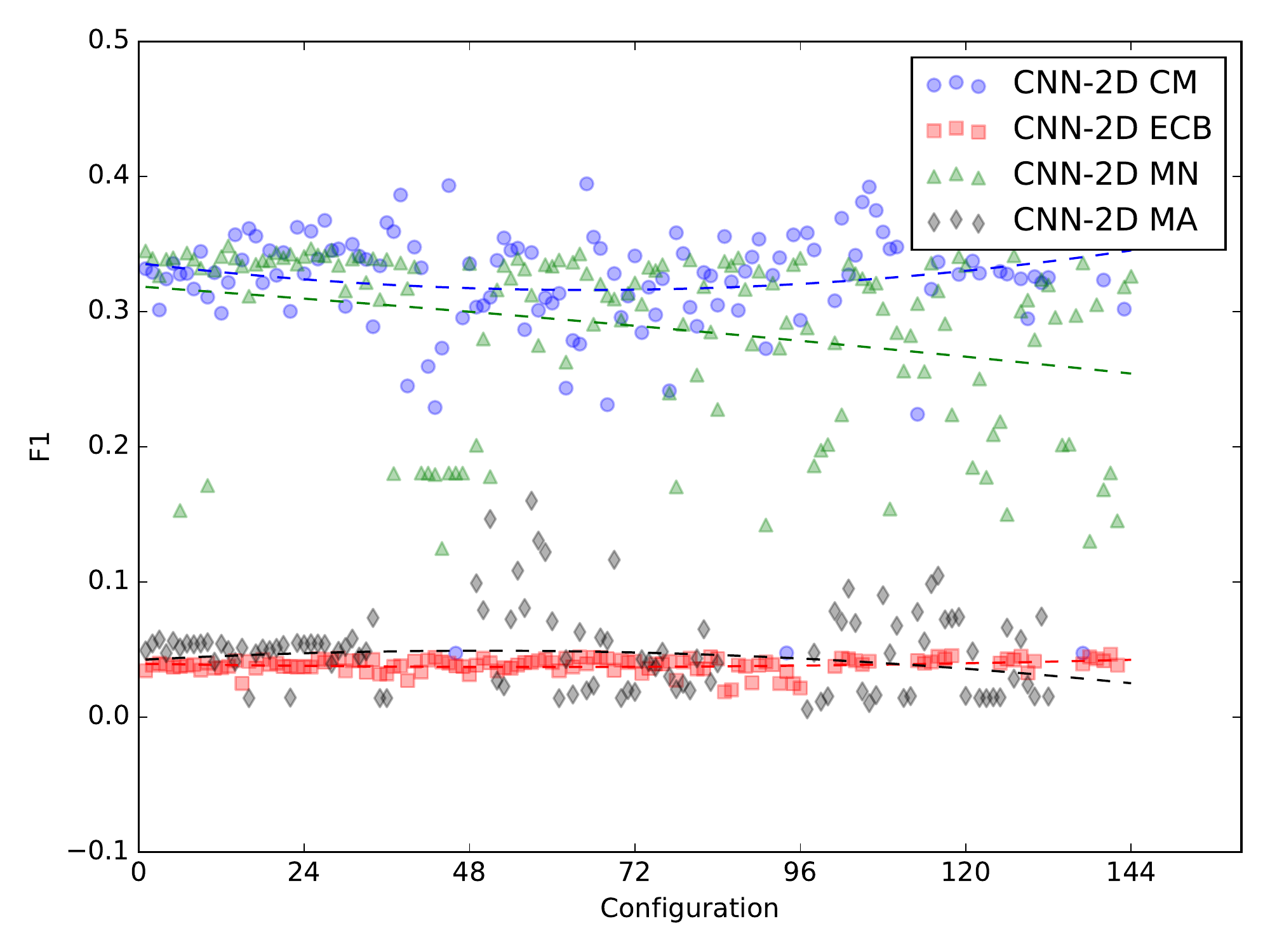}
		\caption{Performance of CNN-2D  }
	\end{subfigure}
	
	\begin{subfigure}[b]{0.5\textwidth}
		\includegraphics[width=\textwidth]{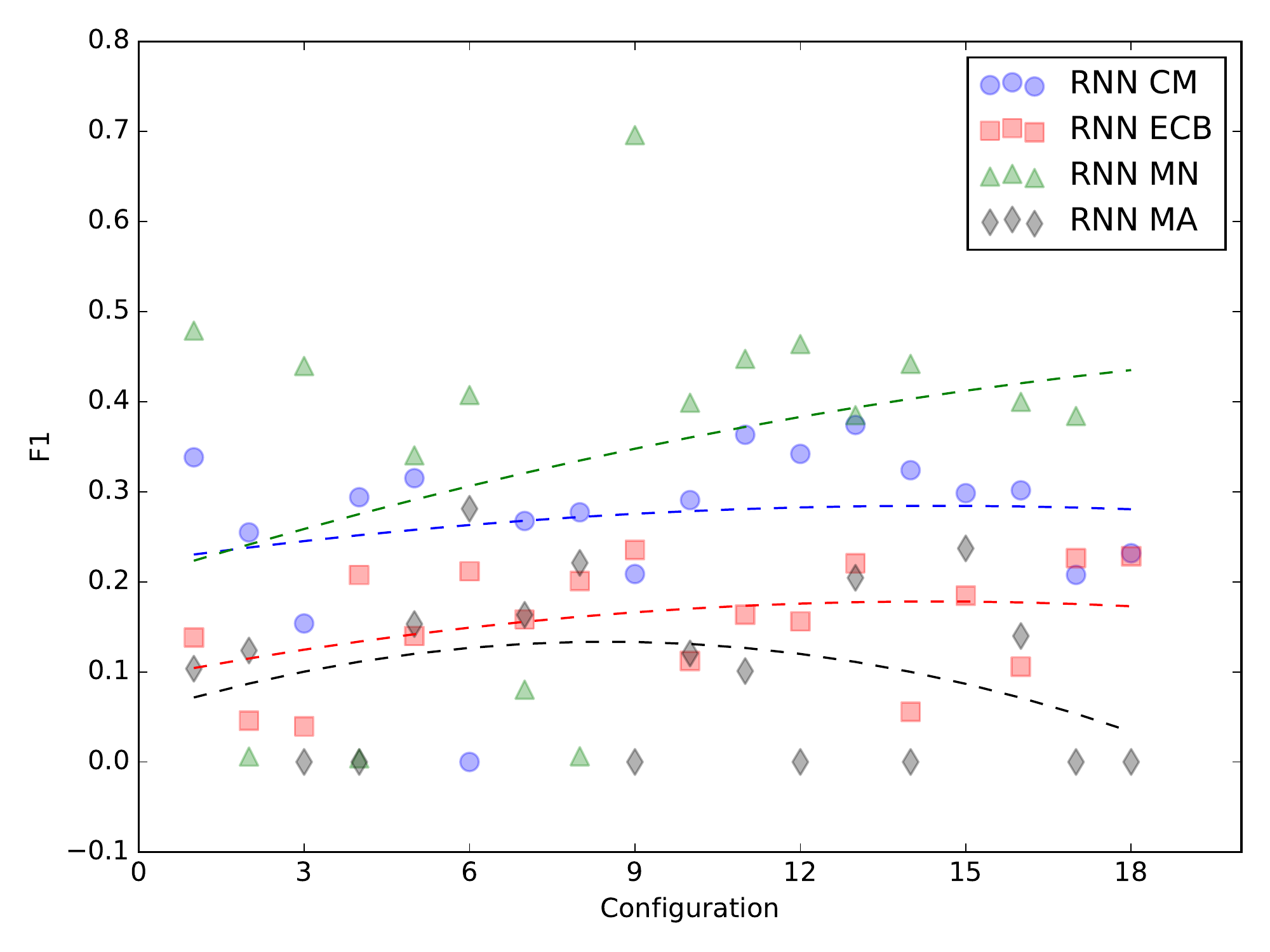}
		\caption{Performance of RNN }
	\end{subfigure}
	
	\caption{Scatter plots of the performance (F1) exhibit by the considered deep learning models along with their corresponding trendline}\label{fig:rq1.1}
\end{figure}

Figure \ref{fig:rq1.1} presents the performance (F1) of the models for the considered smells
for all the configurations that we experimented with.
The results from each model perspective show that performance of the models varies depending on
the smell under analysis.
Another observation from the trendlines shown in the plots is that
performance of the convolution models remains more or less stable and unchanged 
for different configurations while \rnn{} exhibits better performance as the complexity
of the model increases except for \ma{} smell.
It implies that the hyper-parameters that we experimented with do not play
a very significant role for convolution models.

\begin{figure}[h!]
	\centering
	\begin{subfigure}[b]{0.45\textwidth}
		\includegraphics[width=\textwidth]{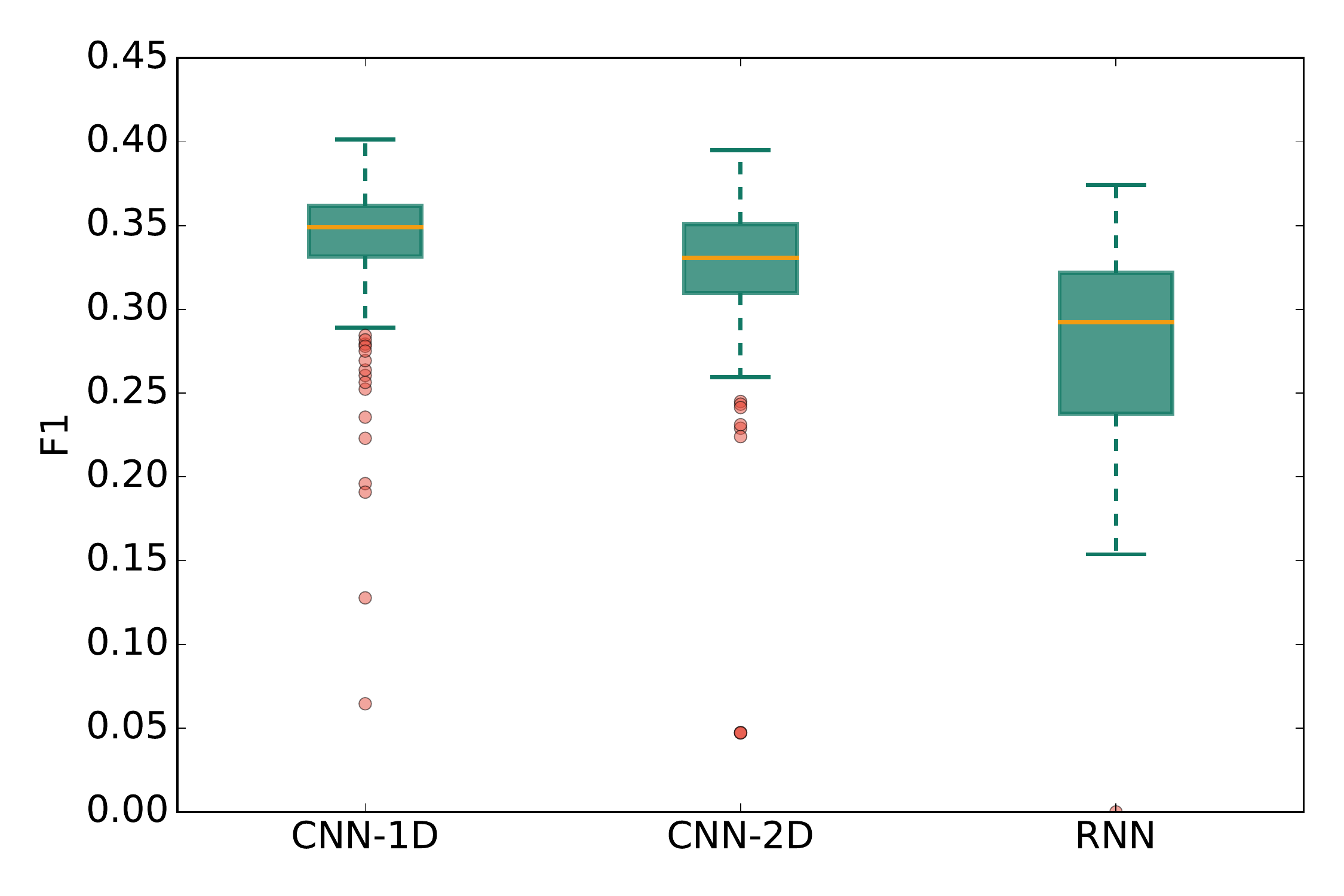}
		\caption{Complex method}
	\end{subfigure}
	~ 
	\begin{subfigure}[b]{0.45\textwidth}
		\includegraphics[width=\textwidth]{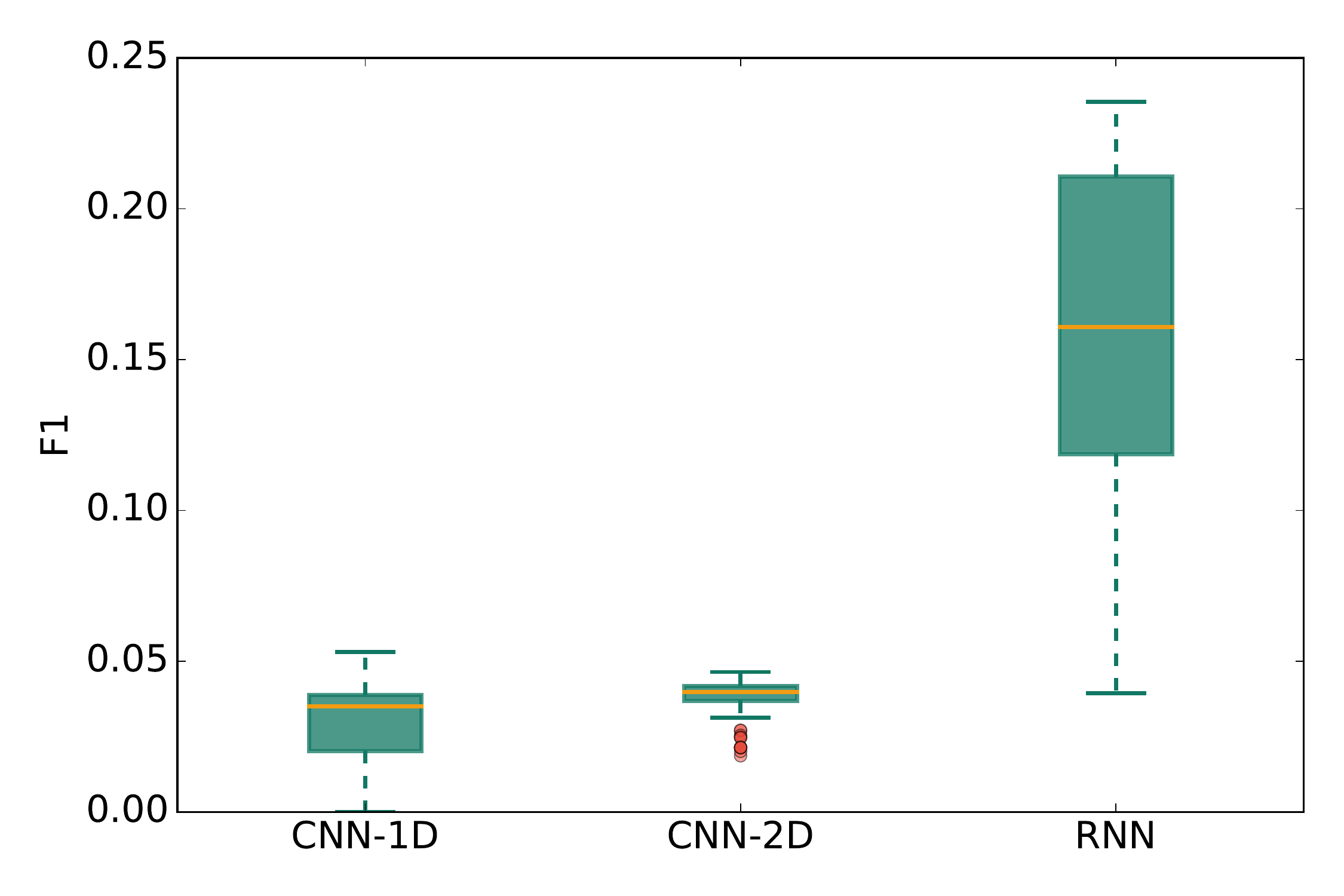}
		\caption{Empty catch block}
	\end{subfigure}
	
	\begin{subfigure}[b]{0.45\textwidth}
		\includegraphics[width=\textwidth]{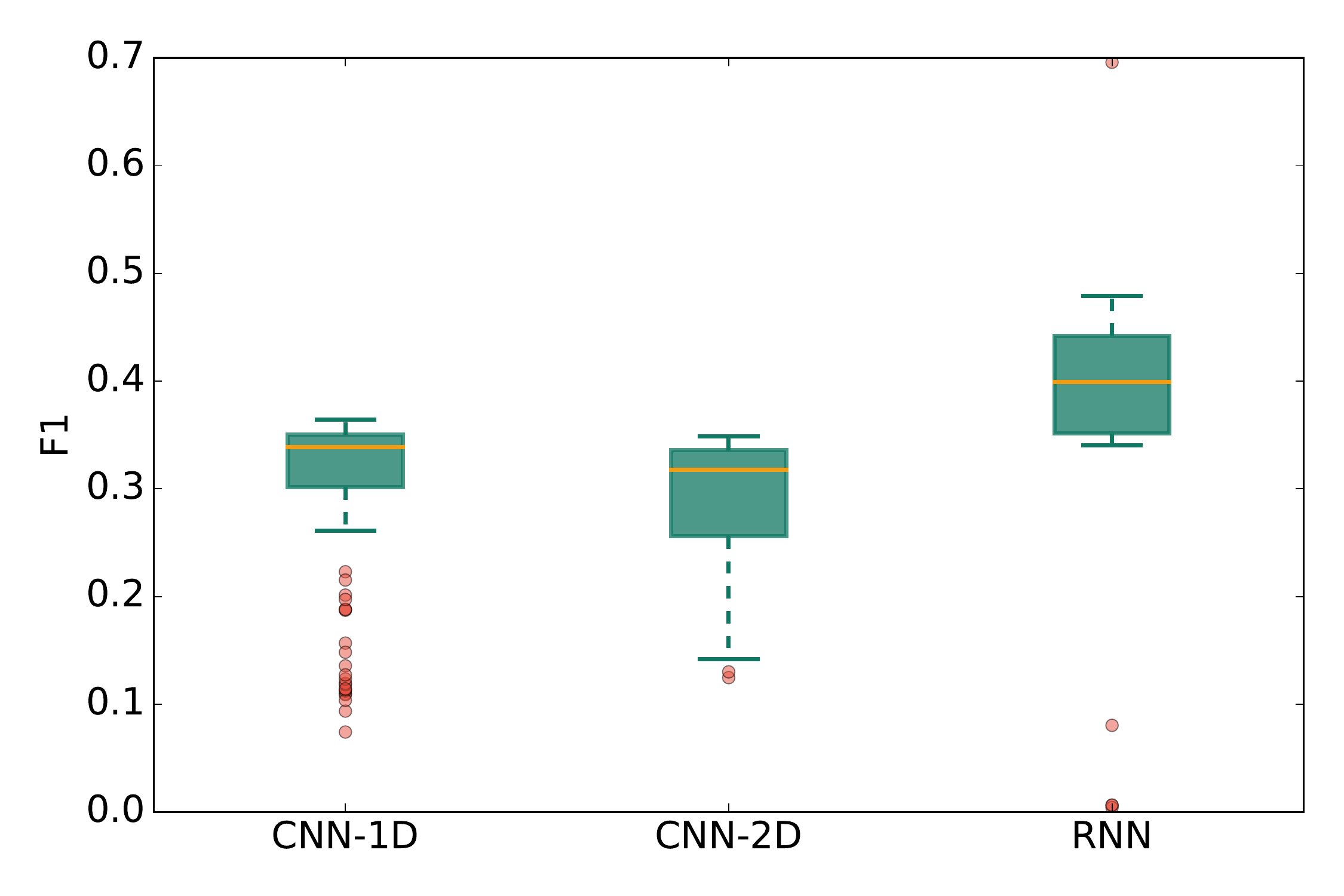}
		\caption{Magic number }
	\end{subfigure}
	\begin{subfigure}[b]{0.45\textwidth}
		\includegraphics[width=\textwidth]{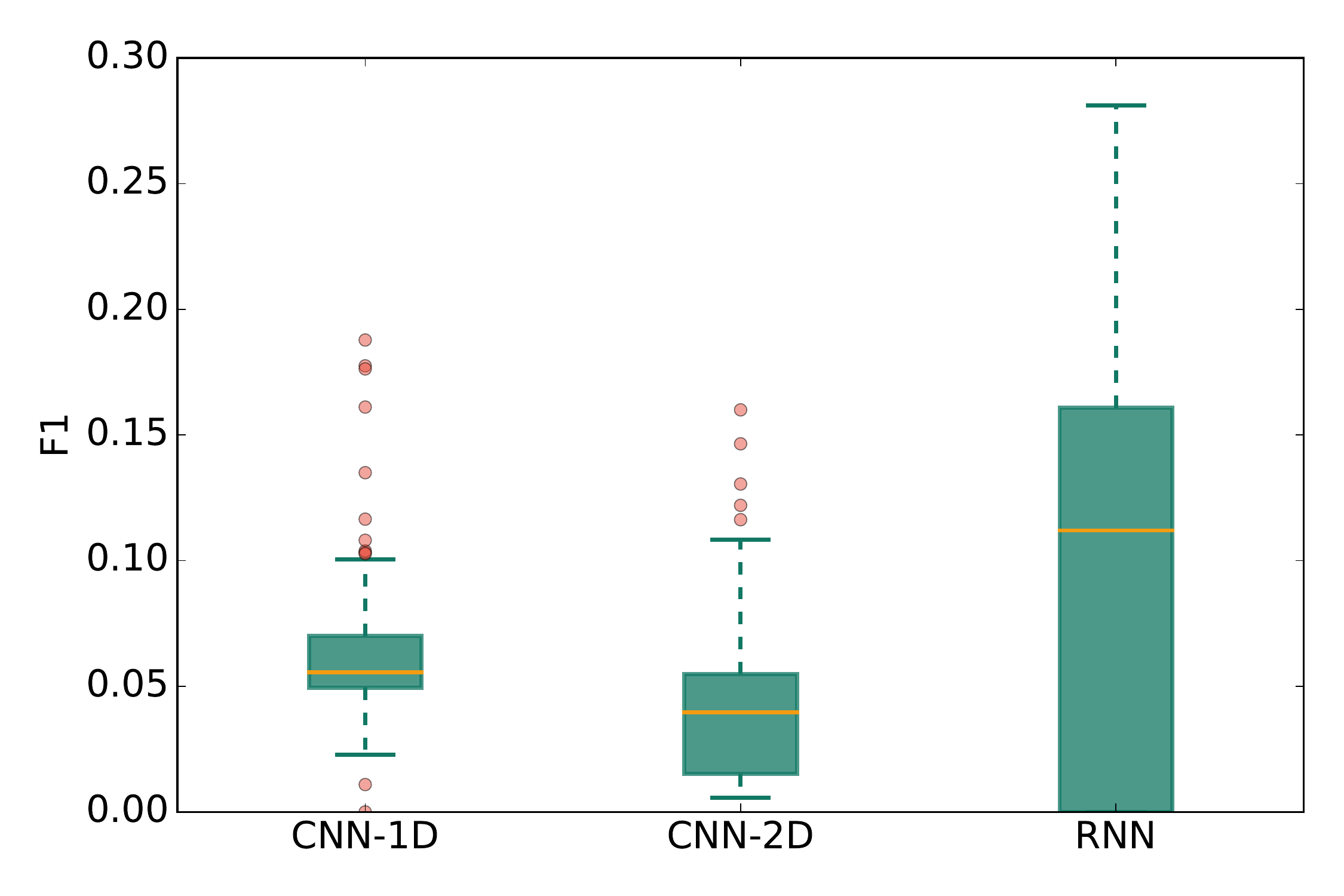}
		\caption{Multifaceted abstraction}
	\end{subfigure}
	
	\caption{Boxplots of the performance (F1) exhibit by the considered deep learning models for all the four smells}\label{fig:rq1.3}
\end{figure}

Figure~\ref{fig:rq1.3} presents the boxplots comparing for each smell performance of all trained models, 
under all configurations.
For \cm{} smell, both convolution models outperform the \rnn{}.
In between the convolution models, overall the various configurations of the \cnn{} model 
appear accumulated around  the mean, whereas \cnntwo{} shows higher variance 
among the F1 scores obtained at different configurations. 
Though, \cnn{} shows lower variance, the model has higher number of outliers compared
to \cnntwo{} model.
\rnn{} model performs significantly superior compared to convolution models for \ecb{} smell with an F1 score 
of $0.22$ versus $0.04$ and $0.02$ achieved by \cnn{} and \cnntwo{} respectively;
the performance of the model, however, shows a wide variation depending on the chosen hyper-parameters.
For \mn{} smell, most of the \rnn{} configurations do better than the best of the convolution-based
configurations.
\rnn{} exhibits a very high variance in the performance compared to convolution models
for \ma{} smell.

Equipped with experiment results, we attempt to validate the hypotheses. 
We present {\sc auc}, precision, recall, and F1 to show the performance
of the analyzed deep learning models.
We attempt to validate each of the addressed hypotheses in the rest of the section.

\begin{description}
	\item[RQ1.H1]\textit{It is feasible to detect smells using deep learning methods.}
\end{description}

\begin{figure}[h!]
	\centering
	\includegraphics[width=0.6\linewidth]{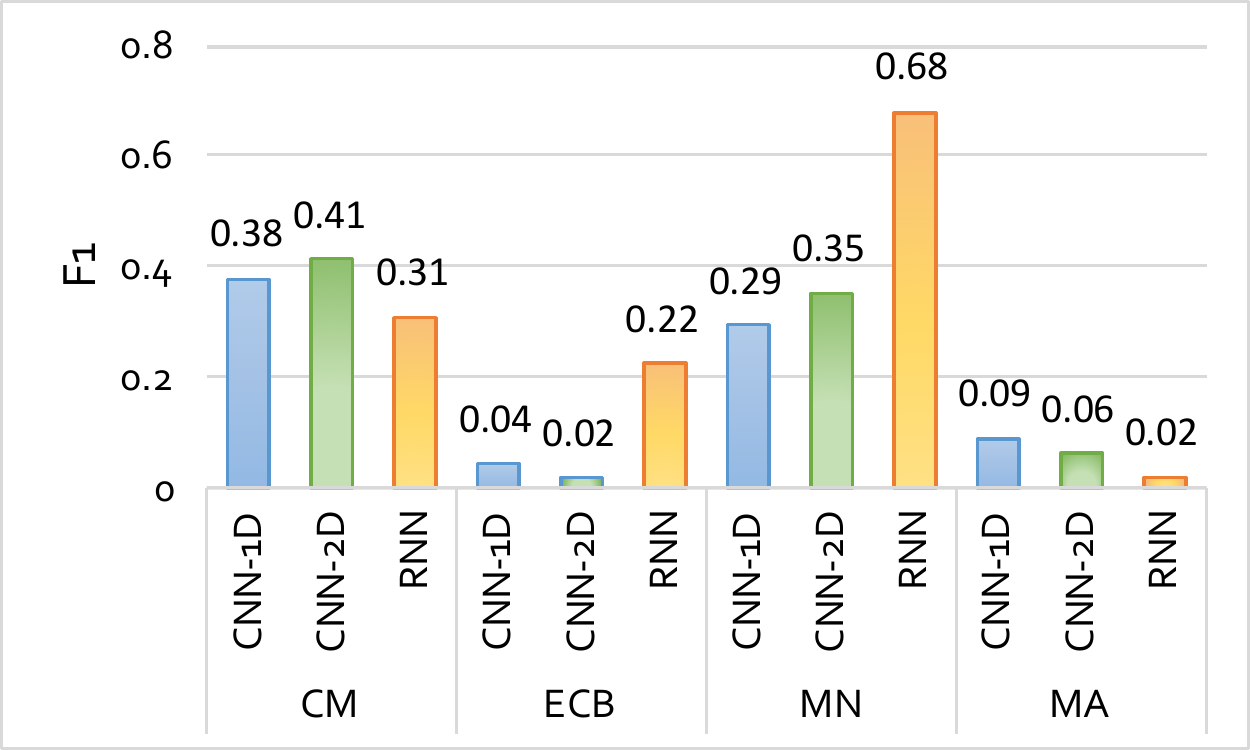}
	\caption{Comparative performance of the deep learning models for each considered smell}
	\label{fig:rq1.h1.f1}
\end{figure}

Table \ref{table:rq1.results.1} lists performance metrics ({\sc auc}, precision, recall, and F1)
for the optimal configuration for each smell, comparing all three deep learning models.
It also lists the hyper-parameters associated with the optimal configuration for each smell.
Figure \ref{fig:rq1.h1.f1} presents the performance (F1) of the deep learning models corresponding
to each smell considered in this exploration.

 \begin{table}[h!]
 	\caption{Performance of all three models with configuration corresponding to the optimal performance. L refers to deep learning layers, F refers to number of filters, K refers to kernel size, MPW refers to maximum pooling window size, ED refers to embedding dimension, LSTM refers to number of LSTM units, and E refers to number of epochs}
\begin{tabular}{l|lrrrr|rrrrrrr}
	&&\multicolumn{4}{c}{Performance}&\multicolumn{7}{c}{Configuration}\\
	& Smells     & AUC    & Precision & Recall & F1 & {\sc l} & {\sc f}  & {\sc k} & {\sc mpw} & {\sc ed} & {\sc lstm} & {\sc e}    \\ \hline
		\multirow{4}{*}{\cnn{} } & {\sc cm}      & 0.82& 0.26 & 	0.69	& 0.38 &  2 &  16 &  7 &  4 & -- & -- &25\\
	& {\sc ecb}     & 0.59 & 0.02	& 0.31	& 0.04 &  2 &  64 &  11 &  4  & -- & -- &40\\
	& {\sc mn}      & 0.68 & 0.18	& 0.77	& 0.29 &  2 &  16 &  5 &  5   & -- & -- &17\\
	& {\sc ma}     & 0.83 & 0.05 &	0.75 &	0.09 &  3 &  16 &  11 &  5  & -- & -- &36\\ \hline
	\multirow{4}{*}{\cnntwo{} } & {\sc cm}      & 0.82 & 0.30 &	0.68 &	0.41 &  3 &  64 &  5 &  4   & -- & -- & 17\\
	& {\sc ecb}     & 0.50 & 0.01 &	1	& 0.02 &  3 &  64 &  7 &  2   & -- & -- & 32\\
	& {\sc mn}   & 0.65 & 0.31 & 0.41	& 0.35 &  1 &  16 &  11 &  2  & -- & -- & 50\\
	& {\sc ma}       & 0.87 & 0.03	& 0.95	& 0.06 &  2 &  8 &  7 &  2   & -- & -- & 19\\ \hline
	\multirow{4}{*}{\rnn{}}    & {\sc cm}       & 0.85 & 0.19 &	0.80 &	0.31 &  3 & & -- & -- & 16 &  32 & 8  \\
	& {\sc ecb}      & 0.86 & 0.13	& 0.76 &	0.22 &  2 & & -- & -- & 16 &  128  & 15   \\
	& {\sc mn}      & 0.91 & 0.55 & 0.91& 0.68& 2 & & -- & -- & 16 &  128 & 19  \\
	& {\sc ma}      & 0.69 & 0.01& 0.86& 0.02& 1 & & -- & -- & 32 &  128 & 9\\ \hline
	\end{tabular}
 \label{table:rq1.results.1}
 \end{table}

For \cm{} smell, \cnntwo{} performs the best; though, performance of \cnn{} is comparable. 
This could be an implication of the fact that the smell is exhibited through the structure 
of a method; hence, {\sc cnn} models, in this case, 
could identify the related structural features for classifying the smells correctly. 
On the other hand, {\sc cnn} models perform significantly poorer than \rnn{} in identifying 
\ecb{} smells.
The smell is characterized by a micro-structure where catch block of a try-catch statement is empty.
\rnn{} model identifies the sequence of tokens (\textit{i.e., }opening and closing braces), 
following the tokens of a try block, 
whereas {\sc cnn} models fail to achieve that and thus \rnn{}
performs significantly better than the {\sc cnn} models.
Also, the \rnn{} model performs remarkably better than {\sc cnn} models for \mn{} smell.
The smell is characterized by a specific range of tokens and the \rnn{} does well in spotting them.
\MA{} is a non-trivial smell that requires analysis of method interactions to observe
incohesiveness of a class.
None of the employed deep learning models could capture the complex characteristics of the smell, 
implying that the token--level representation of the data may not be appropriate for 
capturing higher--level features required for detecting the smell. 
It is evident from the above discussion that all the  employed models are capable of
detecting smells in general; however, their smell-specific performances differ significantly.
\textbf{Therefore, the hypothesis exploring the feasibility of detecting smells using deep learning
	models holds true.}

\begin{description}	
	\item [RQ1.H2] \textit{\cnntwo{} performs better than \cnn{} in
		the context of detecting smells.}
\end{description}
Table \ref{table:rq1.results.1} shows that \cnn{} performs better than \cnntwo{} model
for \ecb{} and \ma{}
smells with optimal configuration. 
On the other hand, \cnntwo{} performs slightly better than its one dimension counterpart
for detecting \cm{} and \mn{}  smells.
In summary, there is no universal superior model for detecting all four smells; 
their performance varies depending on the smell under analysis.
\textbf{Therefore, we reject the hypothesis that \cnntwo{} performs overall better than \cnn{} 
	as none of the models is clearly
	superior to another in all the cases.}

\begin{description}
	\item [RQ1.H3] \textit{\rnn{} model performs better than {\sc cnn} models in the smell detection context.}
\end{description}
Table \ref{table.rq1.h3} presents the comparison of \rnn{} with \cnn{} and \cnntwo{}
by comparing pairwise F1 measure differences in percentages, where the F1 values are 
obtained by the optimal configuration in each case.
Here, the performance difference in percentage is calculated by
$(F1_{RNN} - F1_{CNN})/F1_{RNN} \times 100$.
\rnn{} performs far better for \ecb{} and \mn{} smells
against both convolution models.
However, the performance of \rnn{} is lower for \cm{} and \ma{} smells.

\begin{table}[h!]
	\caption{Performance (F1) comparison of RNN with CNN-1D and CNN-2D}
	\begin{tabular}{l|rr}
	Smell	& \rnn{} vs \cnn{}      & \rnn{} vs \cnntwo{}      \\ \hline
		{\sc cm}     & -22.94\%                         & -33.81\%                                              \\
		{\sc ecb}    & 80.23\%                        & 91.94\%                           \\
		{\sc mn}     & 57.19\%                        & 48.48\%                            \\
		{\sc ma}     & -353.15\%                        & -208.00\%                                
	\end{tabular}
	\label{table.rq1.h3}
\end{table}

\textbf{The analysis suggests that performance of the deep learning models is smell-specific.
	Therefore, we reject the hypothesis that \rnn{} models perform better than {\sc cnn} models for all
	considered smells.}
	


\subsubsection{Implications}
This is the first attempt in the software engineering literature to show the feasibility
of detecting smells using deep learning models 
from the tokenized source code without extensive feature engineering.
It may motivate researchers and developers to explore this direction and build over it.
For instance, \textit{context} plays an important role in deciding whether a reported smell
is actually a quality issue for the development team.
One of the future works that the community may explore is to combine
the models trained using samples classified by the existing smell detection tools
with the developer's feedback to identify more relevant smells considering the context. 

Our results show that, though both convolution methods perform superior for specific  smells,
their performance is comparable for each smell.
This imply that we may use one-dimensional or two-dimensional {\sc cnn} interchangeably
without compromising the performance significantly.

The comparative results on applying diverse deep learning models for detecting different  
types of smells suggest that there exists no universal optimal model for detecting all smells 
under consideration. 
The performance of the model is highly dependent on the kind of smell that the model is trying to classify.
This observation provides grounds for further investigation, encouraging the software engineering community to propose improvements on smell-specific deep learning models.

\subsection{Results of RQ2}
\begin{description}
	\item[RQ2] Is transfer-learning feasible in the context of detecting smells?
	If yes, which deep learning model exhibits superior performance in
	detecting smells when applied in transfer-learning setting?
\end{description}
\subsubsection{Approach}
In the case of direct-learning,
the training and evaluation samples belong to
the same programming language whereas in the transfer-learning case, the
training and evaluation samples come from two similar but different programming languages.
This research question inquires the feasibility of applying transfer-learning \textit{i.e.,} train 
neural networks by using C\# samples and employ the trained model to classify
code fragments written in Java.

For the transfer learning experiment we keep the training samples exactly 
the same as the ones we used in RQ1. 
For evaluation, we download repositories containing Java source code and 
preprocess the samples as described in Section \ref{data_curation}. 
Similar to RQ1, evaluation is performed on a
realistic scenario, \textit{i.e.,} we use all the positive and negative samples
from the selected repositories.
This arrangement ensures that the models would perform as reported if employed in a real-world application.
Table \ref{table:rq2.samples} shows the number of samples used for training and evaluation
for this research question.
\begin{table}[h!]
	\caption{Positive (P) and negative (N) number of samples used for training and evaluation for RQ2}
	\begin{tabular}{l|rrr|rrr}
		&\multicolumn{3}{c}{\cnn{} and \rnn{}}&\multicolumn{3}{c}{\cnntwo{}}\\
		& Training & \multicolumn{2}{c}{Evaluation}  
		& Training & \multicolumn{2}{c}{Evaluation}                              \\
		& {\sc p} and {\sc n}     & {\sc p} & {\sc n} & {\sc p} and {\sc n}     & {\sc p} & {\sc n}\\\hline
		{\sc cm}           & 3,472    & 2,163    & 48,633 & 2,641 & 2001 & 30,215  \\
		{\sc ecb}        & 1,200      & 597     & 50,199   & 982 & 538 & 31,678       \\
		{\sc mn}             & 5,000     & 42,037 & 50,905 & 5,000 & 7,778 & 24,438         \\
		{\sc ma}  & 290      & 25    & 13,110 & 284 & 23 & 11,812 \\\hline         
	\end{tabular}
	\label{table:rq2.samples}
\end{table}

\subsubsection{Results}
\begin{figure}
	\centering
	\begin{subfigure}[b]{0.3\textwidth}
		\includegraphics[width=\textwidth]{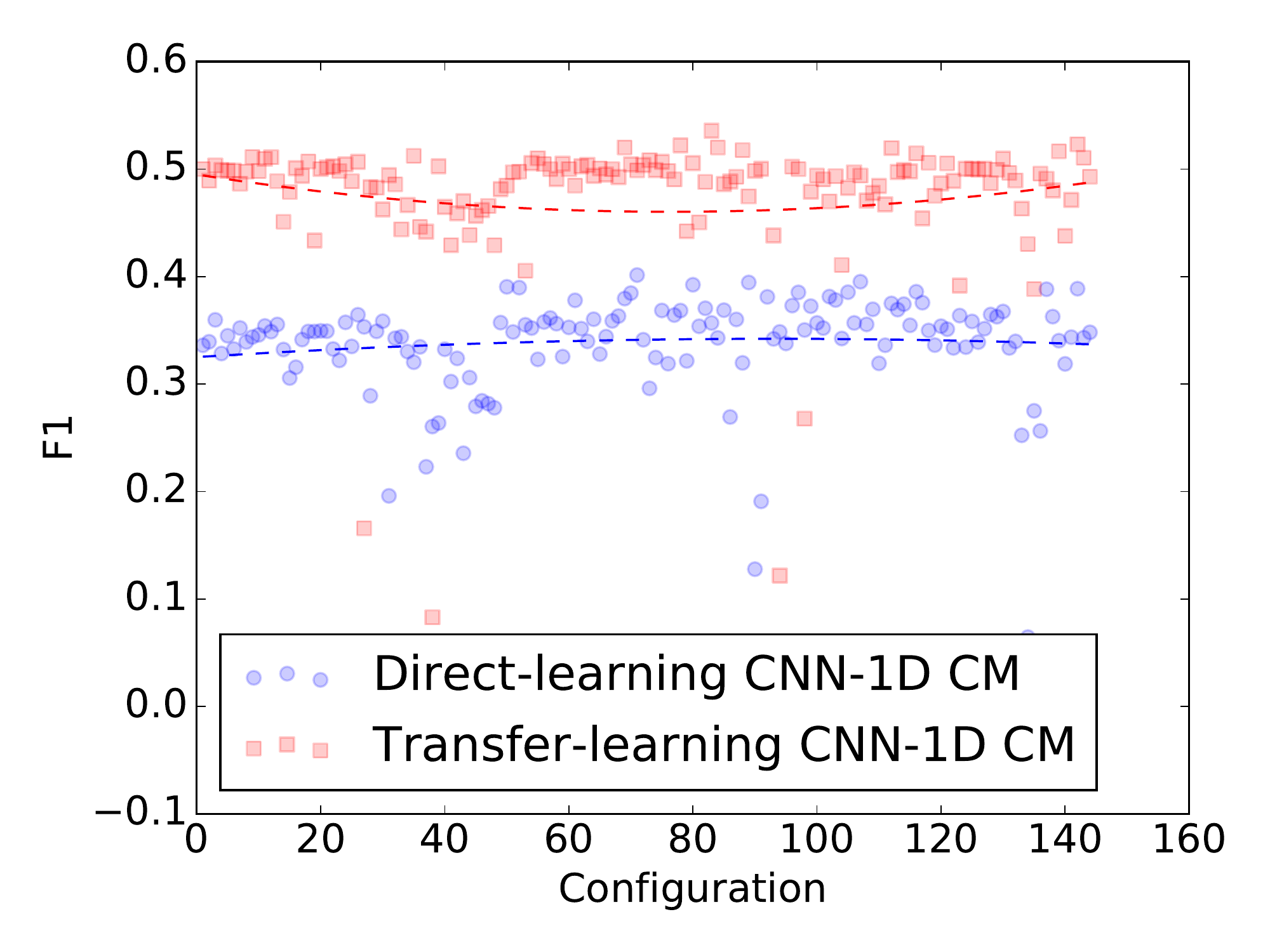}
		\caption{CNN-1D for \cm{} smell}
	\end{subfigure}
	~ 
	\begin{subfigure}[b]{0.3\textwidth}
		\includegraphics[width=\textwidth]{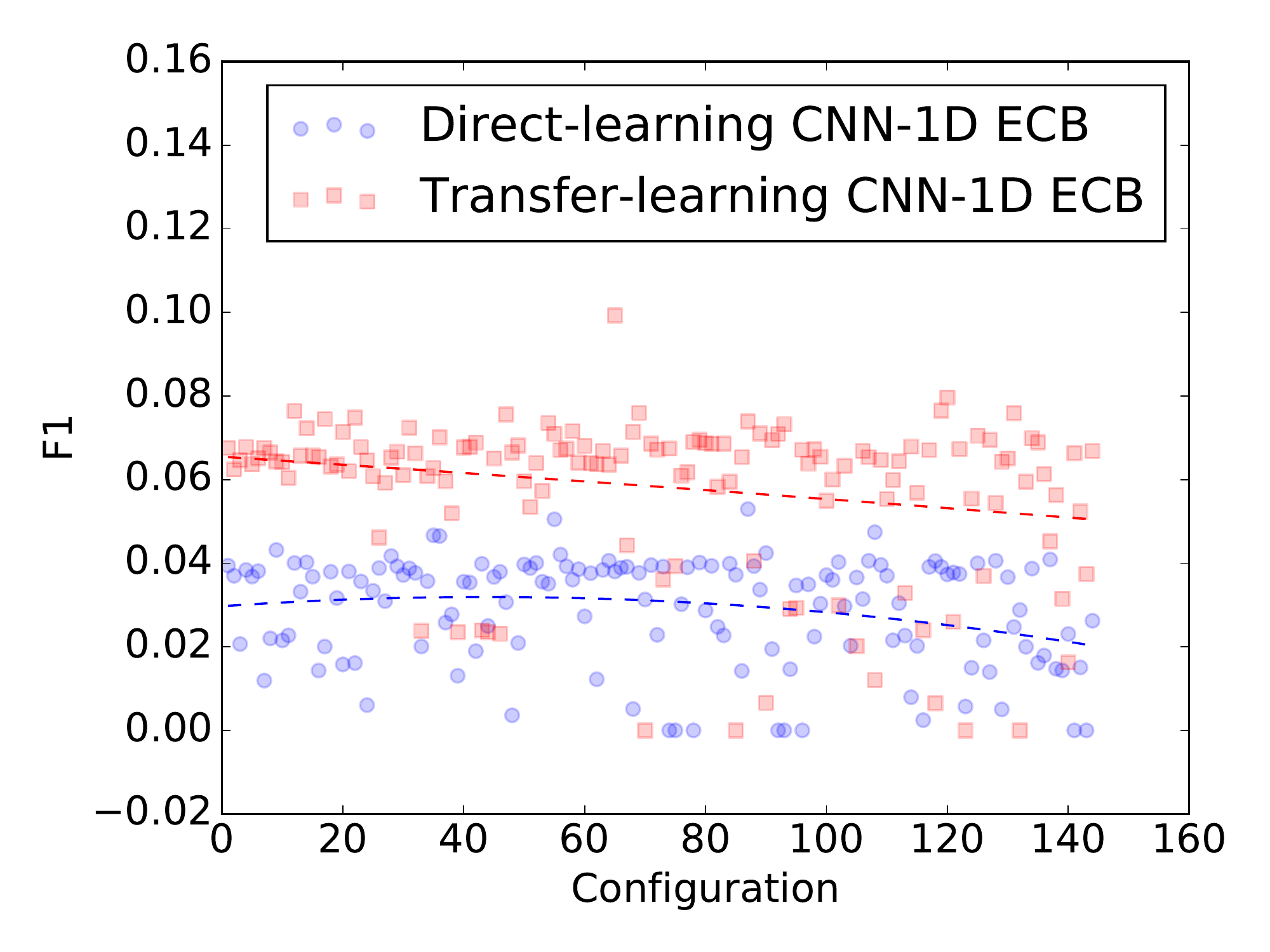}
		\caption{CNN-1D for \ecb{} smell}
	\end{subfigure}
	\begin{subfigure}[b]{0.3\textwidth}
		\includegraphics[width=\textwidth]{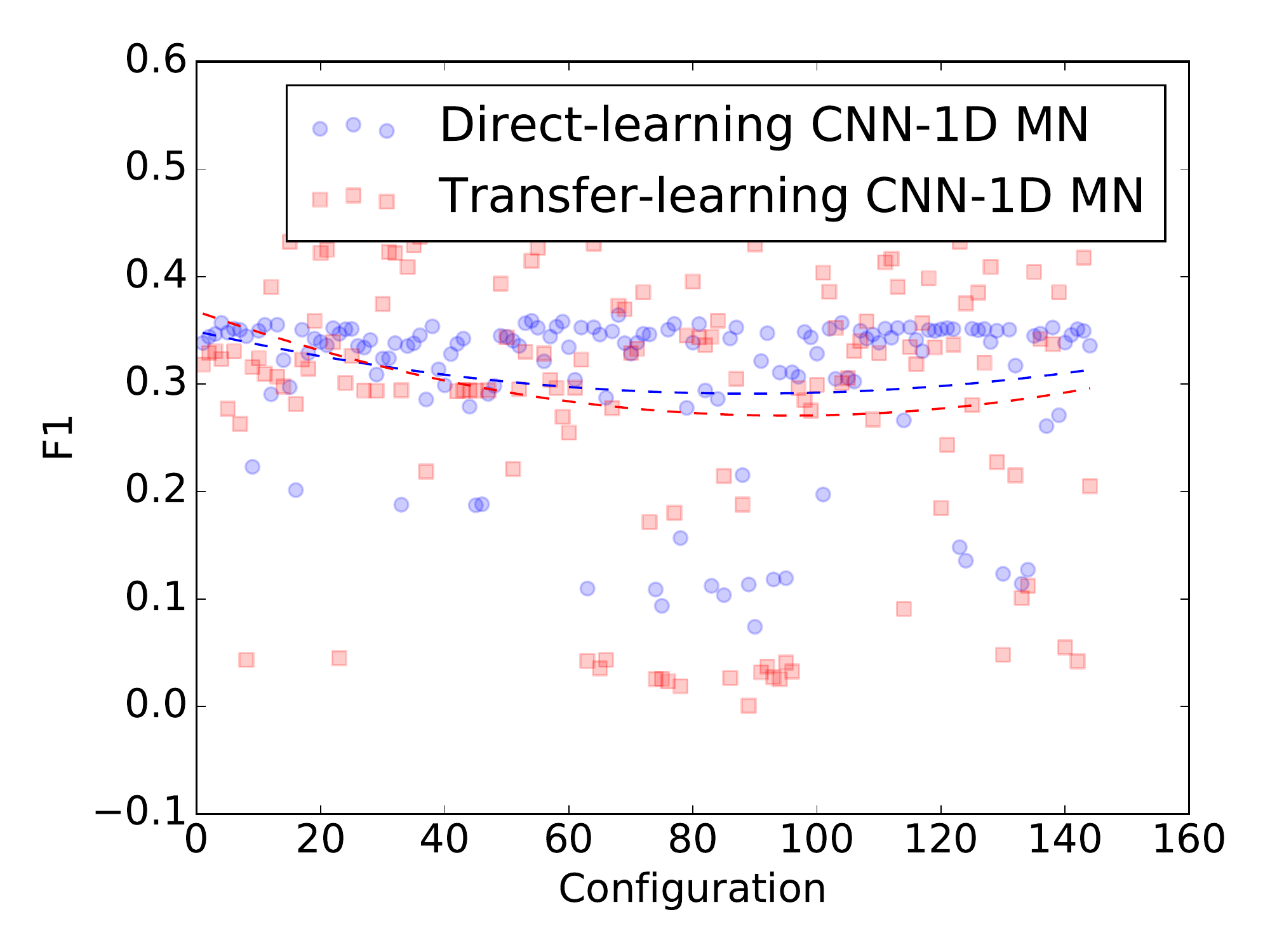}
		\caption{CNN-1D for \mn{} smell}
	\end{subfigure}
	
	\begin{subfigure}[b]{0.3\textwidth}
		\includegraphics[width=\textwidth]{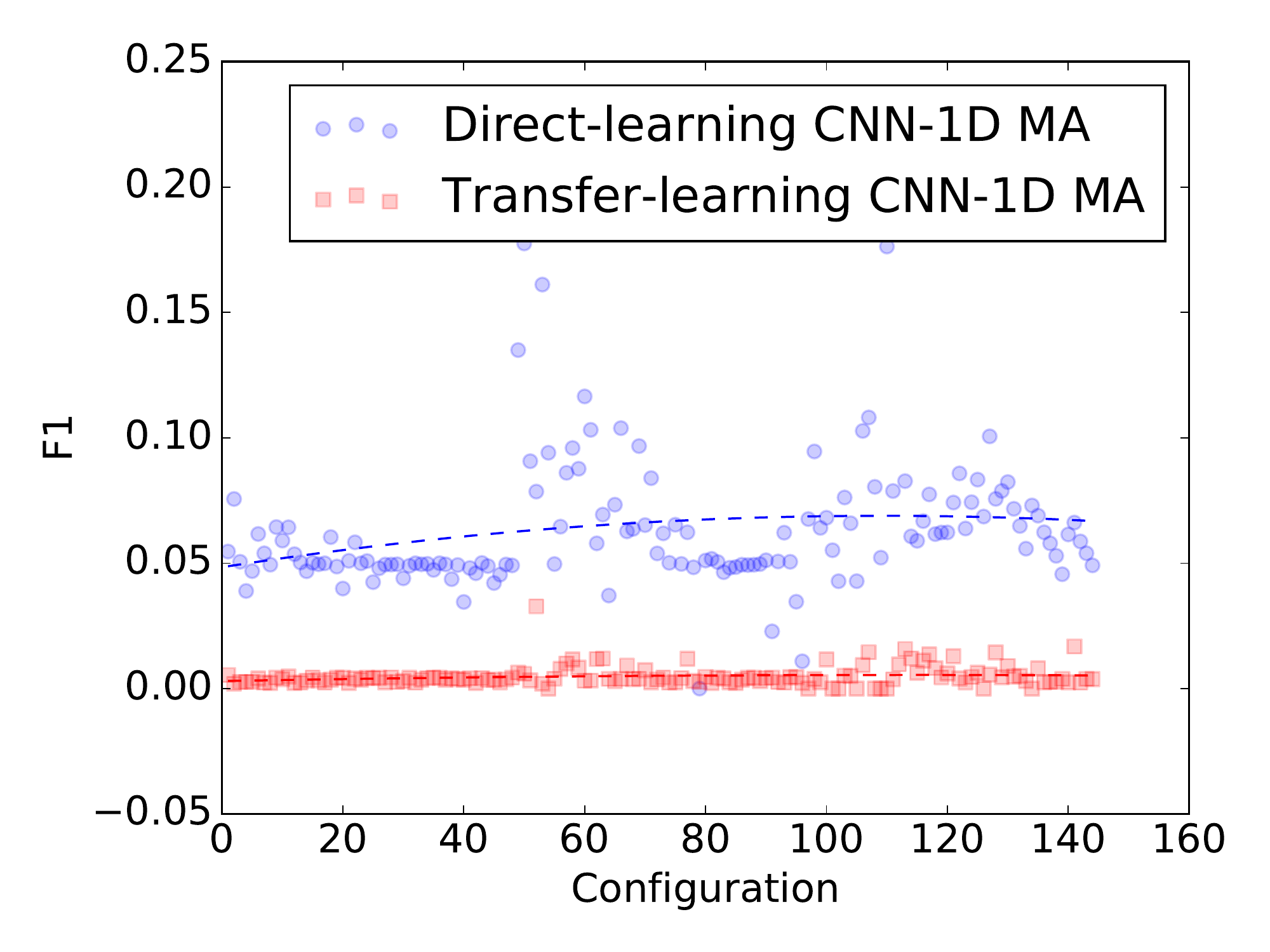}
		\caption{CNN-1D for \ma{} smell}
	\end{subfigure}
	\begin{subfigure}[b]{0.3\textwidth}
		\includegraphics[width=\textwidth]{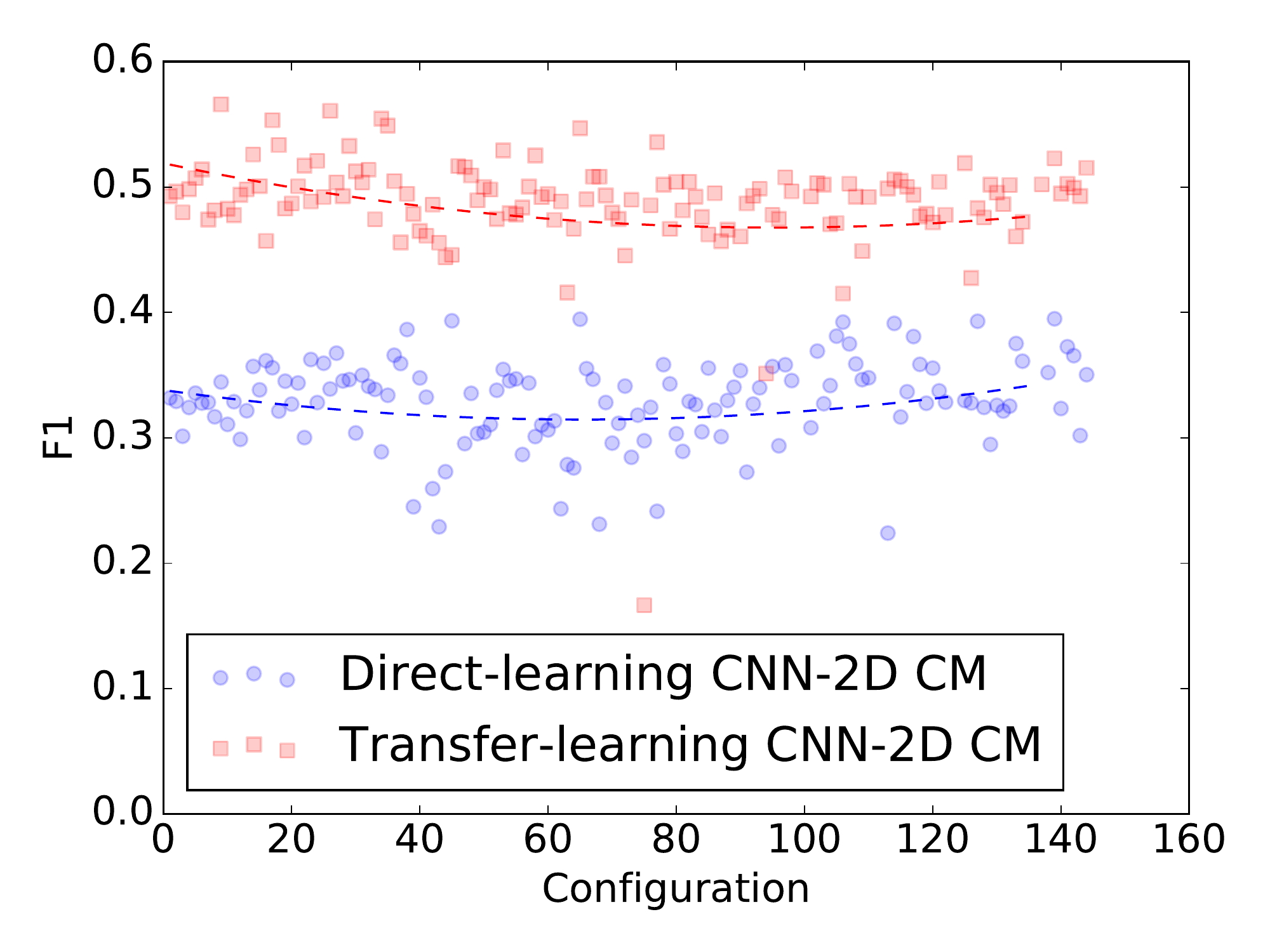}
		\caption{CNN-2D  for \cm{} smell}
	\end{subfigure}
	\begin{subfigure}[b]{0.3\textwidth}
		\includegraphics[width=\textwidth]{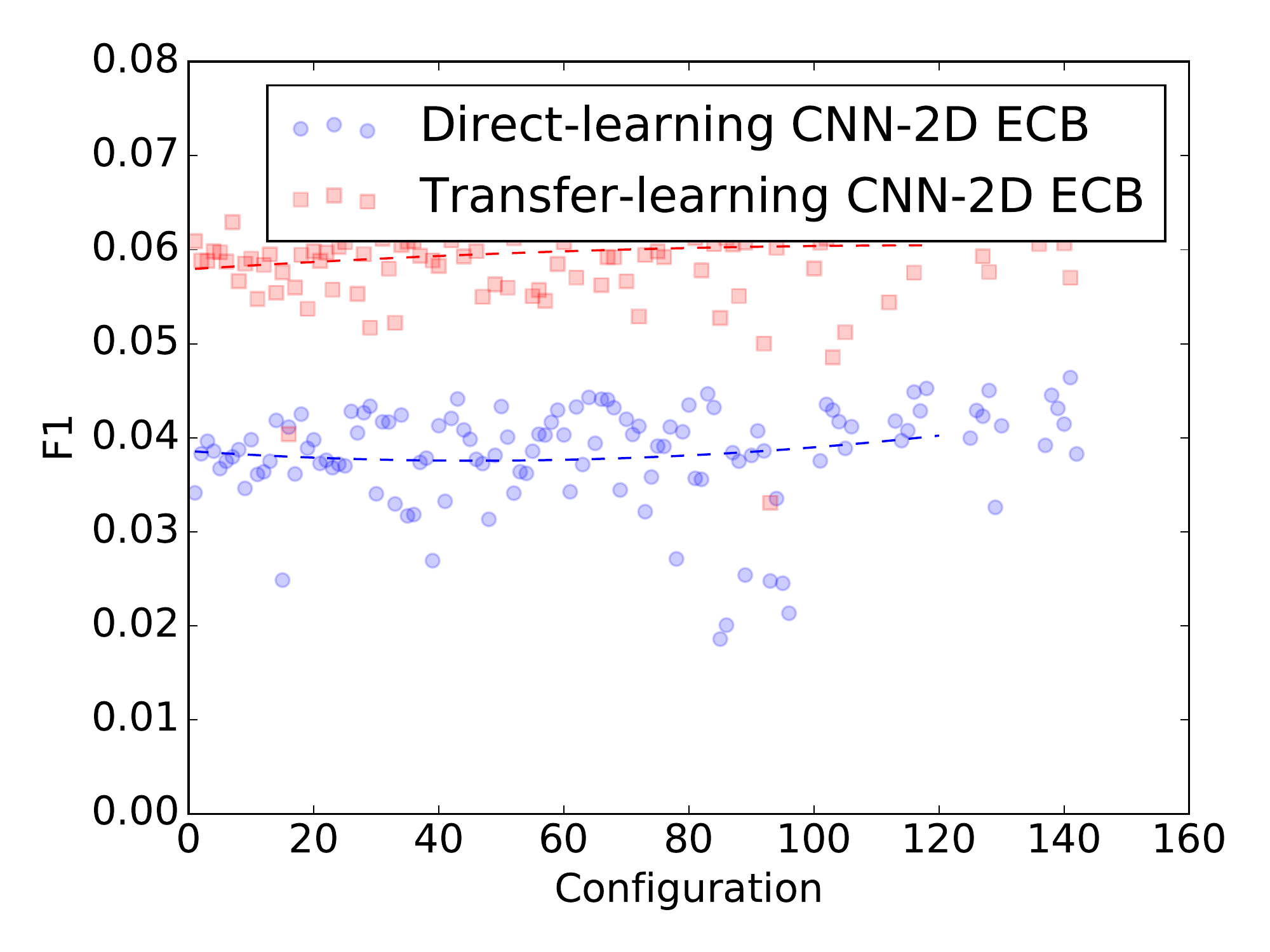}
		\caption{CNN-2D  for \ecb{} smell}
	\end{subfigure}
	
	\begin{subfigure}[b]{0.3\textwidth}
		\includegraphics[width=\textwidth]{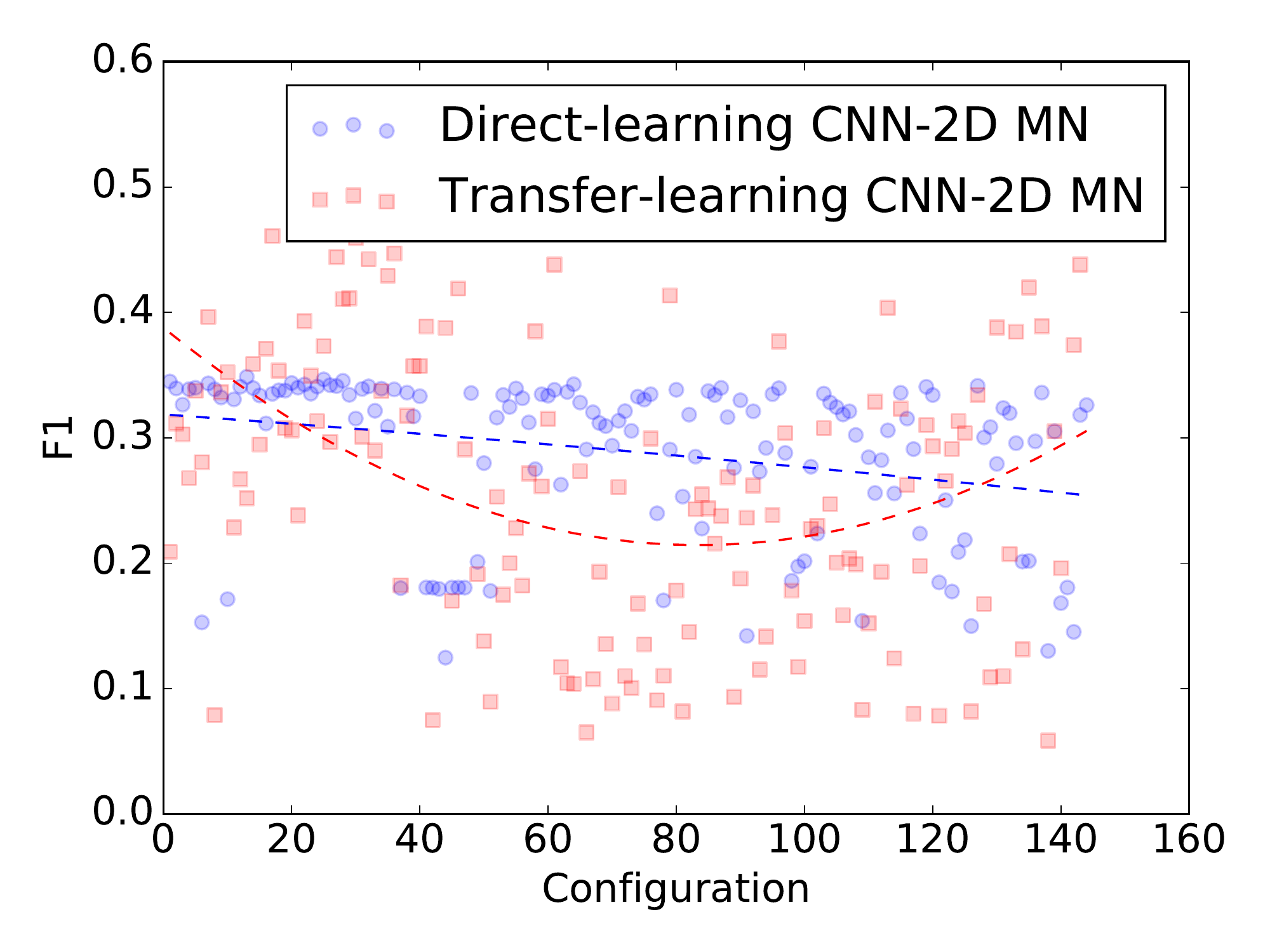}
		\caption{CNN-2D  for \mn{} smell}
	\end{subfigure}
	\begin{subfigure}[b]{0.3\textwidth}
		\includegraphics[width=\textwidth]{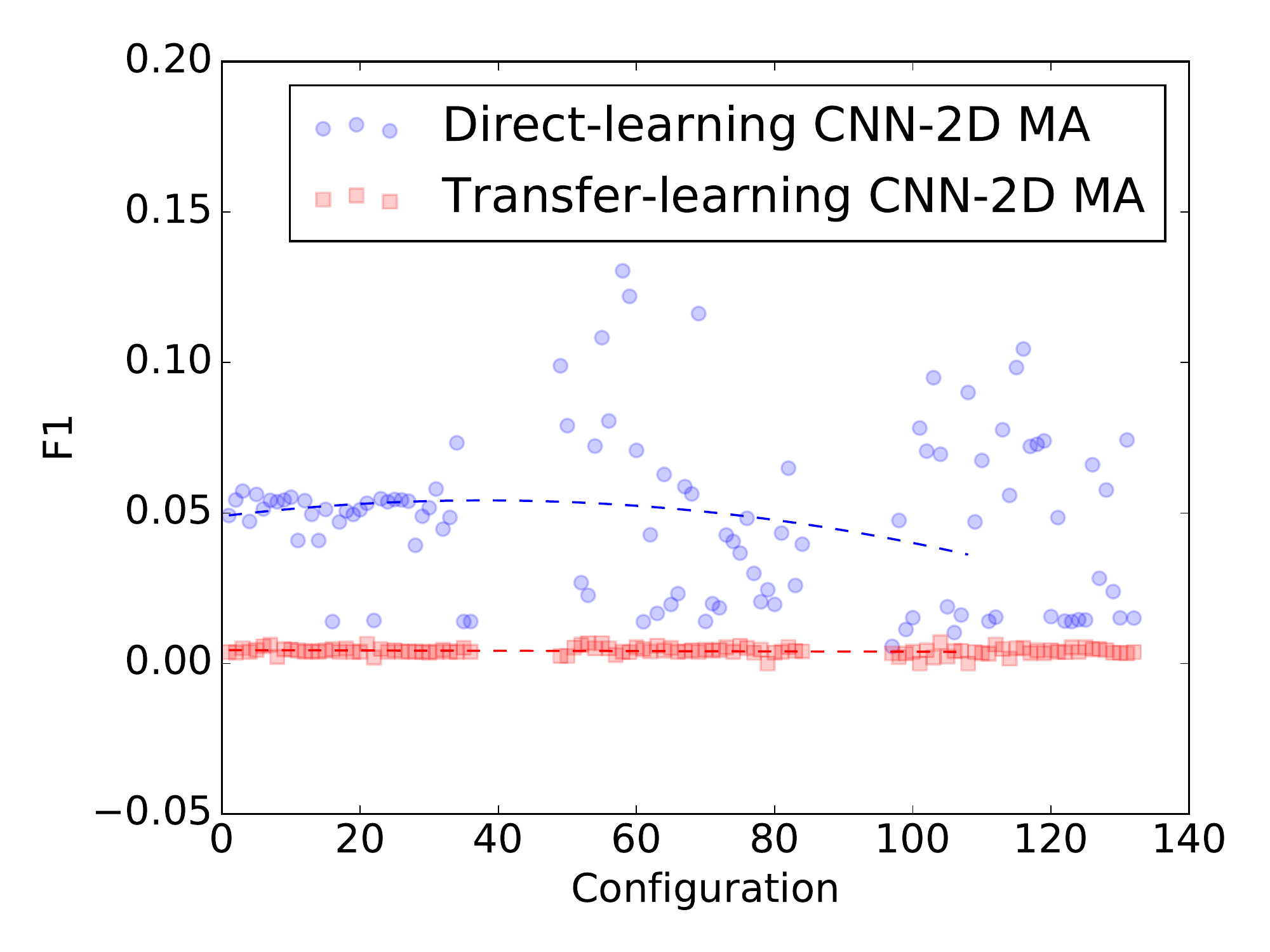}
		\caption{CNN-2D  for \ma{} smell}
	\end{subfigure}
	\begin{subfigure}[b]{0.3\textwidth}
		\includegraphics[width=\textwidth]{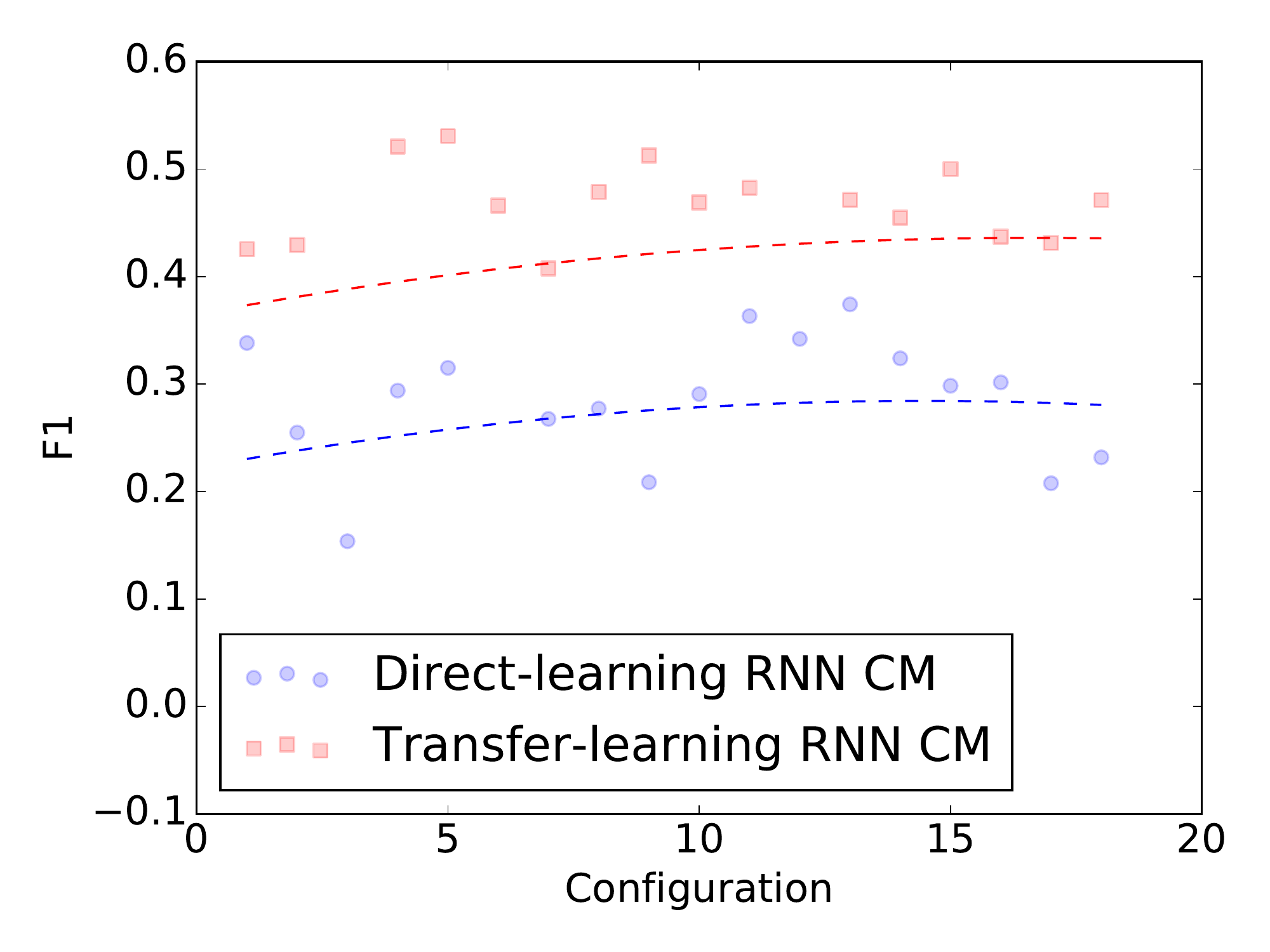}
		\caption{RNN for \cm{} smell}
	\end{subfigure}
	
	\begin{subfigure}[b]{0.3\textwidth}
		\includegraphics[width=\textwidth]{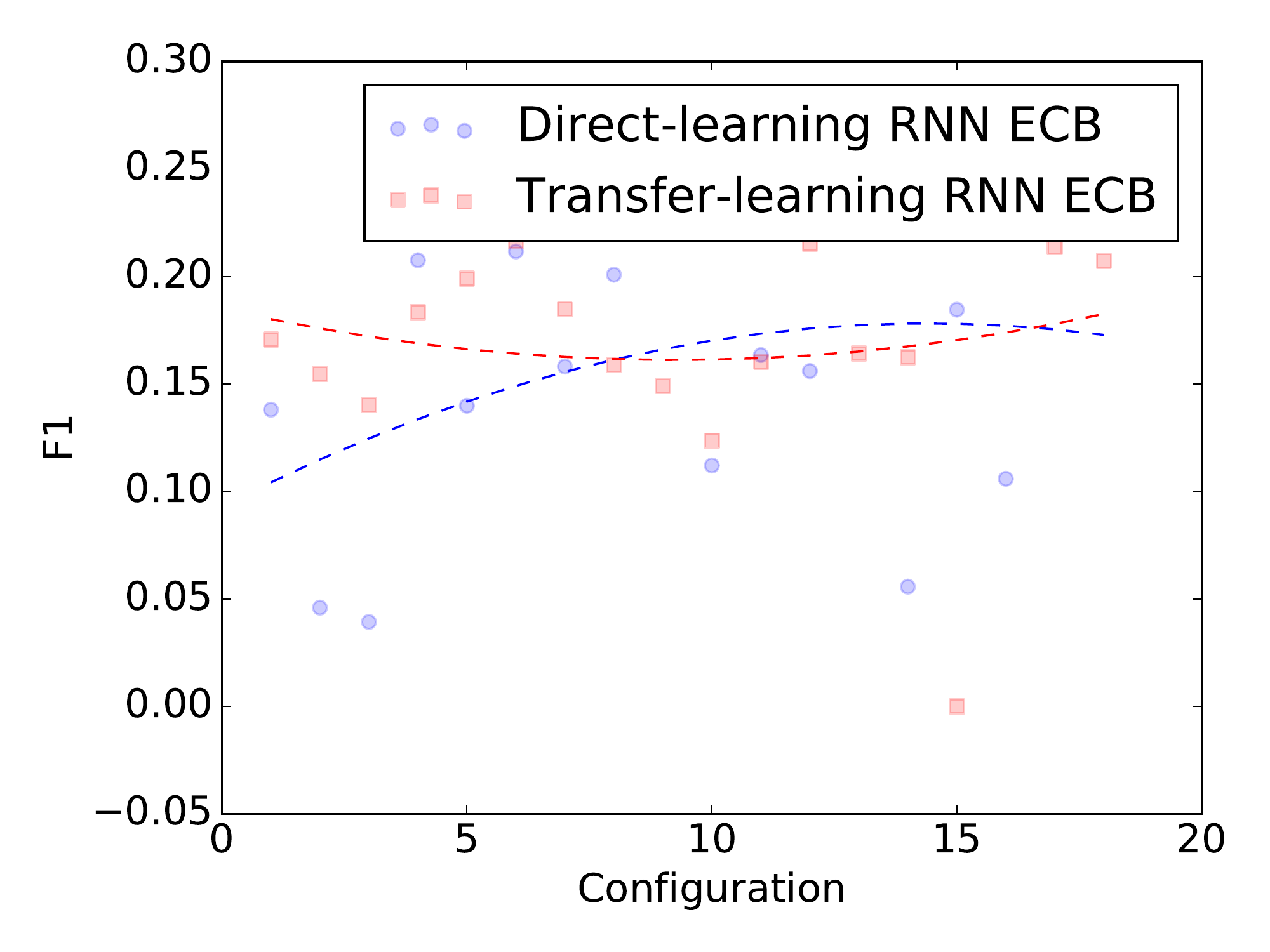}
		\caption{RNN for \ecb{} smell}
	\end{subfigure}
	\begin{subfigure}[b]{0.3\textwidth}
		\includegraphics[width=\textwidth]{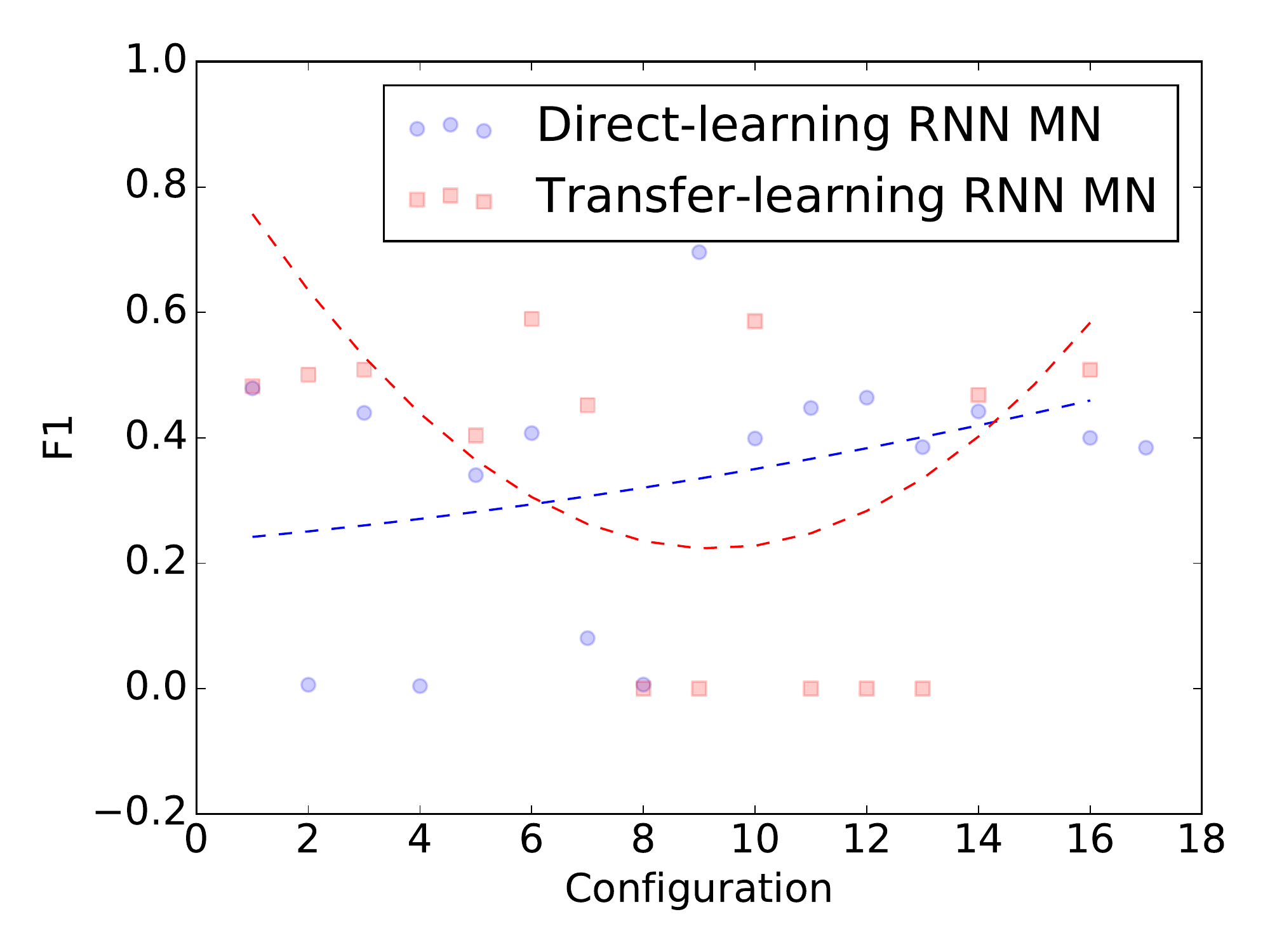}
		\caption{RNN for \mn{} smell}
	\end{subfigure}
	\begin{subfigure}[b]{0.3\textwidth}
		\includegraphics[width=\textwidth]{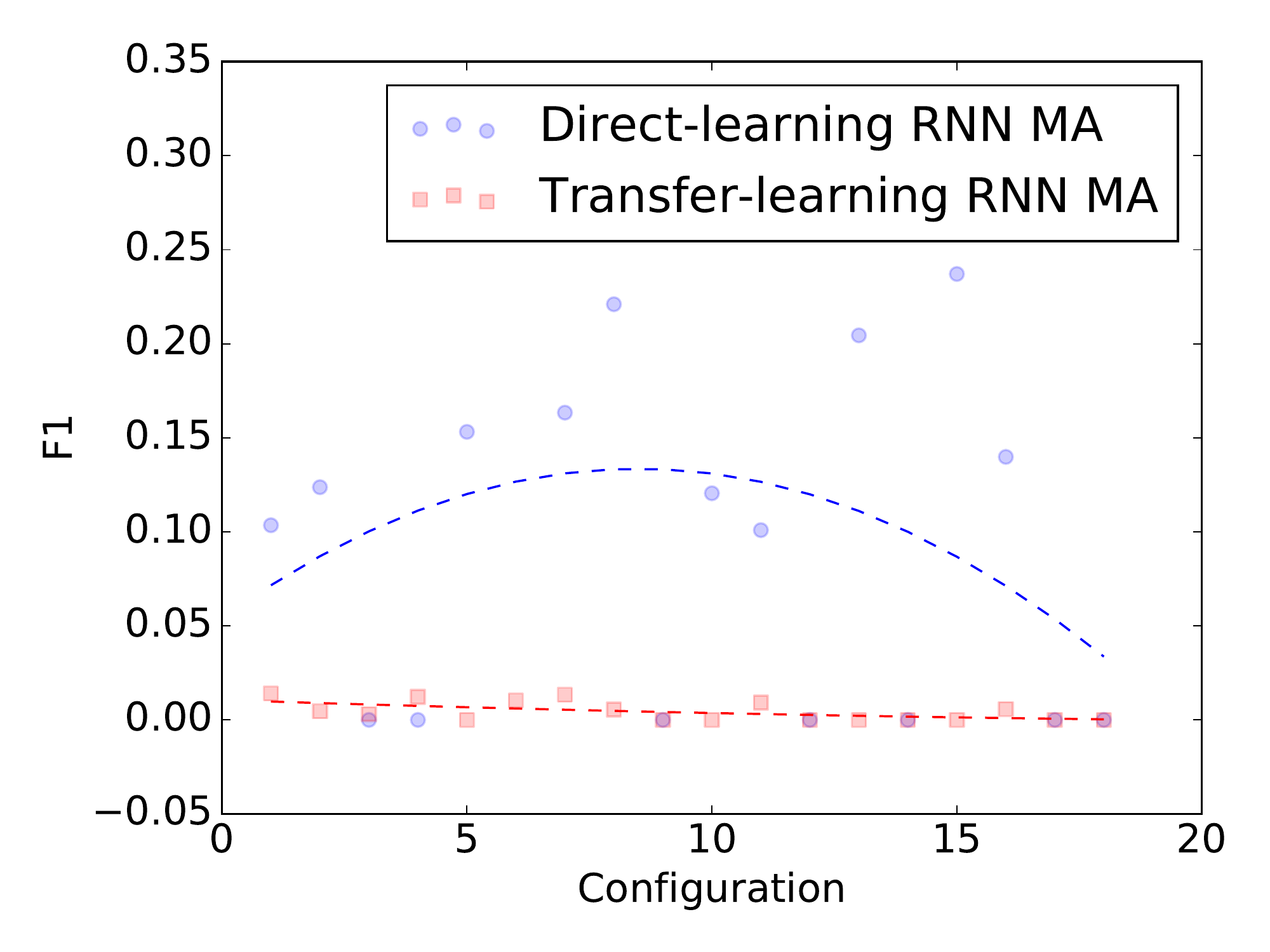}
		\caption{RNN for \ma{} smell}
	\end{subfigure}
	\caption{Scatter plots for each model and for each considered smell comparing F1  of direct-learning and transfer-learning along with corresponding trendline}\label{fig:rq2.2}
\end{figure}

As an overview, 
Figure \ref{fig:rq2.2} shows the scatter plots for each deep learning model comparing
the performance (F1) of both the direct-learning and transfer-learning for all the considered smells for all the configurations.
These plots outline the performance exhibited by the models in both the cases with 
trend lines distinguishing the compared series.
The plots imply that the models perform better in the transfer-learning case for all except \ma{} design smell.

In the rest of the section, we report quantitative results on applying transfer 
learning between C\# to Java. 
The results are based on the optimal configuration of each model for each smell. 

\begin{description}
	\item[RQ2.H1]\textit{ It is feasible to apply transfer-learning in the context of code smells detection.}
\end{description}

Table \ref{table:rq2.results.1} presents the performance of the models for all the
considered smells demonstrating strong evidence on the feasibility of applying transfer-learning for  smell detection. 
The performance pattern is in alignment to that in the direct-learning case;
Spearman correlation between the performance produced by direct-learning and
transfer-learning is $0.98$ (with p-value = $1.309\times 10^{-8}$).
\textbf{Therefore, we accept the hypothesis that transfer-learning is feasible  in the context of code smells detection.}

\begin{table}[h!]
 	\caption{Performance of all three models with configuration corresponding to the optimal performance. L refers to deep learning layers, F refers to number of filters, K refers to kernel size, MPW refers to maximum pooling window size, ED refers to embedding dimension, LSTM refers to number of LSTM units, and E refers to number of epochs}
\begin{tabular}{l|lrrrr|rrrrrrr}
	&&\multicolumn{4}{c}{Performance}&\multicolumn{7}{c}{Configuration}\\
	& Smells     & AUC    & Precision & Recall & F1 & {\sc l} & {\sc f}  & {\sc k} & {\sc mpw} & {\sc ed} & {\sc lstm} & {\sc e}    \\ \hline
	\multirow{4}{*}{\cnn{} } & {\sc cm}      & 0.87& 0.38 & 	0.79	& 0.51 & 2 & 32 & 7 & 4 & --  & --  & 23\\
	& {\sc ecb}     & 0.56 & 0.05	& 0.15	& 0.08 & 3 & 8 & 5 & 5 & --  & --  & 27\\
	& {\sc mn}      & 0.64 & 0.48	& 0.37	& 0.42 &  1 &  32 &  11 &  3   & --  & --  & 12\\
	& {\sc ma}     & 0.52 & 0.01 &	0.04 &	0.02 &  2 &  8 &  11 &  5  & --  & --  & 13\\ \hline
	\multirow{4}{*}{\cnntwo{} } & {\sc cm}      & 0.88 & 0.43 &	0.84 &	0.57 &  1 &  8 &  7 &  2   & --  & --  & 37\\
	& {\sc ecb}     & 0.54 & 0.04 &	0.12	& 0.06 &  3 &  16 &  5 &  4  & --  & --  & 19\\
	& {\sc mn}   & 0.65 & 0.43 & 0.54	& 0.48 &  1 &  64 &  5 &  4  & --  & --  & 8\\
	& {\sc ma}       & 0.50 & 0.0	& 0.0	& 0.0 & 3  &  8 &  5 &  5   & --  & --  & 17\\ \hline
	\multirow{4}{*}{\rnn{}}    & {\sc cm}       & 0.66 & 0.62 &	0.32 &	0.42 &  1 & & --  & --  &32 &  64  & 8\\
	& {\sc ecb}      & 0.90 & 0.09	& 0.91 &	0.16 &  3 & & --  & --  &32 &  32   & 27 \\
	& {\sc mn}      & 0.95& 0.94 & 0.91& 0.92& 1 & & --  & --  &32 &  32  & 22\\
	& {\sc ma}      & 0.51 & 0.0& 0.08& 0.0& 1 & & --  & --  &32 &  32 & 18\\ \hline
	\end{tabular}
	\label{table:rq2.results.1}
\end{table}

\begin{figure}[h!]
	\centering
	\includegraphics[width=0.6\linewidth]{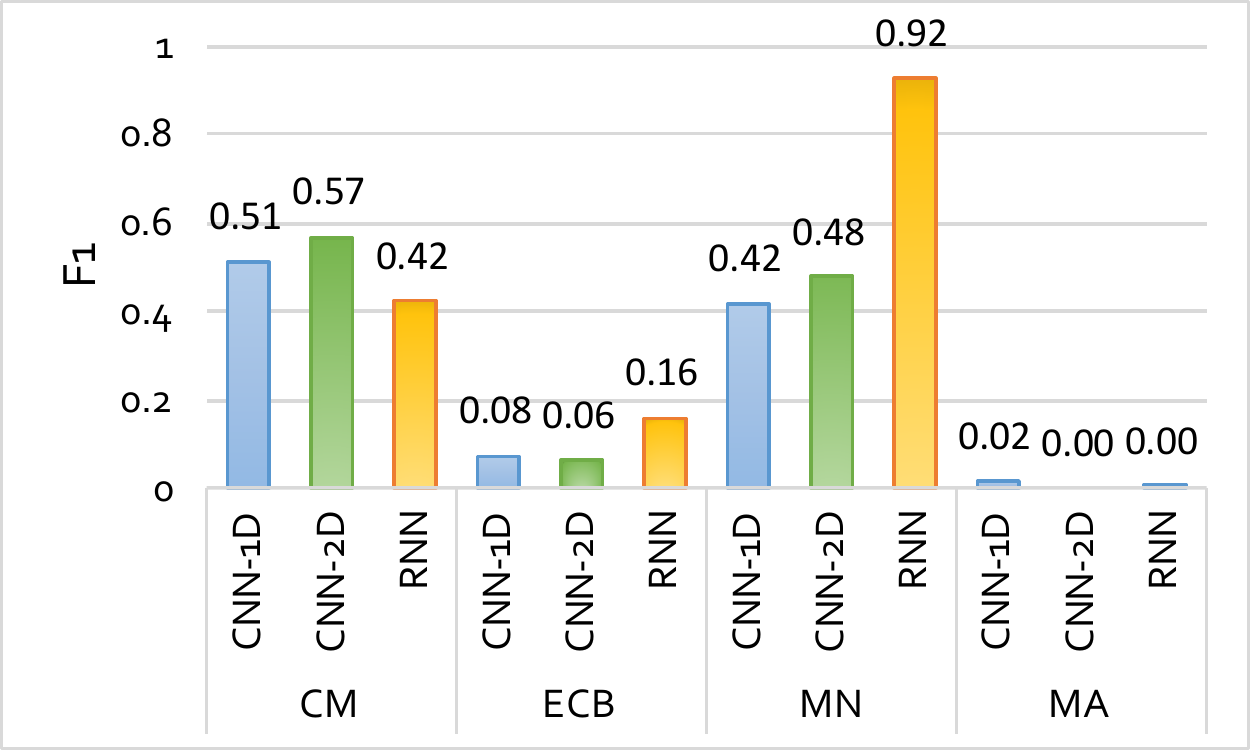}
	\caption{Comparative performance of the deep learning models for each considered smell in transfer-learning settings}
	\label{fig:rq2.h1.f1}
\end{figure} 

Figure \ref{fig:rq2.h1.f1} presents a comparison among the performance (\textit{i.e.,} F1) exhibited by
all the deep learning models for each considered smell.
\rnn{} performs significantly superior for \ecb{} and \mn{} smells following a trend
comparable to direct-training.
For \cm{} smell, \cnntwo{} performs the best followed by \cnn{}.
All the three models perform poorly with \ma{} smell.

\begin{description}
	\item[RQ2.H2] \textit{Transfer-learning performs inferior compared to direct learning.}
\end{description}

\begin{figure}[h!]
	\centering
	\includegraphics[width=0.8\linewidth]{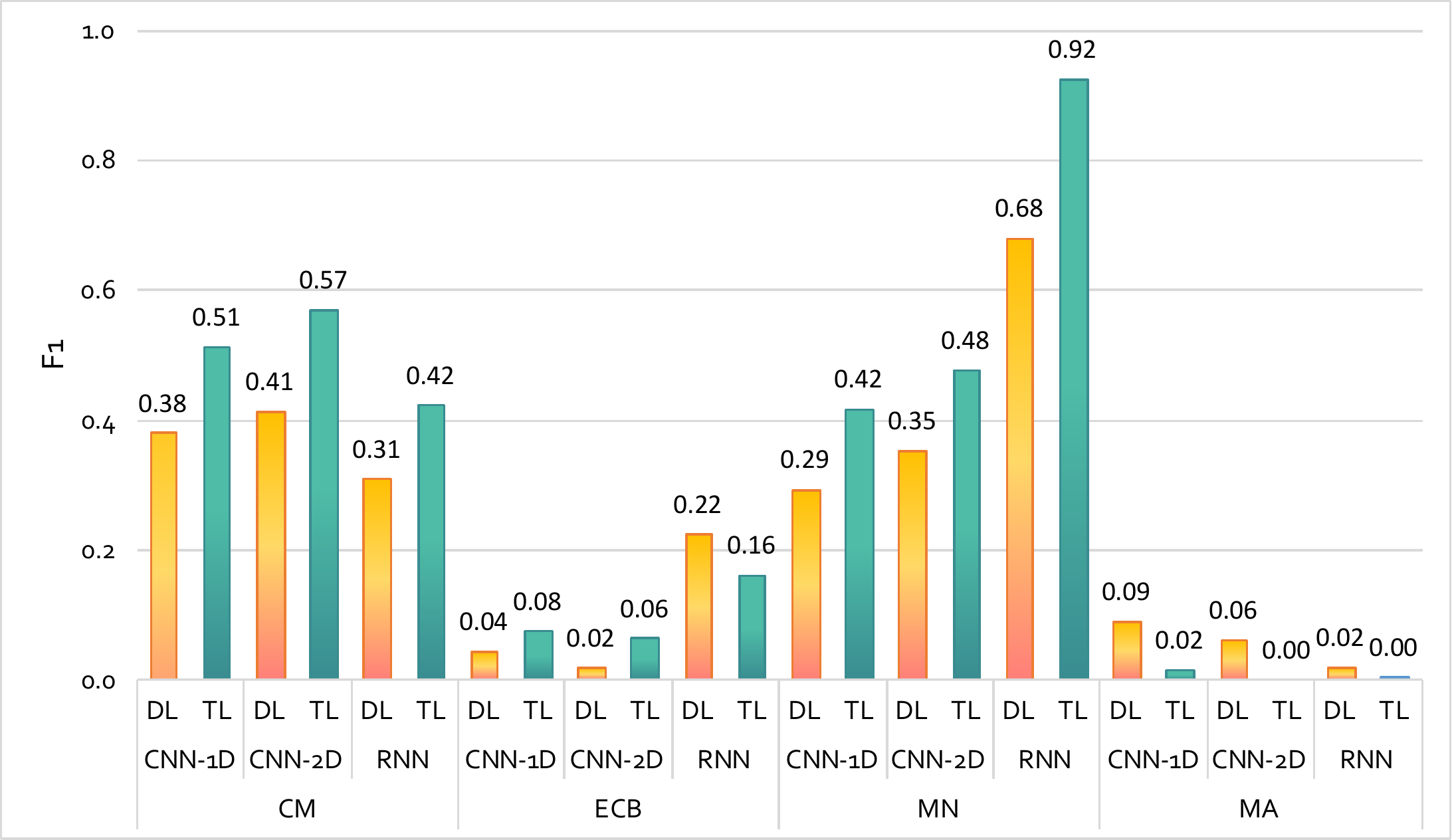}
	\caption{Comparison of performance of the deep learning models between direct-learning (DL) and transfer-learning (TL) settings}
	\label{fig:rq2.h2.f1}
\end{figure} 

Figure \ref{fig:rq2.h2.f1} compares the performance 
of the models at their optimal configurations applied in the transfer-learning and in 
direct-learning.
We observe that, in majority of cases, transfer-learning performs better than the corresponding
direct-learning counterpart models.
The only exception for implementation smells is \rnn{} applied on \ecb{} smell,
where direct-learning shows better results. 
For the only design smell, \textit{i.e.,} \ma{}, all the models perform poorly in both  cases.

To dig deeper into the cause of better performance of deep learning models in the transfer-learning case,
we calculate the ratio of positive and negative evaluation samples in both research questions.
Table \ref{table:rq2.sample.diff} presents the ratio for samples used in both the research questions
as well as percentage
difference of the ratios of positive and negative samples
in RQ2 compared to the sample ratio in RQ1.
The percentage difference is computed as follows:
$(Ratio_{RQ2} - Ratio_{RQ1})/Ratio_{RQ1} \times 100$.
It is evident that Java code samples have higher ratio of positive samples,
up to $188\%$ higher, compared to C\# samples for implementation smells.
We deduce that due to significantly higher number of positive samples,
the deep learning models show better performance statistics in the transfer-learning case.
On the other hand, \ma{} smell occurs significantly lower (up to $72\%$) in Java code compared to C\# code
and this lower ratio further degrades the performance of the models for \ma{} smell.

\begin{table}[h!]
	\caption{Difference in ratio (in percent) of positive and negative evaluation samples in RQ2 compared to sample ratio in RQ1}
	\begin{tabular}{l|rr|rr|rr}
		&\multicolumn{2}{c}{Ratio (RQ1)}&\multicolumn{2}{c}{Ratio (RQ2)}&\multicolumn{2}{c}{Difference \%}
		\\
		Smell&  {\sc 1d} & {\sc 2d}& {\sc 1d} & {\sc 2d}& {\sc 1d} & {\sc 2d}\\\hline
		{\sc cm}& 0.0287& 0.0250& 0.0445& 0.0662& 35.53& 62.19\\
		{\sc ecb} &0.0097& 0.0092& 0.0119& 0.0170& 18.14& 45.88\\
		{\sc mn} &0.1242& 0.1210& 0.2084& 0.3183& 40.40& 61.98\\
		{\sc ma}  & 0.0055& 0.0070& 0.0019& 0.0019& -188.42& -260.87 \\\hline         
	\end{tabular}
	\label{table:rq2.sample.diff}
\end{table}

We perform an additional experiment
to ensure that the better performance exhibited by transfer-learning compared to direct-learning
is due to the higher number of positive instances in the Java samples compared to C\# evaluation samples.
In this experiment, we reverse the direction of transfer-learning \textit{i.e.,}
we train the models using Java samples and
evaluate the trained models on C\# samples.

We download Java repositories as described in Section \ref{data_curation} and select $500$ repositories randomly
to prepare training samples.
Similarly, we select $100$ C\# repositories to create evaluation samples.
We perform the data curation operations mentioned in Section \ref{data_curation} and compile a set of
training and evaluation samples.
Table \ref{table:rq2.2.samples} presents the number of samples that are used for this experiment.

\begin{table}[h!]
	\caption{Number of positive (P) and negative (N) samples used for training (Java samples) and evaluation (C\# samples)}
	\begin{tabular}{l|rrr|rrr}
		&\multicolumn{3}{c}{\cnn{} and \rnn{}}&\multicolumn{3}{c}{\cnntwo{}}\\
		& Training & \multicolumn{2}{c}{Evaluation}  
		& Training & \multicolumn{2}{c}{Evaluation}                              \\
		& {\sc p} and {\sc n}     & {\sc p} & {\sc n} & {\sc p} and {\sc n}     & {\sc p} & {\sc n}\\\hline
		{\sc cm}           & 3,794    & 627    & 28,974 & 3,323 & 586 & 25,486  \\
		{\sc ecb}        & 3,821      & 191     & 29,410   & 3,386 & 181 & 25,891       \\
		{\sc mn}             & 3,547     & 2,224& 27,377 & 3,124 & 2,348 & 23,724         \\
		{\sc ma}  & 237      & 281    & 11,543 & 213 & 277 & 9,321 \\\hline         
	\end{tabular}
	\label{table:rq2.2.samples}
\end{table}

\begin{figure}[h!]
	\centering
	\includegraphics[width=0.7\linewidth]{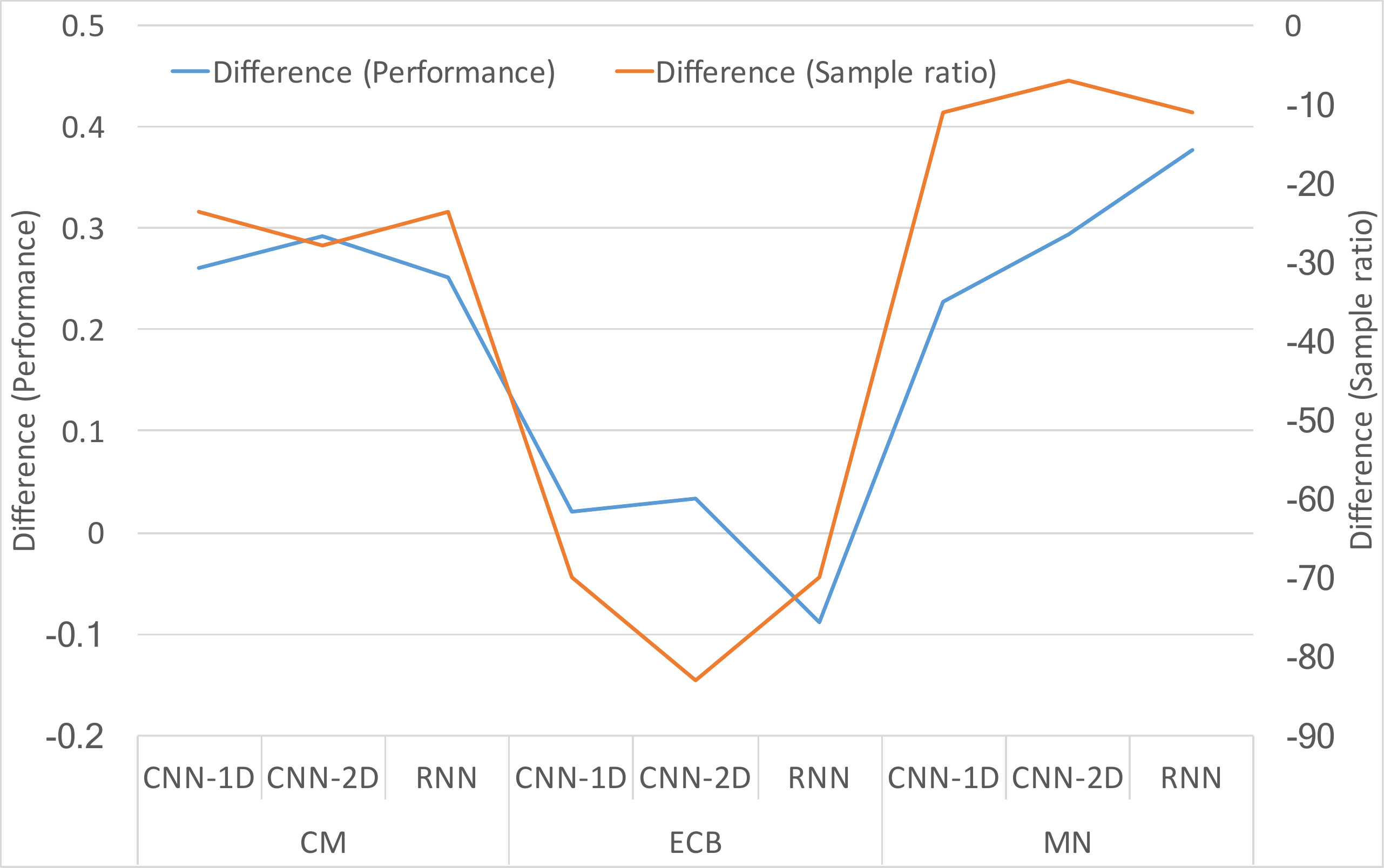}
	\caption{Difference in performance of the deep learning models and the sample ratio for two transfer-learning tasks}
	\label{fig:rq2.h2.diff}
\end{figure}

	We perform the experiment on all configurations, identify optimal configurations for the models,
	and measure the performance
	for the optimal configurations.
	Table \ref{table:rq2.2.results.1} presents the obtained results for all the models with all the smells.
	We observe that the performance of the models is far lower than the performance reported in
	Table \ref{table:rq2.results.1}. 
	We compare the difference in performance of models with
        difference in sample ratios in both the cases---in the first case, models are trained using C\# samples and evaluated on Java samples and in the second case
	models are trained using Java samples and evaluated on C\# samples.
	We perform the above comparison for all but \ma{} smell because the models perform very poorly for the smell.
	Figure \ref{fig:rq2.h2.diff} shows the difference in performance of the deep learning models
	and the difference in the sample ratios of the above-mentioned transfer-learning tasks.
	The figure shows that the difference in performance and the difference in sample ratios are strongly correlated;
	hence, as the number of positive instances in evaluation
        sample increase, the models perform better.
        Indeed, the Spearman correlation between the two is high at $=
        0.7342$ (with p-value  $= 0.024$).

\begin{table}[h!]
	\caption{Performance of all three models with configuration corresponding to the optimal performance. L refers to deep learning layers, F refers to number of filters, K refers to kernel size, MPW refers to maximum pooling window size, ED refers to embedding dimension, LSTM refers to number of LSTM units, and E refers to number of epochs}
	\begin{tabular}{l|lrrrr|rrrrrrr}
		&&\multicolumn{4}{c}{Performance}&\multicolumn{7}{c}{Configuration}\\
		& Smells     & AUC    & Precision & Recall & F1 & {\sc l} & {\sc f}  & {\sc k} & {\sc mpw} & {\sc ed} & {\sc lstm} & {\sc e}    \\ \hline
		\multirow{4}{*}{\cnn{} } & {\sc cm}      & 0.67& 0.21 & 	0.31	& 0.25 & 3 & 8 & 5 & 5 & --  & --  & 5\\
		& {\sc ecb}     & 0.53 & 0.04	& 0.07	& 0.05 & 2 & 32 & 11 & 3 & --  & --  & 38\\
		& {\sc mn}      & 0.56 & 0.22	& 0.16	& 0.19 &  3 &  16 &  11 &  3   & --  & --  & 14\\
		& {\sc ma}     & 0.77 & 0.05 &	0.99 &	0.10 &  1 &  8 &  7 &  3  & --  & --  & 5\\ \hline
		\multirow{4}{*}{\cnntwo{} } & {\sc cm}      & 0.83 & 0.17 &	0.75 &	0.28 &  3 &  16 &  7 &  5   & --  & --  & 9\\
		& {\sc ecb}     & 0.52 & 0.02 &	0.05	& 0.03 &  3 &  16 &  5 &  2  & --  & --  & 7\\
		& {\sc mn}   & 0.55 & 0.35 & 0.12	& 0.18 &  3 &  16 &  7 &  2  & --  & --  & 7\\
		& {\sc ma}       & 0.55 & 0.09	& 0.13	& 0.11 & 2  &  16 &  7 &  2   & --  & --  & 19\\ \hline
		\multirow{4}{*}{\rnn{}}    & {\sc cm}       & 0.82 & 0.09 &	0.81 &	0.17 &  2 & & --  & --  &16 &  64  & 4\\
		& {\sc ecb}      & 0.91 & 0.15	& 0.85 &	0.25 &  3 & & --  & --  &16 &  128   & 13 \\
		& {\sc mn}      & 0.90& 0.39 & 0.92& 0.55& 3 & & --  & --  &32 &  64  & 28\\
		& {\sc ma}      & 0.66 & 0.05& 0.63& 0.09& 1 & & --  & --  &16 &  32 & 23\\ \hline
	\end{tabular}
	\label{table:rq2.2.results.1}
\end{table}

\textbf{Therefore, due to size discrepancies in the samples available 
	for evaluation in direct-learning and transfer-learning, we cannot conclude
	the superiority or inferiority of the results produced by applying transfer-learning compared to those of direct-learning.}

\subsubsection{Implications}
Our results demonstrate that it is feasible to apply transfer-learning in the smell detection context.
Exploiting this approach can lead to a new category
of smell detection tools, specifically for the programming languages
where comprehensive smell detection tools are not available.


\subsection{Discussion}
As is the case with most research, our results are sobering rather than 
sensational.  
Although it is possible to detect some code
smells using deep learning models, the method is by no means universal,
and the outcome is sensitive to the training set composition and the
training time.
In the rest of the section, we elaborate on these observations
emerging from the presented results.

\subsubsection{Is there any silver-bullet?}
In practical settings
one would want to employ 
a universal model architecture that performs consistently
well for all the considered smells would allow the implementation of tools
simpler.

\rnn{} has the reputation to perform well with textual data and sequential patterns 
while {\sc cnn} is considered good for imaging data and visual patterns.
Given the similarity of source code and natural language, it is expected to obtain good
performance from \rnn{}.
Our results show that \rnn{} significantly outperforms both {\sc cnn} models  
in the cases of \ecb{} and \mn{}. 
However, in the case of \cm{}, the {\sc cnn} models outperform the \rnn{} whereas in the case of \ma{}, none of the models yield satisfactory results.
These outcomes suggest that there is not one deep learning model that can be used for all kinds of smells. 
We have a uniform model architecture for each model and we observed that
the performance of the model differs significantly for different smells.
It suggests that it is non-trivial, if not impossible, to propose a universal model architecture that
works for all smells.
Each smell exhibits diverse distinctive features and hence their detection mechanisms  differ significantly.
Therefore, given the nature of the problem, it is unlikely that one universal model architecture
will be the silver-bullet for the detection of a wide range of smells.

\subsubsection{Performance comparison with baseline}
It is not feasible to compare the results presented in this paper with other
attempts \cite{Foutse2009b, Foutse2011, Abdou2012, Abdou2012b, Sergio2010, Barbez2019, Fontana2016}
that use machine learning for smell detection due to the following reasons.
First, the replication packages of the related attempts are not available.
Second, for most of the existing attempts, the ratio of positive and negative evaluation
samples is not known; in the absence of this information, we cannot compare them
with our results fairly since the ratio plays an important role in the performance of machine
learning models.
Furthermore, the existing approaches compute metrics and feed them to machine learning
models while we feed tokenized source code.

We compare our results with the results obtained from two baseline
random classifiers that do not really learn from the data but use only
the distribution of smells in the training set to form their
predictions. 
Table \ref{table:discussion.comparison} presents the
comparison. 
The first random classifier generates predictions by
following the training set’s class distribution: that is, for every
item in the evaluation set it predicts whether it is a smell or not
based on the frequency of smells in the training data.
We did that for both balanced and unbalanced evaluation samples to mimic the learning
process of the actual experiment. 
In the middle three columns, referred to as `\textit{{\sc rc} (frequency)}', of the table
we show the results for the balanced setting, 
as they were better than the results for the unbalanced setting.
The second random classifier predicts always that a smell is present;
this gives perfect recall, but low precision, as you can see in the columns corresponding to
`\textit{{\sc rc} (all smells)}' of the table. 
Overall, our models perform far better than a random
classifier for all but \ma{} smell for both baseline variants.

\begin{table}[h!]
	\caption{Comparison of performance (Precision, Recall, and F1)
          with a random classifier (RC) following the training set
          frequencies or responding always indicating a smell} 
	\begin{tabular}{l|lrrr|rrr|rrr}
		&&\multicolumn{9}{c}{Performance}\\
		&&\multicolumn{3}{c}{Our
                   results}&\multicolumn{3}{c}{RC (frequency)} 
		&\multicolumn{3}{c}{RC (all smells)}\\
		& Smells    & P & R & F1 & P    & R & F1 
		 & P    & R & F1   \\ \hline
		\multirow{4}{*}{\cnn{} } & {\sc cm}      & 0.38 & 	0.79	& 0.51 
		&0.03 & 0.50 & 0.05 
		& 0.03 & 1 & 0.05\\
		& {\sc ecb}     & 0.05	& 0.15	& 0.08 
		&0.01 & 0.50 & 0.02 
		& 0.01 & 1 & 0.02\\
		& {\sc mn}      & 0.48	& 0.37	& 0.42 
		&0.11 & 0.50 & 0.18 
		& 0.11 & 1 & 0.20\\
		& {\sc ma}    & 0.01 &	0.04 &	0.02 
		&0.01 & 0.50 & 0.01 
		& 0.01 & 1 & 0.01\\ \hline
		\multirow{4}{*}{\cnntwo{} } & {\sc cm}      & 0.43 &	0.84 &	0.57 
		&0.02 & 0.50 & 0.05 
		& 0.02 & 1 & 0.05\\
		& {\sc ecb}     & 0.04 &	0.12	& 0.06 
		&0.01 & 0.50 & 0.02 
		& 0.01 & 1 & 0.02\\
		& {\sc mn}   & 0.43 & 0.54	& 0.48 
		&0.11 & 0.50 & 0.18 
		& 0.11 & 1 & 0.19\\
		& {\sc ma}     & 0.0	& 0.0	& 0.0 
		&0.01 & 0.50 & 0.01 
		& 0.01 & 1 & 0.01\\ \hline
		\multirow{4}{*}{\rnn{}}    & {\sc cm}       & 0.62 &	0.32 &	0.42 
		&0.03 & 0.50 & 0.05 
		& 0.03 & 1 & 0.05\\
		& {\sc ecb}      & 0.09	& 0.91 &	0.16 
		&0.01 & 0.50 & 0.02 
		& 0.01 & 1 & 0.02\\
		& {\sc mn}     & 0.94 & 0.91& 0.92
		&0.11 & 0.50 & 0.18 
		& 0.11 & 1 & 0.20\\
		& {\sc ma}      & 0.0& 0.08& 0.0
		&0.01 & 0.50 & 0.01 
		& 0.01 & 1 & 0.01\\ \hline
	\end{tabular}
	\label{table:discussion.comparison}
\end{table}

\subsubsection{Poor performance in detecting a design smell}
The presented neural networks perform very poor when it comes to detecting the sole
design smell \ma{}.
We infer the following two reasons for this under-performance.
First, design smells such as \ma{} are inherently difficult to spot unless a deeper semantic
analysis is performed.
Specifically, in the case of \ma{}, interactions among methods of a class as well as the member fields
are required to observe cohesion among the methods which is a non-trivial aspect and
the neural networks could not spot this aspect with the provided input.
Therefore, we need to provide refined semantics information in the form of engineered  features along with the source code to help
neural networks identify the inherent patterns.
Second, the number of positive training samples were very low, thus significantly  restricting our training set. 
The low number severely impacts the ability of neural networks to infer the responsible aspect
that cause the smell.
This limitation can be addressed by increasing the number of repositories under analysis.

\subsubsection{Trading performance with training-time}
As observed in the results section, \rnn{} performs significantly superior than {\sc cnn}
in some specific cases.
However, we also note that \rnn{} models take  considerable more time to train compared to {\sc cnn} models.
We log the time taken by each experiment for the comparison.
Table \ref{table:discussion} presents the average time taken by each model for each smell
per epoch.
The table shows that the \rnn{} performance is coming from much more intense processing
compared to {\sc cnn}.
Therefore, if the performance of \rnn{} and {\sc cnn} is comparable for a given task,
one should go with {\sc cnn}-based solution for significantly faster training time.

\begin{table}[h!]
	\caption{Average training-time taken by the models to train a single epoch in seconds}
	\begin{tabular}{l|rrr}
		& \cnn{}  & \cnntwo{}  & \rnn{}     \\ \hline
		{\sc cm}   & 1.2   & 1.0   & 2,134  \\
		{\sc ecb}  & 0.8   & 0.5   & 1,038  \\
		{\sc mn}   & 3.2   & 3.9   & 5,737  \\
		{\sc ma}   & 0.8   & 4.6  & 2,208
	\end{tabular}
	\label{table:discussion}
\end{table}

\subsection{Opportunities}

The study may encourage the research community to
explore the deep learning methods as a viable option for addressing the problem 
of smell detection. 
Given that we did not consider the context and developers' opinion on smells
reported by deterministic tools, it would be acutely interesting to combine these aspects
either by considering the developers' opinion (by manually tagging the samples)
while segregating positive and negative
samples or by designing
the models that take such opinions as an input to the model.

We have shown that transfer-learning is feasible in the context of code smells. 
This result introduces new directions for automating smell detection which is particularly useful for programming languages
for which smell detection tools are either not available or not matured.

Though this work shows the feasibility of detecting implementation smells;
however, complex smells such as \ma{} require further exploration and present
many open research challenges.
The research community may build on the results presented in this study and further
explore optimizations to the presented models, alternative models, or innovative model architectures
to address the detection of complex design and architecture smells.

Beyond smell detection, proposing an appropriate refactoring to remove a smell is a non-trivial challenge.
There have been some attempts \cite{Tsantalis2018, Biegel2011}
to separate refactoring changes from bug fixes and feature additions.
One may exploit the information produced from such tools and the power of deep learning methods to construct tools that
propose suitable refactoring mechanism.


\section{Threats to Validity}
\label{sec:threats}
Threats to the validity of our work mainly stem from correctness of the employed tools,
our assumption concerning similarity of both the programming languages, and
generalizability and repeatability of the presented results.

\subsection{Construct Validity}
\textit{Construct validity} measures the degree to which tools and metrics
actually measure the properties that they are supposed to measure.
It concerns the appropriateness of observations and inferences
made on the basis of measurements taken during the study.

In the context of using deep learning techniques for smell detection, we use Designite
and DesigniteJava
to detect smells in C\# and Java code respectively and use these results as 
the ground truth.
Relying on the outcome of two different tools may pose a threat to validity especially
in the case of transfer-learning.
To mitigate the risk, we make sure that both the tools use exactly the same set of 
metrics, thresholds, and heuristics to detect smells.
Also, we ensure the smell detection similarity by employing automated as well as manual testing.

To address potential threats posed by representational discrepancies between the two languages we ensure that Tokenizer generates same tokens for same or similar language constructs.
For instance, all the common reserved words are mapped to the same integer token for both the 
programming languages.

\subsection{Internal Validity}
\textit{Internal validity} refers to the validity of the research findings.
It is primarily concerned with controlling the extraneous variables and outside
influences that may impact the outcome. 

In the context of our investigation, exploring the feasibility of applying transfer-learning for
smell detection, we assume that both programming languages are similar by paradigm,
structure, and language constructs.
It would be interesting to observe how two completely different programming languages
(for example, Java and Python) can be combined in a transfer-learning experiment.

\subsection{External Validity}
\textit{External validity} concerns generalizability and repeatability
of the produced results.
The method presented in the study is programming language agnostic and
thus can be repeated for any other programming language given the availability of appropriate
tool-chain.
To encourage the replication and building over this work, we have made all the tools, scripts,
and data available online.\footnote{\url{https://github.com/tushartushar/DeepLearningSmells}}

\section{Conclusions}
\label{sec:conclusions}
The interest in machine learning-based techniques for processing
source code has gained momentum in the recent years.
Despite existing attempts, the community has identified the immaturity of the
discipline for source code processing, especially when it comes to identifying 
quality aspects such as code smells.
In this paper, we establish that deep learning methods can be used for smell detection.
Specifically, we found that {\sc cnn} and \rnn{} deep learning models can be used for
code smell detection, 
though with varying performance.
We did not find a clearly superior method between {\sc 1d} and {\sc 2d} convolution neural networks;
\cnn{} performed slightly better for the smells \ecb{} and \ma{}, while
\cnntwo{} performed superior than its one dimensional counterpart for \cm{} and \mn{}.
Further, our results indicate that \rnn{} performs far better than convolutional networks
for smells \ecb{} and \mn{}.
Our experiment on applying transfer-learning proves the feasibility of practicing transfer-learning
in the context of smell detection, especially for the implementation smells.

With the results presented in the paper we encourage software engineering 
researchers to build over our work as we identify ample opportunities 
for automating smell detection based on deep learning models. 
There are grounds for extending this work to a wider scope by including 
smells belonging to
design and architecture granularities. 
Furthermore, there exist opportunities for further exploiting results and coupling with  deep learning methods for identifying suitable refactoring candidates.
From the practical side, the tool developers may induct the deep learning methods for
effective smell detection and using transfer-learning to detect smells for programming 
languages where the comprehensive code smell detection tools are not available. 

\begin{acks}
	
	This work is partially funded by the {\sc seneca} project, which is part of the Marie Sk\l{}odowska-Curie Innovative Training
	Networks ({\sc itn-eid}) under grant agreement number 642954 
	and by the {\sc CROSSMINER} project, which has received funding from the European
	Union's Horizon 2020 Research and Innovation Programme under 
	grant agreement No. 732223.
	
	We would like to thank Antonis Gkortzis, Theodore Stassinopoulos, and Alexandra Chaniotakis
	for generously contributing effort to our DesigniteJava project.
	
	This work was supported by computational time granted from the National Infrastructures
	for Research and Technology {\sc s.a.} ({\sc grnet s.a.}) 
	in the National {\sc hpc} facility - {\sc aris} - under project {\sc id} pa180903-smells{\sc dl}.
	
\end{acks}

\bibliographystyle{ACM-Reference-Format}
\bibliography{references}


\begin{thebibliography}{100}


\ifx \showCODEN    \undefined \def \showCODEN     #1{\unskip}     \fi
\ifx \showDOI      \undefined \def \showDOI       #1{#1}\fi
\ifx \showISBNx    \undefined \def \showISBNx     #1{\unskip}     \fi
\ifx \showISBNxiii \undefined \def \showISBNxiii  #1{\unskip}     \fi
\ifx \showISSN     \undefined \def \showISSN      #1{\unskip}     \fi
\ifx \showLCCN     \undefined \def \showLCCN      #1{\unskip}     \fi
\ifx \shownote     \undefined \def \shownote      #1{#1}          \fi
\ifx \showarticletitle \undefined \def \showarticletitle #1{#1}   \fi
\ifx \showURL      \undefined \def \showURL       {\relax}        \fi
\providecommand\bibfield[2]{#2}
\providecommand\bibinfo[2]{#2}
\providecommand\natexlab[1]{#1}
\providecommand\showeprint[2][]{arXiv:#2}

\bibitem[\protect\citeauthoryear{Abdeljaber, Avci, Kiranyaz, Gabbouj, and
  Inman}{Abdeljaber et~al\mbox{.}}{2017}]%
        {Abdeljaber2017}
\bibfield{author}{\bibinfo{person}{Osama Abdeljaber}, \bibinfo{person}{Onur
  Avci}, \bibinfo{person}{Serkan Kiranyaz}, \bibinfo{person}{Moncef Gabbouj},
  {and} \bibinfo{person}{Daniel~J Inman}.} \bibinfo{year}{2017}\natexlab{}.
\newblock \showarticletitle{Real-time vibration-based structural damage
  detection using one-dimensional convolutional neural networks}.
\newblock \bibinfo{journal}{\emph{Journal of Sound and Vibration}}
  \bibinfo{volume}{388} (\bibinfo{year}{2017}), \bibinfo{pages}{154--170}.
\newblock


\bibitem[\protect\citeauthoryear{Alexandru, Panichella, and Gall}{Alexandru
  et~al\mbox{.}}{2017}]%
        {alexandru2017}
\bibfield{author}{\bibinfo{person}{Carol~V Alexandru},
  \bibinfo{person}{Sebastiano Panichella}, {and} \bibinfo{person}{Harald~C
  Gall}.} \bibinfo{year}{2017}\natexlab{}.
\newblock \showarticletitle{Replicating parser behavior using neural machine
  translation}. In \bibinfo{booktitle}{\emph{Proceedings of the 25th
  International Conference on Program Comprehension}}. IEEE Press,
  \bibinfo{pages}{316--319}.
\newblock


\bibitem[\protect\citeauthoryear{Allamanis, Barr, Devanbu, and
  Sutton}{Allamanis et~al\mbox{.}}{2018}]%
        {Allamanis2018}
\bibfield{author}{\bibinfo{person}{Miltiadis Allamanis},
  \bibinfo{person}{Earl~T Barr}, \bibinfo{person}{Premkumar Devanbu}, {and}
  \bibinfo{person}{Charles Sutton}.} \bibinfo{year}{2018}\natexlab{}.
\newblock \showarticletitle{A survey of machine learning for big code and
  naturalness}.
\newblock \bibinfo{journal}{\emph{ACM Computing Surveys (CSUR)}}
  \bibinfo{volume}{51}, \bibinfo{number}{4} (\bibinfo{year}{2018}),
  \bibinfo{pages}{81}.
\newblock


\bibitem[\protect\citeauthoryear{Allamanis, Peng, and Sutton}{Allamanis
  et~al\mbox{.}}{2016}]%
        {Allamanis2016}
\bibfield{author}{\bibinfo{person}{Miltiadis Allamanis}, \bibinfo{person}{Hao
  Peng}, {and} \bibinfo{person}{Charles Sutton}.}
  \bibinfo{year}{2016}\natexlab{}.
\newblock \showarticletitle{A convolutional attention network for extreme
  summarization of source code}. In \bibinfo{booktitle}{\emph{International
  Conference on Machine Learning}}. \bibinfo{pages}{2091--2100}.
\newblock


\bibitem[\protect\citeauthoryear{Arnaoudova, Di~Penta, Antoniol, and
  Gu{\'e}h{\'e}neuc}{Arnaoudova et~al\mbox{.}}{2013}]%
        {Venera2013}
\bibfield{author}{\bibinfo{person}{Venera Arnaoudova},
  \bibinfo{person}{Massimiliano Di~Penta}, \bibinfo{person}{Giuliano Antoniol},
  {and} \bibinfo{person}{Yann-Ga{\"e}l Gu{\'e}h{\'e}neuc}.}
  \bibinfo{year}{2013}\natexlab{}.
\newblock \showarticletitle{{A New Family of Software Anti-patterns: Linguistic
  Anti-patterns}}. In \bibinfo{booktitle}{\emph{CSMR '13: Proceedings of the
  2013 17th European Conference on Software Maintenance and Reengineering}}.
  \bibinfo{publisher}{IEEE Computer Society}, \bibinfo{pages}{187--196}.
\newblock


\bibitem[\protect\citeauthoryear{Barbez, Khomh, and Guéhéneuc}{Barbez
  et~al\mbox{.}}{2019}]%
        {Barbez2019}
\bibfield{author}{\bibinfo{person}{Antoine Barbez}, \bibinfo{person}{Foutse
  Khomh}, {and} \bibinfo{person}{Yann-Gaël Guéhéneuc}.}
  \bibinfo{year}{2019}\natexlab{}.
\newblock \bibinfo{title}{A Machine-learning Based Ensemble Method For
  Anti-patterns Detection}.
\newblock
\newblock
\showeprint[arxiv]{cs.SE/1903.01899}


\bibitem[\protect\citeauthoryear{Baziotis, Pelekis, and Doulkeridis}{Baziotis
  et~al\mbox{.}}{2017}]%
        {Baziotis2017}
\bibfield{author}{\bibinfo{person}{Christos Baziotis}, \bibinfo{person}{Nikos
  Pelekis}, {and} \bibinfo{person}{Christos Doulkeridis}.}
  \bibinfo{year}{2017}\natexlab{}.
\newblock \showarticletitle{Datastories at semeval-2017 task 4: Deep lstm with
  attention for message-level and topic-based sentiment analysis}. In
  \bibinfo{booktitle}{\emph{Proceedings of the 11th International Workshop on
  Semantic Evaluation (SemEval-2017)}}. \bibinfo{pages}{747--754}.
\newblock


\bibitem[\protect\citeauthoryear{Bengio, Courville, and Vincent}{Bengio
  et~al\mbox{.}}{2013}]%
        {Bengio2013}
\bibfield{author}{\bibinfo{person}{Yoshua Bengio}, \bibinfo{person}{Aaron
  Courville}, {and} \bibinfo{person}{Pascal Vincent}.}
  \bibinfo{year}{2013}\natexlab{}.
\newblock \showarticletitle{Representation learning: A review and new
  perspectives}.
\newblock \bibinfo{journal}{\emph{IEEE transactions on pattern analysis and
  machine intelligence}} \bibinfo{volume}{35}, \bibinfo{number}{8}
  (\bibinfo{year}{2013}), \bibinfo{pages}{1798--1828}.
\newblock


\bibitem[\protect\citeauthoryear{Biegel, Soetens, Hornig, Diehl, and
  Demeyer}{Biegel et~al\mbox{.}}{2011}]%
        {Biegel2011}
\bibfield{author}{\bibinfo{person}{Benjamin Biegel},
  \bibinfo{person}{Quinten~David Soetens}, \bibinfo{person}{Willi Hornig},
  \bibinfo{person}{Stephan Diehl}, {and} \bibinfo{person}{Serge Demeyer}.}
  \bibinfo{year}{2011}\natexlab{}.
\newblock \showarticletitle{Comparison of Similarity Metrics for Refactoring
  Detection}. In \bibinfo{booktitle}{\emph{Proceedings of the 8th Working
  Conference on Mining Software Repositories}} \emph{(\bibinfo{series}{MSR
  '11})}. \bibinfo{publisher}{ACM}, \bibinfo{pages}{53--62}.
\newblock
\showISBNx{978-1-4503-0574-7}
\urldef\tempurl%
\url{https://doi.org/10.1145/1985441.1985452}
\showDOI{\tempurl}


\bibitem[\protect\citeauthoryear{Bryton, Brito E~Abreu, and Monteiro}{Bryton
  et~al\mbox{.}}{2010}]%
        {Sergio2010}
\bibfield{author}{\bibinfo{person}{S{\'e}rgio Bryton},
  \bibinfo{person}{Fernando Brito E~Abreu}, {and} \bibinfo{person}{Miguel
  Monteiro}.} \bibinfo{year}{2010}\natexlab{}.
\newblock \showarticletitle{{Reducing subjectivity in code smells detection:
  Experimenting with the Long Method}}. In
  \bibinfo{booktitle}{\emph{Proceedings - 7th International Conference on the
  Quality of Information and Communications Technology, QUATIC 2010}}.
  Faculdade de Ciencias e Tecnologia, New University of Lisbon, Caparica,
  Portugal, \bibinfo{publisher}{IEEE}, \bibinfo{pages}{337--342}.
\newblock


\bibitem[\protect\citeauthoryear{Cho, van Merrienboer, Gulcehre, Bahdanau,
  Bougares, Schwenk, and Bengio}{Cho et~al\mbox{.}}{2014}]%
        {Cho2014}
\bibfield{author}{\bibinfo{person}{Kyunghyun Cho}, \bibinfo{person}{Bart van
  Merrienboer}, \bibinfo{person}{Caglar Gulcehre}, \bibinfo{person}{Dzmitry
  Bahdanau}, \bibinfo{person}{Fethi Bougares}, \bibinfo{person}{Holger
  Schwenk}, {and} \bibinfo{person}{Yoshua Bengio}.}
  \bibinfo{year}{2014}\natexlab{}.
\newblock \showarticletitle{Learning Phrase Representations using RNN
  Encoder--Decoder for Statistical Machine Translation}. In
  \bibinfo{booktitle}{\emph{Proceedings of the 2014 Conference on Empirical
  Methods in Natural Language Processing (EMNLP)}}.
  \bibinfo{pages}{1724--1734}.
\newblock


\bibitem[\protect\citeauthoryear{Chollet}{Chollet}{2017}]%
        {Chollet2017}
\bibfield{author}{\bibinfo{person}{Francois Chollet}.}
  \bibinfo{year}{2017}\natexlab{}.
\newblock \bibinfo{booktitle}{\emph{Deep learning with python}}.
\newblock \bibinfo{publisher}{Manning Publications Co.}
\newblock


\bibitem[\protect\citeauthoryear{Czibula, Marian, and Czibula}{Czibula
  et~al\mbox{.}}{2015}]%
        {Gabriela2015}
\bibfield{author}{\bibinfo{person}{Gabriela Czibula},
  \bibinfo{person}{Zsuzsanna Marian}, {and} \bibinfo{person}{Istvan~Gergely
  Czibula}.} \bibinfo{year}{2015}\natexlab{}.
\newblock \showarticletitle{{Detecting software design defects using relational
  association rule mining}}.
\newblock \bibinfo{journal}{\emph{Knowledge and Information Systems}}
  \bibinfo{volume}{42}, \bibinfo{number}{3} (\bibinfo{date}{March}
  \bibinfo{year}{2015}), \bibinfo{pages}{545--577}.
\newblock


\bibitem[\protect\citeauthoryear{de~Andrade, Almeida, and Crnkovic}{de~Andrade
  et~al\mbox{.}}{2014}]%
        {Hugo2014}
\bibfield{author}{\bibinfo{person}{Hugo~Sica de Andrade},
  \bibinfo{person}{Eduardo Almeida}, {and} \bibinfo{person}{Ivica Crnkovic}.}
  \bibinfo{year}{2014}\natexlab{}.
\newblock \showarticletitle{Architectural Bad Smells in Software Product Lines:
  An Exploratory Study}. In \bibinfo{booktitle}{\emph{Proceedings of the WICSA
  2014 Companion Volume}} \emph{(\bibinfo{series}{WICSA '14 Companion})}.
  \bibinfo{publisher}{ACM}, Article \bibinfo{articleno}{12},
  \bibinfo{numpages}{6}~pages.
\newblock
\showISBNx{978-1-4503-2523-3}
\urldef\tempurl%
\url{https://doi.org/10.1145/2578128.2578237}
\showDOI{\tempurl}


\bibitem[\protect\citeauthoryear{Deng, Dong, Socher, Li, Li, and Fei-Fei}{Deng
  et~al\mbox{.}}{2009}]%
        {imagenet_cvpr09}
\bibfield{author}{\bibinfo{person}{J. Deng}, \bibinfo{person}{W. Dong},
  \bibinfo{person}{R. Socher}, \bibinfo{person}{L.-J. Li}, \bibinfo{person}{K.
  Li}, {and} \bibinfo{person}{L. Fei-Fei}.} \bibinfo{year}{2009}\natexlab{}.
\newblock \showarticletitle{{ImageNet: A Large-Scale Hierarchical Image
  Database}}. In \bibinfo{booktitle}{\emph{CVPR09}}.
\newblock


\bibitem[\protect\citeauthoryear{Ernst}{Ernst}{2017}]%
        {Ernst2017}
\bibfield{author}{\bibinfo{person}{Michael~D Ernst}.}
  \bibinfo{year}{2017}\natexlab{}.
\newblock \showarticletitle{Natural language is a programming language:
  Applying natural language processing to software development}. In
  \bibinfo{booktitle}{\emph{LIPIcs-Leibniz International Proceedings in
  Informatics}}, Vol.~\bibinfo{volume}{71}. Schloss Dagstuhl-Leibniz-Zentrum
  fuer Informatik.
\newblock


\bibitem[\protect\citeauthoryear{Felleman and Van~Essen}{Felleman and
  Van~Essen}{1991}]%
        {Felleman1991}
\bibfield{author}{\bibinfo{person}{Daniel~J Felleman} {and}
  \bibinfo{person}{David~C Van~Essen}.} \bibinfo{year}{1991}\natexlab{}.
\newblock \showarticletitle{Distributed Hierarchical Processing in the Primate
  Cerebral Cortex}.
\newblock \bibinfo{journal}{\emph{Cerebral Cortex}} \bibinfo{volume}{1},
  \bibinfo{number}{1} (\bibinfo{year}{1991}), \bibinfo{pages}{1--47}.
\newblock


\bibitem[\protect\citeauthoryear{Fontana, Pigazzini, Roveda, and
  Zanoni}{Fontana et~al\mbox{.}}{2016}]%
        {Fontana2016}
\bibfield{author}{\bibinfo{person}{Francesca~Arcelli Fontana},
  \bibinfo{person}{Ilaria Pigazzini}, \bibinfo{person}{Riccardo Roveda}, {and}
  \bibinfo{person}{Marco Zanoni}.} \bibinfo{year}{2016}\natexlab{}.
\newblock \showarticletitle{Automatic Detection of Instability Architectural
  Smells}. In \bibinfo{booktitle}{\emph{Software Maintenance and Evolution
  (ICSME), 2016 IEEE International Conference on}}. IEEE,
  \bibinfo{pages}{433--437}.
\newblock


\bibitem[\protect\citeauthoryear{Foster, Griswold, and Lerner}{Foster
  et~al\mbox{.}}{2012}]%
        {Foster2012}
\bibfield{author}{\bibinfo{person}{Stephen~R Foster},
  \bibinfo{person}{William~G Griswold}, {and} \bibinfo{person}{Sorin Lerner}.}
  \bibinfo{year}{2012}\natexlab{}.
\newblock \showarticletitle{WitchDoctor: IDE support for real-time
  auto-completion of refactorings}. In \bibinfo{booktitle}{\emph{Software
  Engineering (ICSE), 2012 34th International Conference on}}. IEEE,
  \bibinfo{pages}{222--232}.
\newblock


\bibitem[\protect\citeauthoryear{Fowler}{Fowler}{1999}]%
        {Fowler1999}
\bibfield{author}{\bibinfo{person}{Martin Fowler}.}
  \bibinfo{year}{1999}\natexlab{}.
\newblock \bibinfo{booktitle}{\emph{{Refactoring: Improving the Design of
  Existing Programs}} (\bibinfo{edition}{1} ed.)}.
\newblock \bibinfo{publisher}{Addison-Wesley Professional}.
\newblock
\showISBNx{978-0-201-48567-7}


\bibitem[\protect\citeauthoryear{Fu and Shen}{Fu and Shen}{2015}]%
        {Shizhe2015}
\bibfield{author}{\bibinfo{person}{Shizhe Fu} {and} \bibinfo{person}{Beijun
  Shen}.} \bibinfo{year}{2015}\natexlab{}.
\newblock \showarticletitle{{Code Bad Smell Detection through Evolutionary Data
  Mining}}. In \bibinfo{booktitle}{\emph{International Symposium on Empirical
  Software Engineering and Measurement}}. Shanghai Jiaotong University,
  Shanghai, China, \bibinfo{publisher}{IEEE}, \bibinfo{pages}{41--49}.
\newblock


\bibitem[\protect\citeauthoryear{Fu and Menzies}{Fu and Menzies}{2017}]%
        {Fu2017}
\bibfield{author}{\bibinfo{person}{Wei Fu} {and} \bibinfo{person}{Tim
  Menzies}.} \bibinfo{year}{2017}\natexlab{}.
\newblock \showarticletitle{Easy over hard: A case study on deep learning}. In
  \bibinfo{booktitle}{\emph{Proceedings of the 2017 11th Joint Meeting on
  Foundations of Software Engineering}}. ACM, \bibinfo{pages}{49--60}.
\newblock


\bibitem[\protect\citeauthoryear{{Gal} and {Ghahramani}}{{Gal} and
  {Ghahramani}}{2015}]%
        {Yarin2015}
\bibfield{author}{\bibinfo{person}{Yarin {Gal}} {and} \bibinfo{person}{Zoubin
  {Ghahramani}}.} \bibinfo{year}{2015}\natexlab{}.
\newblock \showarticletitle{{A Theoretically Grounded Application of Dropout in
  Recurrent Neural Networks}}.
\newblock \bibinfo{journal}{\emph{arXiv e-prints}}, Article
  \bibinfo{articleno}{arXiv:1512.05287} (\bibinfo{date}{Dec}
  \bibinfo{year}{2015}), \bibinfo{numpages}{arXiv:1512.05287}~pages.
\newblock
\showeprint[arxiv]{stat.ML/1512.05287}


\bibitem[\protect\citeauthoryear{Garcia, Popescu, Edwards, and
  Medvidovic}{Garcia et~al\mbox{.}}{2009a}]%
        {Joshua2009b}
\bibfield{author}{\bibinfo{person}{Joshua Garcia}, \bibinfo{person}{Daniel
  Popescu}, \bibinfo{person}{George Edwards}, {and} \bibinfo{person}{Nenad
  Medvidovic}.} \bibinfo{year}{2009}\natexlab{a}.
\newblock \showarticletitle{{Identifying Architectural Bad Smells}}. In
  \bibinfo{booktitle}{\emph{CSMR '09: Proceedings of the 2009 European
  Conference on Software Maintenance and Reengineering}}.
  \bibinfo{publisher}{IEEE Computer Society}, \bibinfo{pages}{255--258}.
\newblock


\bibitem[\protect\citeauthoryear{Garcia, Popescu, Edwards, and
  Medvidovic}{Garcia et~al\mbox{.}}{2009b}]%
        {Joshua2009}
\bibfield{author}{\bibinfo{person}{Joshua Garcia}, \bibinfo{person}{Daniel
  Popescu}, \bibinfo{person}{George Edwards}, {and} \bibinfo{person}{Nenad
  Medvidovic}.} \bibinfo{year}{2009}\natexlab{b}.
\newblock \showarticletitle{Toward a Catalogue of Architectural Bad Smells}. In
  \bibinfo{booktitle}{\emph{Proceedings of the 5th International Conference on
  the Quality of Software Architectures: Architectures for Adaptive Software
  Systems}} \emph{(\bibinfo{series}{QoSA '09})}.
  \bibinfo{publisher}{Springer-Verlag}, \bibinfo{pages}{146--162}.
\newblock
\showISBNx{978-3-642-02350-7}
\urldef\tempurl%
\url{https://doi.org/10.1007/978-3-642-02351-4_10}
\showDOI{\tempurl}


\bibitem[\protect\citeauthoryear{Goodfellow, Bengio, Courville, and
  Bengio}{Goodfellow et~al\mbox{.}}{2016}]%
        {Goodfellow2016}
\bibfield{author}{\bibinfo{person}{Ian Goodfellow}, \bibinfo{person}{Yoshua
  Bengio}, \bibinfo{person}{Aaron Courville}, {and} \bibinfo{person}{Yoshua
  Bengio}.} \bibinfo{year}{2016}\natexlab{}.
\newblock \bibinfo{booktitle}{\emph{Deep learning}}. Vol.~\bibinfo{volume}{1}.
\newblock \bibinfo{publisher}{MIT press Cambridge}.
\newblock


\bibitem[\protect\citeauthoryear{Graves, Jaitly, and Mohamed}{Graves
  et~al\mbox{.}}{2013}]%
        {Graves2013}
\bibfield{author}{\bibinfo{person}{Alex Graves}, \bibinfo{person}{Navdeep
  Jaitly}, {and} \bibinfo{person}{Abdel-rahman Mohamed}.}
  \bibinfo{year}{2013}\natexlab{}.
\newblock \showarticletitle{Hybrid speech recognition with deep bidirectional
  LSTM}. In \bibinfo{booktitle}{\emph{Automatic Speech Recognition and
  Understanding (ASRU), 2013 IEEE Workshop on}}. IEEE,
  \bibinfo{pages}{273--278}.
\newblock


\bibitem[\protect\citeauthoryear{Greff, Srivastava, Koutn{\'\i}k, Steunebrink,
  and Schmidhuber}{Greff et~al\mbox{.}}{2017}]%
        {Greff2017}
\bibfield{author}{\bibinfo{person}{Klaus Greff}, \bibinfo{person}{Rupesh~K
  Srivastava}, \bibinfo{person}{Jan Koutn{\'\i}k}, \bibinfo{person}{Bas~R
  Steunebrink}, {and} \bibinfo{person}{J{\"u}rgen Schmidhuber}.}
  \bibinfo{year}{2017}\natexlab{}.
\newblock \showarticletitle{LSTM: A search space odyssey}.
\newblock \bibinfo{journal}{\emph{IEEE transactions on neural networks and
  learning systems}} \bibinfo{volume}{28}, \bibinfo{number}{10}
  (\bibinfo{year}{2017}), \bibinfo{pages}{2222--2232}.
\newblock


\bibitem[\protect\citeauthoryear{Gupta, Pal, Kanade, and Shevade}{Gupta
  et~al\mbox{.}}{2017}]%
        {Gupta2017}
\bibfield{author}{\bibinfo{person}{Rahul Gupta}, \bibinfo{person}{Soham Pal},
  \bibinfo{person}{Aditya Kanade}, {and} \bibinfo{person}{Shirish Shevade}.}
  \bibinfo{year}{2017}\natexlab{}.
\newblock \showarticletitle{DeepFix: Fixing Common C Language Errors by Deep
  Learning.}. In \bibinfo{booktitle}{\emph{AAAI}}. \bibinfo{pages}{1345--1351}.
\newblock


\bibitem[\protect\citeauthoryear{Hellendoorn and Devanbu}{Hellendoorn and
  Devanbu}{2017}]%
        {Hellendoorn2017}
\bibfield{author}{\bibinfo{person}{Vincent~J Hellendoorn} {and}
  \bibinfo{person}{Premkumar Devanbu}.} \bibinfo{year}{2017}\natexlab{}.
\newblock \showarticletitle{Are deep neural networks the best choice for
  modeling source code?}. In \bibinfo{booktitle}{\emph{Proceedings of the 2017
  11th Joint Meeting on Foundations of Software Engineering}}. ACM,
  \bibinfo{pages}{763--773}.
\newblock


\bibitem[\protect\citeauthoryear{Hindle, Barr, Su, Gabel, and Devanbu}{Hindle
  et~al\mbox{.}}{2012}]%
        {Hindle2012}
\bibfield{author}{\bibinfo{person}{Abram Hindle}, \bibinfo{person}{Earl~T
  Barr}, \bibinfo{person}{Zhendong Su}, \bibinfo{person}{Mark Gabel}, {and}
  \bibinfo{person}{Premkumar Devanbu}.} \bibinfo{year}{2012}\natexlab{}.
\newblock \showarticletitle{On the naturalness of software}. In
  \bibinfo{booktitle}{\emph{Software Engineering (ICSE), 2012 34th
  International Conference on}}. IEEE, \bibinfo{pages}{837--847}.
\newblock


\bibitem[\protect\citeauthoryear{Hinton, Osindero, and Teh}{Hinton
  et~al\mbox{.}}{2006}]%
        {Hinton2006}
\bibfield{author}{\bibinfo{person}{Geoffrey~E Hinton}, \bibinfo{person}{Simon
  Osindero}, {and} \bibinfo{person}{Yee-Whye Teh}.}
  \bibinfo{year}{2006}\natexlab{}.
\newblock \showarticletitle{A fast learning algorithm for deep belief nets}.
\newblock \bibinfo{journal}{\emph{Neural computation}} \bibinfo{volume}{18},
  \bibinfo{number}{7} (\bibinfo{year}{2006}), \bibinfo{pages}{1527--1554}.
\newblock


\bibitem[\protect\citeauthoryear{Hochreiter and Schmidhuber}{Hochreiter and
  Schmidhuber}{1997}]%
        {Hochreiter1997}
\bibfield{author}{\bibinfo{person}{Sepp Hochreiter} {and}
  \bibinfo{person}{J{\"u}rgen Schmidhuber}.} \bibinfo{year}{1997}\natexlab{}.
\newblock \showarticletitle{Long short-term memory}.
\newblock \bibinfo{journal}{\emph{Neural computation}} \bibinfo{volume}{9},
  \bibinfo{number}{8} (\bibinfo{year}{1997}), \bibinfo{pages}{1735--1780}.
\newblock


\bibitem[\protect\citeauthoryear{Hubel and Wiesel}{Hubel and Wiesel}{1962}]%
        {Hubel1962}
\bibfield{author}{\bibinfo{person}{David~H Hubel} {and}
  \bibinfo{person}{Torsten~N Wiesel}.} \bibinfo{year}{1962}\natexlab{}.
\newblock \showarticletitle{Receptive fields, binocular interaction and
  functional architecture in the cat's visual cortex}.
\newblock \bibinfo{journal}{\emph{The Journal of physiology}}
  \bibinfo{volume}{160}, \bibinfo{number}{1} (\bibinfo{year}{1962}),
  \bibinfo{pages}{106--154}.
\newblock


\bibitem[\protect\citeauthoryear{Huo, Li, and Zhou}{Huo et~al\mbox{.}}{2016}]%
        {Huo2016}
\bibfield{author}{\bibinfo{person}{Xuan Huo}, \bibinfo{person}{Ming Li}, {and}
  \bibinfo{person}{Zhi-Hua Zhou}.} \bibinfo{year}{2016}\natexlab{}.
\newblock \showarticletitle{Learning Unified Features from Natural and
  Programming Languages for Locating Buggy Source Code.}. In
  \bibinfo{booktitle}{\emph{IJCAI}}. \bibinfo{pages}{1606--1612}.
\newblock


\bibitem[\protect\citeauthoryear{Ioffe and Szegedy}{Ioffe and Szegedy}{2015}]%
        {Ioffe2015}
\bibfield{author}{\bibinfo{person}{Sergey Ioffe} {and}
  \bibinfo{person}{Christian Szegedy}.} \bibinfo{year}{2015}\natexlab{}.
\newblock \showarticletitle{Batch normalization: accelerating deep network
  training by reducing internal covariate shift}. In
  \bibinfo{booktitle}{\emph{Proceedings of the 32nd International Conference on
  International Conference on Machine Learning-Volume 37}}. JMLR. org,
  \bibinfo{pages}{448--456}.
\newblock


\bibitem[\protect\citeauthoryear{Iyer, Konstas, Cheung, and Zettlemoyer}{Iyer
  et~al\mbox{.}}{2016}]%
        {Iyer2016}
\bibfield{author}{\bibinfo{person}{Srinivasan Iyer}, \bibinfo{person}{Ioannis
  Konstas}, \bibinfo{person}{Alvin Cheung}, {and} \bibinfo{person}{Luke
  Zettlemoyer}.} \bibinfo{year}{2016}\natexlab{}.
\newblock \showarticletitle{Summarizing source code using a neural attention
  model}. In \bibinfo{booktitle}{\emph{Proceedings of the 54th Annual Meeting
  of the Association for Computational Linguistics (Volume 1: Long Papers)}},
  Vol.~\bibinfo{volume}{1}. \bibinfo{pages}{2073--2083}.
\newblock


\bibitem[\protect\citeauthoryear{Johnson and Zhang}{Johnson and Zhang}{2015}]%
        {Johnson2015}
\bibfield{author}{\bibinfo{person}{Rie Johnson} {and} \bibinfo{person}{Tong
  Zhang}.} \bibinfo{year}{2015}\natexlab{}.
\newblock \showarticletitle{Effective Use of Word Order for Text Categorization
  with Convolutional Neural Networks}. In \bibinfo{booktitle}{\emph{Proceedings
  of the 2015 Conference of the North American Chapter of the Association for
  Computational Linguistics: Human Language Technologies}}.
  \bibinfo{pages}{103--112}.
\newblock


\bibitem[\protect\citeauthoryear{Kessentini, Kessentini, Sahraoui, Bechikh, and
  Ouni}{Kessentini et~al\mbox{.}}{2014}]%
        {Wael2014}
\bibfield{author}{\bibinfo{person}{Wael Kessentini}, \bibinfo{person}{Marouane
  Kessentini}, \bibinfo{person}{Houari Sahraoui}, \bibinfo{person}{Slim
  Bechikh}, {and} \bibinfo{person}{Ali Ouni}.} \bibinfo{year}{2014}\natexlab{}.
\newblock \showarticletitle{{A Cooperative Parallel Search-Based Software
  Engineering Approach for Code-Smells Detection}}.
\newblock \bibinfo{journal}{\emph{IEEE Transactions on Software Engineering}}
  \bibinfo{volume}{40}, \bibinfo{number}{9} (\bibinfo{year}{2014}),
  \bibinfo{pages}{841--861}.
\newblock


\bibitem[\protect\citeauthoryear{Khomh, Vaucher, Gu{\'e}h{\'e}neuc, and
  Sahraoui}{Khomh et~al\mbox{.}}{2009}]%
        {Foutse2009b}
\bibfield{author}{\bibinfo{person}{Foutse Khomh}, \bibinfo{person}{St{\'e}phane
  Vaucher}, \bibinfo{person}{Yann-Ga{\"e}l Gu{\'e}h{\'e}neuc}, {and}
  \bibinfo{person}{Houari Sahraoui}.} \bibinfo{year}{2009}\natexlab{}.
\newblock \showarticletitle{{A Bayesian Approach for the Detection of Code and
  Design Smells}}. In \bibinfo{booktitle}{\emph{QSIC '09: Proceedings of the
  2009 Ninth International Conference on Quality Software}}.
  \bibinfo{publisher}{IEEE Computer Society}, \bibinfo{pages}{305--314}.
\newblock


\bibitem[\protect\citeauthoryear{Khomh, Vaucher, Gu{\'e}h{\'e}neuc, and
  Sahraoui}{Khomh et~al\mbox{.}}{2011}]%
        {Foutse2011}
\bibfield{author}{\bibinfo{person}{Foutse Khomh}, \bibinfo{person}{St{\'e}phane
  Vaucher}, \bibinfo{person}{Yann-Ga{\"e}l Gu{\'e}h{\'e}neuc}, {and}
  \bibinfo{person}{Houari Sahraoui}.} \bibinfo{year}{2011}\natexlab{}.
\newblock \showarticletitle{{BDTEX: A GQM-based Bayesian approach for the
  detection of antipatterns}}. In \bibinfo{booktitle}{\emph{Journal of Systems
  and Software}}. Ecole Polytechnique de Montreal, Montreal, Canada,
  \bibinfo{pages}{559--572}.
\newblock


\bibitem[\protect\citeauthoryear{Kingma and Ba}{Kingma and Ba}{2014}]%
        {Kingma2014}
\bibfield{author}{\bibinfo{person}{Diederik~P Kingma} {and}
  \bibinfo{person}{Jimmy Ba}.} \bibinfo{year}{2014}\natexlab{}.
\newblock \showarticletitle{Adam: A Method for Stochastic Optimization}.
\newblock \bibinfo{journal}{\emph{arXiv preprint arXiv:1412.6980}}
  (\bibinfo{year}{2014}).
\newblock


\bibitem[\protect\citeauthoryear{Kraus, Ba, and Frey}{Kraus
  et~al\mbox{.}}{2016}]%
        {Kraus2016}
\bibfield{author}{\bibinfo{person}{Oren~Z Kraus}, \bibinfo{person}{Jimmy~Lei
  Ba}, {and} \bibinfo{person}{Brendan~J Frey}.}
  \bibinfo{year}{2016}\natexlab{}.
\newblock \showarticletitle{Classifying and segmenting microscopy images with
  deep multiple instance learning}.
\newblock \bibinfo{journal}{\emph{Bioinformatics}} \bibinfo{volume}{32},
  \bibinfo{number}{12} (\bibinfo{year}{2016}), \bibinfo{pages}{i52--i59}.
\newblock


\bibitem[\protect\citeauthoryear{Krizhevsky and Hinton}{Krizhevsky and
  Hinton}{2009}]%
        {cifar2009}
\bibfield{author}{\bibinfo{person}{Alex Krizhevsky} {and}
  \bibinfo{person}{Geoffrey Hinton}.} \bibinfo{year}{2009}\natexlab{}.
\newblock \bibinfo{booktitle}{\emph{Learning multiple layers of features from
  tiny images}}.
\newblock \bibinfo{type}{{T}echnical {R}eport}.
  \bibinfo{institution}{Citeseer}.
\newblock


\bibitem[\protect\citeauthoryear{Krizhevsky, Sutskever, and Hinton}{Krizhevsky
  et~al\mbox{.}}{2012}]%
        {Krizhevsky2012}
\bibfield{author}{\bibinfo{person}{Alex Krizhevsky}, \bibinfo{person}{Ilya
  Sutskever}, {and} \bibinfo{person}{Geoffrey~E Hinton}.}
  \bibinfo{year}{2012}\natexlab{}.
\newblock \showarticletitle{ImageNet classification with deep convolutional
  neural networks}. In \bibinfo{booktitle}{\emph{Advances in neural information
  processing systems}}. \bibinfo{pages}{1097--1105}.
\newblock


\bibitem[\protect\citeauthoryear{Kruchten, Nord, and Ozkaya}{Kruchten
  et~al\mbox{.}}{2012}]%
        {Philippe2012}
\bibfield{author}{\bibinfo{person}{Philippe Kruchten},
  \bibinfo{person}{Robert~L. Nord}, {and} \bibinfo{person}{Ipek Ozkaya}.}
  \bibinfo{year}{2012}\natexlab{}.
\newblock \showarticletitle{Technical Debt: From Metaphor to Theory and
  Practice}.
\newblock \bibinfo{journal}{\emph{IEEE Software}} \bibinfo{volume}{29},
  \bibinfo{number}{6} (\bibinfo{year}{2012}), \bibinfo{pages}{18--21}.
\newblock
\showISSN{0740-7459}
\urldef\tempurl%
\url{https://doi.org/10.1109/MS.2012.167}
\showDOI{\tempurl}


\bibitem[\protect\citeauthoryear{Lawrence, Giles, Tsoi, and Back}{Lawrence
  et~al\mbox{.}}{1997}]%
        {Lawrence1997}
\bibfield{author}{\bibinfo{person}{Steve Lawrence}, \bibinfo{person}{C~Lee
  Giles}, \bibinfo{person}{Ah~Chung Tsoi}, {and} \bibinfo{person}{Andrew~D
  Back}.} \bibinfo{year}{1997}\natexlab{}.
\newblock \showarticletitle{Face recognition: A convolutional neural-network
  approach}.
\newblock \bibinfo{journal}{\emph{IEEE transactions on neural networks}}
  \bibinfo{volume}{8}, \bibinfo{number}{1} (\bibinfo{year}{1997}),
  \bibinfo{pages}{98--113}.
\newblock


\bibitem[\protect\citeauthoryear{LeCun, Bengio, and Hinton}{LeCun
  et~al\mbox{.}}{2015}]%
        {Lecun2015}
\bibfield{author}{\bibinfo{person}{Yann LeCun}, \bibinfo{person}{Yoshua
  Bengio}, {and} \bibinfo{person}{Geoffrey Hinton}.}
  \bibinfo{year}{2015}\natexlab{}.
\newblock \showarticletitle{Deep learning}.
\newblock \bibinfo{journal}{\emph{nature}} \bibinfo{volume}{521},
  \bibinfo{number}{7553} (\bibinfo{year}{2015}), \bibinfo{pages}{436}.
\newblock


\bibitem[\protect\citeauthoryear{LeCun, Bottou, Bengio, and Haffner}{LeCun
  et~al\mbox{.}}{1998}]%
        {Lecun1998}
\bibfield{author}{\bibinfo{person}{Yann LeCun}, \bibinfo{person}{L{\'e}on
  Bottou}, \bibinfo{person}{Yoshua Bengio}, {and} \bibinfo{person}{Patrick
  Haffner}.} \bibinfo{year}{1998}\natexlab{}.
\newblock \showarticletitle{Gradient-based learning applied to document
  recognition}.
\newblock \bibinfo{journal}{\emph{Proc. IEEE}} \bibinfo{volume}{86},
  \bibinfo{number}{11} (\bibinfo{year}{1998}), \bibinfo{pages}{2278--2324}.
\newblock


\bibitem[\protect\citeauthoryear{LeCun, Cortes, and Burges}{LeCun
  et~al\mbox{.}}{2010}]%
        {mnist2010}
\bibfield{author}{\bibinfo{person}{Yann LeCun}, \bibinfo{person}{Corinna
  Cortes}, {and} \bibinfo{person}{CJ Burges}.} \bibinfo{year}{2010}\natexlab{}.
\newblock \showarticletitle{MNIST handwritten digit database}.
\newblock \bibinfo{journal}{\emph{AT\&T Labs [Online]. Available: http://yann.
  lecun. com/exdb/mnist}}  \bibinfo{volume}{2} (\bibinfo{year}{2010}).
\newblock


\bibitem[\protect\citeauthoryear{Lee, Yoon, and Cho}{Lee et~al\mbox{.}}{2017}]%
        {Lee2017}
\bibfield{author}{\bibinfo{person}{Song-Mi Lee}, \bibinfo{person}{Sang~Min
  Yoon}, {and} \bibinfo{person}{Heeryon Cho}.} \bibinfo{year}{2017}\natexlab{}.
\newblock \showarticletitle{Human activity recognition from accelerometer data
  using Convolutional Neural Network}. In \bibinfo{booktitle}{\emph{Big Data
  and Smart Computing (BigComp), 2017 IEEE International Conference on}}. IEEE,
  \bibinfo{pages}{131--134}.
\newblock


\bibitem[\protect\citeauthoryear{Li, He, Zhu, and Lyu}{Li
  et~al\mbox{.}}{2017}]%
        {Li2017}
\bibfield{author}{\bibinfo{person}{Jian Li}, \bibinfo{person}{Pinjia He},
  \bibinfo{person}{Jieming Zhu}, {and} \bibinfo{person}{Michael~R Lyu}.}
  \bibinfo{year}{2017}\natexlab{}.
\newblock \showarticletitle{Software defect prediction via convolutional neural
  network}. In \bibinfo{booktitle}{\emph{Software Quality, Reliability and
  Security (QRS), 2017 IEEE International Conference on}}. IEEE,
  \bibinfo{pages}{318--328}.
\newblock


\bibitem[\protect\citeauthoryear{Ling, Blunsom, Grefenstette, Hermann,
  Ko{\v{c}}isk{\`y}, Wang, and Senior}{Ling et~al\mbox{.}}{2016}]%
        {Ling2016}
\bibfield{author}{\bibinfo{person}{Wang Ling}, \bibinfo{person}{Phil Blunsom},
  \bibinfo{person}{Edward Grefenstette}, \bibinfo{person}{Karl~Moritz Hermann},
  \bibinfo{person}{Tom{\'a}{\v{s}} Ko{\v{c}}isk{\`y}}, \bibinfo{person}{Fumin
  Wang}, {and} \bibinfo{person}{Andrew Senior}.}
  \bibinfo{year}{2016}\natexlab{}.
\newblock \showarticletitle{Latent Predictor Networks for Code Generation}. In
  \bibinfo{booktitle}{\emph{Proceedings of the 54th Annual Meeting of the
  Association for Computational Linguistics (Volume 1: Long Papers)}},
  Vol.~\bibinfo{volume}{1}. \bibinfo{pages}{599--609}.
\newblock


\bibitem[\protect\citeauthoryear{Lippert and Roock}{Lippert and Roock}{2006}]%
        {Lippert2006}
\bibfield{author}{\bibinfo{person}{Martin Lippert} {and}
  \bibinfo{person}{Stephen Roock}.} \bibinfo{year}{2006}\natexlab{}.
\newblock \bibinfo{booktitle}{\emph{Refactoring in large software projects:
  performing complex restructurings successfully}}.
\newblock \bibinfo{publisher}{John Wiley \& Sons}.
\newblock


\bibitem[\protect\citeauthoryear{Luong, Pham, and Manning}{Luong
  et~al\mbox{.}}{2015}]%
        {Luong2015}
\bibfield{author}{\bibinfo{person}{Thang Luong}, \bibinfo{person}{Hieu Pham},
  {and} \bibinfo{person}{Christopher~D Manning}.}
  \bibinfo{year}{2015}\natexlab{}.
\newblock \showarticletitle{Effective Approaches to Attention-based Neural
  Machine Translation}. In \bibinfo{booktitle}{\emph{Proceedings of the 2015
  Conference on Empirical Methods in Natural Language Processing}}.
  \bibinfo{pages}{1412--1421}.
\newblock


\bibitem[\protect\citeauthoryear{Maas, Daly, Pham, Huang, Ng, and Potts}{Maas
  et~al\mbox{.}}{2011}]%
        {imdb2011}
\bibfield{author}{\bibinfo{person}{Andrew~L. Maas}, \bibinfo{person}{Raymond~E.
  Daly}, \bibinfo{person}{Peter~T. Pham}, \bibinfo{person}{Dan Huang},
  \bibinfo{person}{Andrew~Y. Ng}, {and} \bibinfo{person}{Christopher Potts}.}
  \bibinfo{year}{2011}\natexlab{}.
\newblock \showarticletitle{Learning Word Vectors for Sentiment Analysis}. In
  \bibinfo{booktitle}{\emph{Proceedings of the 49th Annual Meeting of the
  Association for Computational Linguistics: Human Language Technologies}}.
  \bibinfo{publisher}{Association for Computational Linguistics},
  \bibinfo{address}{Portland, Oregon, USA}, \bibinfo{pages}{142--150}.
\newblock
\urldef\tempurl%
\url{http://www.aclweb.org/anthology/P11-1015}
\showURL{%
\tempurl}


\bibitem[\protect\citeauthoryear{Maiga, Ali, Bhattacharya, Saban{\'e},
  Gu{\'e}h{\'e}neuc, and A{\"\i}meur}{Maiga et~al\mbox{.}}{2012a}]%
        {Abdou2012b}
\bibfield{author}{\bibinfo{person}{Abdou Maiga}, \bibinfo{person}{Nasir Ali},
  \bibinfo{person}{Neelesh Bhattacharya}, \bibinfo{person}{Aminata Saban{\'e}},
  \bibinfo{person}{Yann-Ga{\"e}l Gu{\'e}h{\'e}neuc}, {and}
  \bibinfo{person}{Esma A{\"\i}meur}.} \bibinfo{year}{2012}\natexlab{a}.
\newblock \showarticletitle{{SMURF: A SVM-based incremental anti-pattern
  detection approach}}. In \bibinfo{booktitle}{\emph{Proceedings - Working
  Conference on Reverse Engineering, WCRE}}. Ptidej Team,
  \bibinfo{publisher}{IEEE}, \bibinfo{pages}{466--475}.
\newblock


\bibitem[\protect\citeauthoryear{Maiga, Ali, Bhattacharya, Saban{\'e},
  Gu{\'e}h{\'e}neuc, Antoniol, and A{\"\i}meur}{Maiga et~al\mbox{.}}{2012b}]%
        {Abdou2012}
\bibfield{author}{\bibinfo{person}{Abdou Maiga}, \bibinfo{person}{Nasir Ali},
  \bibinfo{person}{Neelesh Bhattacharya}, \bibinfo{person}{Aminata Saban{\'e}},
  \bibinfo{person}{Yann-Ga{\"e}l Gu{\'e}h{\'e}neuc}, \bibinfo{person}{Giuliano
  Antoniol}, {and} \bibinfo{person}{Esma A{\"\i}meur}.}
  \bibinfo{year}{2012}\natexlab{b}.
\newblock \showarticletitle{{Support vector machines for anti-pattern
  detection}}. In \bibinfo{booktitle}{\emph{ASE 2012: Proceedings of the 27th
  IEEE/ACM International Conference on Automated Software Engineering}}.
  Polytechnic School of Montreal, \bibinfo{publisher}{ACM},
  \bibinfo{pages}{278--281}.
\newblock


\bibitem[\protect\citeauthoryear{Marinescu}{Marinescu}{2004}]%
        {Radu2004}
\bibfield{author}{\bibinfo{person}{Radu Marinescu}.}
  \bibinfo{year}{2004}\natexlab{}.
\newblock \showarticletitle{Detection Strategies: Metrics-Based Rules for
  Detecting Design Flaws}. In \bibinfo{booktitle}{\emph{Proceedings of the 20th
  IEEE International Conference on Software Maintenance}}
  \emph{(\bibinfo{series}{ICSM '04})}. \bibinfo{publisher}{IEEE Computer
  Society}, \bibinfo{pages}{350--359}.
\newblock
\showISBNx{0-7695-2213-0}


\bibitem[\protect\citeauthoryear{Marinescu}{Marinescu}{2005}]%
        {Radu2005}
\bibfield{author}{\bibinfo{person}{R Marinescu}.}
  \bibinfo{year}{2005}\natexlab{}.
\newblock \showarticletitle{{Measurement and quality in object-oriented
  design}}. In \bibinfo{booktitle}{\emph{21st IEEE International Conference on
  Software Maintenance (ICSM'05)}}. Universitatea Politehnica din Timisoara,
  Timisoara, Romania, \bibinfo{publisher}{IEEE}, \bibinfo{pages}{701--704}.
\newblock


\bibitem[\protect\citeauthoryear{Martens}{Martens}{2010}]%
        {Martens2010}
\bibfield{author}{\bibinfo{person}{James Martens}.}
  \bibinfo{year}{2010}\natexlab{}.
\newblock \showarticletitle{Deep learning via Hessian-free optimization.}. In
  \bibinfo{booktitle}{\emph{ICML}}, Vol.~\bibinfo{volume}{27}.
  \bibinfo{pages}{735--742}.
\newblock


\bibitem[\protect\citeauthoryear{Moha, Gu{\'{e}}h{\'{e}}neuc, Duchien, and
  Meur}{Moha et~al\mbox{.}}{2010}]%
        {Naouel2010}
\bibfield{author}{\bibinfo{person}{Naouel Moha},
  \bibinfo{person}{Yann{-}Ga{\"{e}}l Gu{\'{e}}h{\'{e}}neuc},
  \bibinfo{person}{Laurence Duchien}, {and}
  \bibinfo{person}{Anne{-}Fran{\c{c}}oise~Le Meur}.}
  \bibinfo{year}{2010}\natexlab{}.
\newblock \showarticletitle{{DECOR:} {A} Method for the Specification and
  Detection of Code and Design Smells}.
\newblock \bibinfo{journal}{\emph{{IEEE} Trans. Software Eng.}}
  \bibinfo{volume}{36}, \bibinfo{number}{1} (\bibinfo{year}{2010}),
  \bibinfo{pages}{20--36}.
\newblock
\urldef\tempurl%
\url{https://doi.org/10.1109/TSE.2009.50}
\showDOI{\tempurl}


\bibitem[\protect\citeauthoryear{Mou, Li, Zhang, Wang, and Jin}{Mou
  et~al\mbox{.}}{2016}]%
        {Mou2016}
\bibfield{author}{\bibinfo{person}{Lili Mou}, \bibinfo{person}{Ge Li},
  \bibinfo{person}{Lu Zhang}, \bibinfo{person}{Tao Wang}, {and}
  \bibinfo{person}{Zhi Jin}.} \bibinfo{year}{2016}\natexlab{}.
\newblock \showarticletitle{Convolutional Neural Networks over Tree Structures
  for Programming Language Processing.}. In \bibinfo{booktitle}{\emph{AAAI}},
  Vol.~\bibinfo{volume}{2}. \bibinfo{pages}{4}.
\newblock


\bibitem[\protect\citeauthoryear{Munaiah, Kroh, Cabrey, and Nagappan}{Munaiah
  et~al\mbox{.}}{2017}]%
        {Munaiah2017}
\bibfield{author}{\bibinfo{person}{Nuthan Munaiah}, \bibinfo{person}{Steven
  Kroh}, \bibinfo{person}{Craig Cabrey}, {and} \bibinfo{person}{Meiyappan
  Nagappan}.} \bibinfo{year}{2017}\natexlab{}.
\newblock \showarticletitle{Curating GitHub for engineered software projects}.
\newblock \bibinfo{journal}{\emph{Empirical Software Engineering}}
  \bibinfo{volume}{22}, \bibinfo{number}{6} (\bibinfo{date}{01 Dec}
  \bibinfo{year}{2017}), \bibinfo{pages}{3219--3253}.
\newblock
\showISSN{1573-7616}
\urldef\tempurl%
\url{https://doi.org/10.1007/s10664-017-9512-6}
\showDOI{\tempurl}


\bibitem[\protect\citeauthoryear{Nguyen, Nguyen, and Nguyen}{Nguyen
  et~al\mbox{.}}{2013}]%
        {Nguyen2013}
\bibfield{author}{\bibinfo{person}{Anh~Tuan Nguyen},
  \bibinfo{person}{Tung~Thanh Nguyen}, {and} \bibinfo{person}{Tien~N Nguyen}.}
  \bibinfo{year}{2013}\natexlab{}.
\newblock \showarticletitle{Lexical statistical machine translation for
  language migration}. In \bibinfo{booktitle}{\emph{Proceedings of the 2013 9th
  Joint Meeting on Foundations of Software Engineering}}. ACM,
  \bibinfo{pages}{651--654}.
\newblock


\bibitem[\protect\citeauthoryear{Nucci, Palomba, Tamburri, Serebrenik, and
  Lucia}{Nucci et~al\mbox{.}}{2018}]%
        {Nucci2018}
\bibfield{author}{\bibinfo{person}{D.~Di Nucci}, \bibinfo{person}{F. Palomba},
  \bibinfo{person}{D.~A. Tamburri}, \bibinfo{person}{A. Serebrenik}, {and}
  \bibinfo{person}{A.~De Lucia}.} \bibinfo{year}{2018}\natexlab{}.
\newblock \showarticletitle{Detecting code smells using machine learning
  techniques: Are we there yet?}. In \bibinfo{booktitle}{\emph{2018 IEEE 25th
  International Conference on Software Analysis, Evolution and Reengineering
  (SANER)}}, Vol.~\bibinfo{volume}{00}. \bibinfo{pages}{612--621}.
\newblock
\urldef\tempurl%
\url{https://doi.org/10.1109/SANER.2018.8330266}
\showDOI{\tempurl}


\bibitem[\protect\citeauthoryear{Oda, Fudaba, Neubig, Hata, Sakti, Toda, and
  Nakamura}{Oda et~al\mbox{.}}{2015}]%
        {Oda2015}
\bibfield{author}{\bibinfo{person}{Yusuke Oda}, \bibinfo{person}{Hiroyuki
  Fudaba}, \bibinfo{person}{Graham Neubig}, \bibinfo{person}{Hideaki Hata},
  \bibinfo{person}{Sakriani Sakti}, \bibinfo{person}{Tomoki Toda}, {and}
  \bibinfo{person}{Satoshi Nakamura}.} \bibinfo{year}{2015}\natexlab{}.
\newblock \showarticletitle{Learning to generate pseudo-code from source code
  using statistical machine translation (t)}. In
  \bibinfo{booktitle}{\emph{Automated Software Engineering (ASE), 2015 30th
  IEEE/ACM International Conference on}}. IEEE, \bibinfo{pages}{574--584}.
\newblock


\bibitem[\protect\citeauthoryear{Ott, Atchison, Harnack, Best, Anderson,
  Firmani, and Linstead}{Ott et~al\mbox{.}}{2018}]%
        {Ott2018}
\bibfield{author}{\bibinfo{person}{Jordan Ott}, \bibinfo{person}{Abigail
  Atchison}, \bibinfo{person}{Paul Harnack}, \bibinfo{person}{Natalie Best},
  \bibinfo{person}{Haley Anderson}, \bibinfo{person}{Cristiano Firmani}, {and}
  \bibinfo{person}{Erik Linstead}.} \bibinfo{year}{2018}\natexlab{}.
\newblock \showarticletitle{Learning Lexical Features of Programming Languages
  from Imagery Using Convolutional Neural Networks}.
\newblock  (\bibinfo{year}{2018}), \bibinfo{pages}{336--339}.
\newblock
\showISBNx{978-1-4503-5714-2}
\urldef\tempurl%
\url{https://doi.org/10.1145/3196321.3196359}
\showDOI{\tempurl}


\bibitem[\protect\citeauthoryear{Ouni, Kula, Kessentini, and Inoue}{Ouni
  et~al\mbox{.}}{2015}]%
        {Ali2015}
\bibfield{author}{\bibinfo{person}{Ali Ouni}, \bibinfo{person}{Raula~Gaikovina
  Kula}, \bibinfo{person}{Marouane Kessentini}, {and} \bibinfo{person}{Katsuro
  Inoue}.} \bibinfo{year}{2015}\natexlab{}.
\newblock \showarticletitle{{Web Service Antipatterns Detection Using Genetic
  Programming}}. In \bibinfo{booktitle}{\emph{GECCO '15: Proceedings of the
  2015 Annual Conference on Genetic and Evolutionary Computation}}. Osaka
  University, \bibinfo{publisher}{ACM}, \bibinfo{pages}{1351--1358}.
\newblock


\bibitem[\protect\citeauthoryear{Palomba, Bavota, Di~Penta, Oliveto,
  Poshyvanyk, and De~Lucia}{Palomba et~al\mbox{.}}{2015}]%
        {Fabio2015c}
\bibfield{author}{\bibinfo{person}{Fabio Palomba}, \bibinfo{person}{Gabriele
  Bavota}, \bibinfo{person}{Massimiliano Di~Penta}, \bibinfo{person}{Rocco
  Oliveto}, \bibinfo{person}{Denys Poshyvanyk}, {and} \bibinfo{person}{Andrea
  De~Lucia}.} \bibinfo{year}{2015}\natexlab{}.
\newblock \showarticletitle{{Mining version histories for detecting code
  smells}}.
\newblock \bibinfo{journal}{\emph{IEEE Transactions on Software Engineering}}
  \bibinfo{volume}{41}, \bibinfo{number}{5} (\bibinfo{date}{May}
  \bibinfo{year}{2015}), \bibinfo{pages}{462--489}.
\newblock


\bibitem[\protect\citeauthoryear{Palomba, Panichella, De~Lucia, Oliveto, and
  Zaidman}{Palomba et~al\mbox{.}}{2016}]%
        {Palomba2016}
\bibfield{author}{\bibinfo{person}{Fabio Palomba}, \bibinfo{person}{Annibale
  Panichella}, \bibinfo{person}{Andrea De~Lucia}, \bibinfo{person}{Rocco
  Oliveto}, {and} \bibinfo{person}{Andy Zaidman}.}
  \bibinfo{year}{2016}\natexlab{}.
\newblock \showarticletitle{{A textual-based technique for Smell Detection}}.
  In \bibinfo{booktitle}{\emph{2016 IEEE 24th International Conference on
  Program Comprehension (ICPC)}}. Universita di Salerno, Salerno, Italy,
  \bibinfo{publisher}{IEEE}, \bibinfo{pages}{1--10}.
\newblock


\bibitem[\protect\citeauthoryear{Parkhi, Vedaldi, Zisserman,
  et~al\mbox{.}}{Parkhi et~al\mbox{.}}{2015}]%
        {Parkhi2015}
\bibfield{author}{\bibinfo{person}{Omkar~M Parkhi}, \bibinfo{person}{Andrea
  Vedaldi}, \bibinfo{person}{Andrew Zisserman}, {et~al\mbox{.}}}
  \bibinfo{year}{2015}\natexlab{}.
\newblock \showarticletitle{Deep face recognition.}. In
  \bibinfo{booktitle}{\emph{BMVC}}, Vol.~\bibinfo{volume}{1}.
  \bibinfo{pages}{6}.
\newblock


\bibitem[\protect\citeauthoryear{Piech, Huang, Nguyen, Phulsuksombati, Sahami,
  and Guibas}{Piech et~al\mbox{.}}{2015}]%
        {Piech2015}
\bibfield{author}{\bibinfo{person}{Chris Piech}, \bibinfo{person}{Jonathan
  Huang}, \bibinfo{person}{Andy Nguyen}, \bibinfo{person}{Mike Phulsuksombati},
  \bibinfo{person}{Mehran Sahami}, {and} \bibinfo{person}{Leonidas Guibas}.}
  \bibinfo{year}{2015}\natexlab{}.
\newblock \showarticletitle{Learning Program Embeddings to Propagate Feedback
  on Student Code}. In \bibinfo{booktitle}{\emph{International Conference on
  Machine Learning}}. \bibinfo{pages}{1093--1102}.
\newblock


\bibitem[\protect\citeauthoryear{Pu, Narasimhan, Solar-Lezama, and Barzilay}{Pu
  et~al\mbox{.}}{2016}]%
        {Pu2016}
\bibfield{author}{\bibinfo{person}{Yewen Pu}, \bibinfo{person}{Karthik
  Narasimhan}, \bibinfo{person}{Armando Solar-Lezama}, {and}
  \bibinfo{person}{Regina Barzilay}.} \bibinfo{year}{2016}\natexlab{}.
\newblock \showarticletitle{sk\_p: a neural program corrector for MOOCs}. In
  \bibinfo{booktitle}{\emph{Companion Proceedings of the 2016 ACM SIGPLAN
  International Conference on Systems, Programming, Languages and Applications:
  Software for Humanity}}. ACM, \bibinfo{pages}{39--40}.
\newblock


\bibitem[\protect\citeauthoryear{Robles}{Robles}{2010}]%
        {Robles2010}
\bibfield{author}{\bibinfo{person}{Gregorio Robles}.}
  \bibinfo{year}{2010}\natexlab{}.
\newblock \showarticletitle{Replicating MSR: A study of the potential
  replicability of papers published in the Mining Software Repositories
  proceedings}. In \bibinfo{booktitle}{\emph{Mining Software Repositories
  (MSR), 2010 7th IEEE Working Conference on}}. IEEE,
  \bibinfo{pages}{171--180}.
\newblock


\bibitem[\protect\citeauthoryear{Rumelhart, Hinton, and Williams}{Rumelhart
  et~al\mbox{.}}{1986}]%
        {Rumelhart1986}
\bibfield{author}{\bibinfo{person}{David~E Rumelhart},
  \bibinfo{person}{Geoffrey~E Hinton}, {and} \bibinfo{person}{Ronald~J
  Williams}.} \bibinfo{year}{1986}\natexlab{}.
\newblock \showarticletitle{Learning representations by back-propagating
  errors}.
\newblock \bibinfo{journal}{\emph{nature}} \bibinfo{volume}{323},
  \bibinfo{number}{6088} (\bibinfo{year}{1986}), \bibinfo{pages}{533}.
\newblock


\bibitem[\protect\citeauthoryear{Sahin, Kessentini, Bechikh, and Deb}{Sahin
  et~al\mbox{.}}{2014}]%
        {Dilan2014}
\bibfield{author}{\bibinfo{person}{Dilan Sahin}, \bibinfo{person}{Marouane
  Kessentini}, \bibinfo{person}{Slim Bechikh}, {and} \bibinfo{person}{Kalyanmoy
  Deb}.} \bibinfo{year}{2014}\natexlab{}.
\newblock \showarticletitle{{Code-Smell Detection as a Bilevel Problem}}.
\newblock \bibinfo{journal}{\emph{ACM Transactions on Software Engineering and
  Methodology (TOSEM)}} \bibinfo{volume}{24}, \bibinfo{number}{1}
  (\bibinfo{date}{Oct.} \bibinfo{year}{2014}), \bibinfo{pages}{6--44}.
\newblock


\bibitem[\protect\citeauthoryear{Sainath, Kingsbury, Saon, Soltau, Mohamed,
  Dahl, and Ramabhadran}{Sainath et~al\mbox{.}}{2015}]%
        {Sainath2015}
\bibfield{author}{\bibinfo{person}{Tara~N Sainath}, \bibinfo{person}{Brian
  Kingsbury}, \bibinfo{person}{George Saon}, \bibinfo{person}{Hagen Soltau},
  \bibinfo{person}{Abdel-rahman Mohamed}, \bibinfo{person}{George Dahl}, {and}
  \bibinfo{person}{Bhuvana Ramabhadran}.} \bibinfo{year}{2015}\natexlab{}.
\newblock \showarticletitle{Deep convolutional neural networks for large-scale
  speech tasks}.
\newblock \bibinfo{journal}{\emph{Neural Networks}}  \bibinfo{volume}{64}
  (\bibinfo{year}{2015}), \bibinfo{pages}{39--48}.
\newblock


\bibitem[\protect\citeauthoryear{Salehie, Li, and Tahvildari}{Salehie
  et~al\mbox{.}}{2006}]%
        {Mazeiar2006}
\bibfield{author}{\bibinfo{person}{Mazeiar Salehie}, \bibinfo{person}{Shimin
  Li}, {and} \bibinfo{person}{Ladan Tahvildari}.}
  \bibinfo{year}{2006}\natexlab{}.
\newblock \showarticletitle{{A Metric-Based Heuristic Framework to Detect
  Object-Oriented Design Flaws}}. In \bibinfo{booktitle}{\emph{ICPC '06:
  Proceedings of the 14th IEEE International Conference on Program
  Comprehension (ICPC'06)}}. University of Waterloo, \bibinfo{publisher}{IEEE
  Computer Society}, \bibinfo{pages}{159--168}.
\newblock


\bibitem[\protect\citeauthoryear{Sharma}{Sharma}{2016}]%
        {Sharma2016}
\bibfield{author}{\bibinfo{person}{Tushar Sharma}.}
  \bibinfo{year}{2016}\natexlab{}.
\newblock \bibinfo{title}{{Designite - A Software Design Quality Assessment
  Tool}}.
\newblock
\newblock
\urldef\tempurl%
\url{https://doi.org/10.5281/zenodo.2566832}
\showDOI{\tempurl}
\newblock
\shownote{http://www.designite-tools.com.}


\bibitem[\protect\citeauthoryear{Sharma}{Sharma}{2018}]%
        {Sharma2018c}
\bibfield{author}{\bibinfo{person}{Tushar Sharma}.}
  \bibinfo{year}{2018}\natexlab{}.
\newblock \bibinfo{title}{DesigniteJava}.
\newblock
\newblock
\urldef\tempurl%
\url{https://doi.org/10.5281/zenodo.2566861}
\showDOI{\tempurl}
\newblock
\shownote{https://github.com/tushartushar/DesigniteJava.}


\bibitem[\protect\citeauthoryear{Sharma}{Sharma}{2019a}]%
        {Sharma2019b}
\bibfield{author}{\bibinfo{person}{Tushar Sharma}.}
  \bibinfo{year}{2019}\natexlab{a}.
\newblock \bibinfo{title}{CodeSplit for C\#}.
\newblock
\newblock
\urldef\tempurl%
\url{https://doi.org/10.5281/zenodo.2566905}
\showDOI{\tempurl}


\bibitem[\protect\citeauthoryear{Sharma}{Sharma}{2019b}]%
        {Sharma2019}
\bibfield{author}{\bibinfo{person}{Tushar Sharma}.}
  \bibinfo{year}{2019}\natexlab{b}.
\newblock \bibinfo{title}{CodeSplitJava}.
\newblock
\newblock
\urldef\tempurl%
\url{https://doi.org/10.5281/zenodo.2566865}
\showDOI{\tempurl}
\newblock
\shownote{https://github.com/tushartushar/CodeSplitJava.}


\bibitem[\protect\citeauthoryear{Sharma, Mishra, and Tiwari}{Sharma
  et~al\mbox{.}}{2016}]%
        {Tushar2016}
\bibfield{author}{\bibinfo{person}{Tushar Sharma}, \bibinfo{person}{Pratibha
  Mishra}, {and} \bibinfo{person}{Rohit Tiwari}.}
  \bibinfo{year}{2016}\natexlab{}.
\newblock \showarticletitle{{Designite --- A Software Design Quality Assessment
  Tool}}. In \bibinfo{booktitle}{\emph{Proceedings of the First International
  Workshop on Bringing Architecture Design Thinking into Developers' Daily
  Activities}} \emph{(\bibinfo{series}{BRIDGE '16})}. \bibinfo{publisher}{ACM}.
\newblock
\urldef\tempurl%
\url{https://doi.org/10.1145/2896935.2896938}
\showDOI{\tempurl}


\bibitem[\protect\citeauthoryear{Sharma and Spinellis}{Sharma and
  Spinellis}{2018}]%
        {Sharma2018b}
\bibfield{author}{\bibinfo{person}{Tushar Sharma} {and}
  \bibinfo{person}{Diomidis Spinellis}.} \bibinfo{year}{2018}\natexlab{}.
\newblock \showarticletitle{A survey on software smells}.
\newblock \bibinfo{journal}{\emph{Journal of Systems and Software}}
  \bibinfo{volume}{138} (\bibinfo{year}{2018}), \bibinfo{pages}{158 -- 173}.
\newblock
\showISSN{0164-1212}
\urldef\tempurl%
\url{https://doi.org/10.1016/j.jss.2017.12.034}
\showDOI{\tempurl}


\bibitem[\protect\citeauthoryear{Spinellis}{Spinellis}{2019}]%
        {Spinellis2019}
\bibfield{author}{\bibinfo{person}{Diomidis Spinellis}.}
  \bibinfo{year}{2019}\natexlab{}.
\newblock \bibinfo{title}{dspinellis/tokenizer: Version 1.1}.
\newblock
\newblock
\urldef\tempurl%
\url{https://doi.org/10.5281/zenodo.2558420}
\showDOI{\tempurl}
\newblock
\shownote{https://github.com/dspinellis/tokenizer.}


\bibitem[\protect\citeauthoryear{Srivastava, Hinton, Krizhevsky, Sutskever, and
  Salakhutdinov}{Srivastava et~al\mbox{.}}{2014}]%
        {Srivastava2014}
\bibfield{author}{\bibinfo{person}{Nitish Srivastava},
  \bibinfo{person}{Geoffrey Hinton}, \bibinfo{person}{Alex Krizhevsky},
  \bibinfo{person}{Ilya Sutskever}, {and} \bibinfo{person}{Ruslan
  Salakhutdinov}.} \bibinfo{year}{2014}\natexlab{}.
\newblock \showarticletitle{Dropout: a simple way to prevent neural networks
  from overfitting}.
\newblock \bibinfo{journal}{\emph{The Journal of Machine Learning Research}}
  \bibinfo{volume}{15}, \bibinfo{number}{1} (\bibinfo{year}{2014}),
  \bibinfo{pages}{1929--1958}.
\newblock


\bibitem[\protect\citeauthoryear{Sundermeyer, Schl{\"u}ter, and
  Ney}{Sundermeyer et~al\mbox{.}}{2012}]%
        {Sundermeyer2012}
\bibfield{author}{\bibinfo{person}{Martin Sundermeyer}, \bibinfo{person}{Ralf
  Schl{\"u}ter}, {and} \bibinfo{person}{Hermann Ney}.}
  \bibinfo{year}{2012}\natexlab{}.
\newblock \showarticletitle{LSTM neural networks for language modeling}. In
  \bibinfo{booktitle}{\emph{Thirteenth annual conference of the international
  speech communication association}}.
\newblock


\bibitem[\protect\citeauthoryear{Suryanarayana, Samarthyam, and
  Sharma}{Suryanarayana et~al\mbox{.}}{2014}]%
        {Girish2014}
\bibfield{author}{\bibinfo{person}{Girish Suryanarayana},
  \bibinfo{person}{Ganesh Samarthyam}, {and} \bibinfo{person}{Tushar Sharma}.}
  \bibinfo{year}{2014}\natexlab{}.
\newblock \bibinfo{booktitle}{\emph{{Refactoring for Software Design Smells:
  Managing Technical Debt}} (\bibinfo{edition}{1} ed.)}.
\newblock \bibinfo{publisher}{Morgan Kaufmann}.
\newblock
\showISBNx{0128013974}


\bibitem[\protect\citeauthoryear{Szegedy, Liu, Jia, Sermanet, Reed, Anguelov,
  Erhan, Vanhoucke, and Rabinovich}{Szegedy et~al\mbox{.}}{2015}]%
        {Szegedy2015}
\bibfield{author}{\bibinfo{person}{Christian Szegedy}, \bibinfo{person}{Wei
  Liu}, \bibinfo{person}{Yangqing Jia}, \bibinfo{person}{Pierre Sermanet},
  \bibinfo{person}{Scott Reed}, \bibinfo{person}{Dragomir Anguelov},
  \bibinfo{person}{Dumitru Erhan}, \bibinfo{person}{Vincent Vanhoucke}, {and}
  \bibinfo{person}{Andrew Rabinovich}.} \bibinfo{year}{2015}\natexlab{}.
\newblock \showarticletitle{Going deeper with convolutions}. In
  \bibinfo{booktitle}{\emph{Proceedings of the IEEE conference on computer
  vision and pattern recognition}}. \bibinfo{pages}{1--9}.
\newblock


\bibitem[\protect\citeauthoryear{Tsantalis and Chatzigeorgiou}{Tsantalis and
  Chatzigeorgiou}{2011}]%
        {Nikolaos2011}
\bibfield{author}{\bibinfo{person}{Nikolaos Tsantalis} {and}
  \bibinfo{person}{Alexander Chatzigeorgiou}.} \bibinfo{year}{2011}\natexlab{}.
\newblock \showarticletitle{Identification of Extract Method Refactoring
  Opportunities for the Decomposition of Methods}.
\newblock \bibinfo{journal}{\emph{Journal of Systems \& Software}}
  \bibinfo{volume}{84}, \bibinfo{number}{10} (\bibinfo{date}{Oct.}
  \bibinfo{year}{2011}), \bibinfo{pages}{1757--1782}.
\newblock
\showISSN{0164-1212}
\urldef\tempurl%
\url{https://doi.org/10.1016/j.jss.2011.05.016}
\showDOI{\tempurl}


\bibitem[\protect\citeauthoryear{Tsantalis, Mansouri, Eshkevari, Mazinanian,
  and Dig}{Tsantalis et~al\mbox{.}}{2018}]%
        {Tsantalis2018}
\bibfield{author}{\bibinfo{person}{Nikolaos Tsantalis}, \bibinfo{person}{Matin
  Mansouri}, \bibinfo{person}{Laleh~M. Eshkevari}, \bibinfo{person}{Davood
  Mazinanian}, {and} \bibinfo{person}{Danny Dig}.}
  \bibinfo{year}{2018}\natexlab{}.
\newblock \showarticletitle{Accurate and Efficient Refactoring Detection in
  Commit History}. In \bibinfo{booktitle}{\emph{Proceedings of the 40th
  International Conference on Software Engineering}}
  \emph{(\bibinfo{series}{ICSE '18})}. \bibinfo{publisher}{ACM},
  \bibinfo{pages}{483--494}.
\newblock
\showISBNx{978-1-4503-5638-1}
\urldef\tempurl%
\url{https://doi.org/10.1145/3180155.3180206}
\showDOI{\tempurl}


\bibitem[\protect\citeauthoryear{Vasilescu, Casalnuovo, and Devanbu}{Vasilescu
  et~al\mbox{.}}{2017}]%
        {Vasilescu2017}
\bibfield{author}{\bibinfo{person}{Bogdan Vasilescu}, \bibinfo{person}{Casey
  Casalnuovo}, {and} \bibinfo{person}{Premkumar Devanbu}.}
  \bibinfo{year}{2017}\natexlab{}.
\newblock \showarticletitle{Recovering clear, natural identifiers from
  obfuscated JS names}. In \bibinfo{booktitle}{\emph{Proceedings of the 2017
  11th Joint Meeting on Foundations of Software Engineering}}. ACM,
  \bibinfo{pages}{683--693}.
\newblock


\bibitem[\protect\citeauthoryear{Vidal, Marcos, and D{\'\i}az-Pace}{Vidal
  et~al\mbox{.}}{2014}]%
        {Santiago2014}
\bibfield{author}{\bibinfo{person}{Santiago~A Vidal}, \bibinfo{person}{Claudia
  Marcos}, {and} \bibinfo{person}{J~Andr{\'e}s D{\'\i}az-Pace}.}
  \bibinfo{year}{2014}\natexlab{}.
\newblock \showarticletitle{{An approach to prioritize code smells for
  refactoring}}.
\newblock \bibinfo{journal}{\emph{Automated Software Engineering}}
  \bibinfo{volume}{23}, \bibinfo{number}{3} (\bibinfo{year}{2014}),
  \bibinfo{pages}{501--532}.
\newblock


\bibitem[\protect\citeauthoryear{Wang, Huang, Zhao, et~al\mbox{.}}{Wang
  et~al\mbox{.}}{2016}]%
        {Wang2016}
\bibfield{author}{\bibinfo{person}{Yequan Wang}, \bibinfo{person}{Minlie
  Huang}, \bibinfo{person}{Li Zhao}, {et~al\mbox{.}}}
  \bibinfo{year}{2016}\natexlab{}.
\newblock \showarticletitle{Attention-based lstm for aspect-level sentiment
  classification}. In \bibinfo{booktitle}{\emph{Proceedings of the 2016
  conference on empirical methods in natural language processing}}.
  \bibinfo{pages}{606--615}.
\newblock


\bibitem[\protect\citeauthoryear{Wei and Li}{Wei and Li}{2017}]%
        {Wei2017}
\bibfield{author}{\bibinfo{person}{Huihui Wei} {and} \bibinfo{person}{Ming
  Li}.} \bibinfo{year}{2017}\natexlab{}.
\newblock \showarticletitle{Supervised Deep Features for Software Functional
  Clone Detection by Exploiting Lexical and Syntactical Information in Source
  Code.}. In \bibinfo{booktitle}{\emph{IJCAI}}. \bibinfo{pages}{3034--3040}.
\newblock


\bibitem[\protect\citeauthoryear{Wen, Gasic, Mrk{\v{s}}i{\'c}, Su, Vandyke, and
  Young}{Wen et~al\mbox{.}}{2015}]%
        {Wen2015}
\bibfield{author}{\bibinfo{person}{Tsung-Hsien Wen}, \bibinfo{person}{Milica
  Gasic}, \bibinfo{person}{Nikola Mrk{\v{s}}i{\'c}}, \bibinfo{person}{Pei-Hao
  Su}, \bibinfo{person}{David Vandyke}, {and} \bibinfo{person}{Steve Young}.}
  \bibinfo{year}{2015}\natexlab{}.
\newblock \showarticletitle{Semantically Conditioned LSTM-based Natural
  Language Generation for Spoken Dialogue Systems}. In
  \bibinfo{booktitle}{\emph{Proceedings of the 2015 Conference on Empirical
  Methods in Natural Language Processing}}. \bibinfo{pages}{1711--1721}.
\newblock


\bibitem[\protect\citeauthoryear{White, Tufano, Vendome, and Poshyvanyk}{White
  et~al\mbox{.}}{2016}]%
        {White2016}
\bibfield{author}{\bibinfo{person}{Martin White}, \bibinfo{person}{Michele
  Tufano}, \bibinfo{person}{Christopher Vendome}, {and} \bibinfo{person}{Denys
  Poshyvanyk}.} \bibinfo{year}{2016}\natexlab{}.
\newblock \showarticletitle{Deep learning code fragments for code clone
  detection}. In \bibinfo{booktitle}{\emph{Proceedings of the 31st IEEE/ACM
  International Conference on Automated Software Engineering}}. ACM,
  \bibinfo{pages}{87--98}.
\newblock


\bibitem[\protect\citeauthoryear{White, Vendome, Linares-V{\'a}squez, and
  Poshyvanyk}{White et~al\mbox{.}}{2015}]%
        {White2015}
\bibfield{author}{\bibinfo{person}{Martin White}, \bibinfo{person}{Christopher
  Vendome}, \bibinfo{person}{Mario Linares-V{\'a}squez}, {and}
  \bibinfo{person}{Denys Poshyvanyk}.} \bibinfo{year}{2015}\natexlab{}.
\newblock \showarticletitle{Toward deep learning software repositories}. In
  \bibinfo{booktitle}{\emph{Proceedings of the 12th Working Conference on
  Mining Software Repositories}}. IEEE Press, \bibinfo{pages}{334--345}.
\newblock


\bibitem[\protect\citeauthoryear{Yin and Neubig}{Yin and Neubig}{2017}]%
        {Yin2017}
\bibfield{author}{\bibinfo{person}{Pengcheng Yin} {and} \bibinfo{person}{Graham
  Neubig}.} \bibinfo{year}{2017}\natexlab{}.
\newblock \showarticletitle{A Syntactic Neural Model for General-Purpose Code
  Generation}. In \bibinfo{booktitle}{\emph{Proceedings of the 55th Annual
  Meeting of the Association for Computational Linguistics (Volume 1: Long
  Papers)}}, Vol.~\bibinfo{volume}{1}. \bibinfo{pages}{440--450}.
\newblock


\end{thebibliography}

\end{document}